\documentclass[times]{elsarticle}
\makeatletter
\def\ps@pprintTitle{%
 \let\@oddhead\@empty
 \let\@evenhead\@empty
 \def\@oddfoot{\centerline{\thepage}}%
 \let\@evenfoot\@oddfoot}
\makeatother
\usepackage[utf8]{inputenc}
\usepackage{amsmath}
\usepackage{indentfirst}
\usepackage{subfig}
\usepackage{courier}
\usepackage{graphicx}
\usepackage{algorithm,algorithmic}
\usepackage{geometry}
\usepackage{dsfont}
\usepackage[monochrome]{color}
\usepackage{multirow}
\usepackage{verbatim}
\usepackage[justification=centering]{caption}
\usepackage{siunitx}
\usepackage{longtable,tabularx}
\usepackage{threeparttable}
\usepackage{makecell}

\usepackage{multirow}
\usepackage{hhline}
\usepackage{textcomp}
\usepackage{gensymb}

\newcommand{\beq}{\begin{eqnarray}}
\newcommand{\eeq}{\end{eqnarray}}
\newcommand{\bal}{\begin{eqnarray}\begin{aligned}}
\newcommand{\eal}{\end{aligned}\end{eqnarray}}

\newcommand{\re}{\mathds{R}}

\newcommand{\RNum}[1]{\uppercase\expandafter{\romannumeral #1\relax}}
\newcommand{\eqr}[1]{Eq.~\eqref{#1}}

\makeatletter
\newcommand\notsotiny{\@setfontsize\notsotiny{5.7}{6.7}}
\makeatother

\newcommand{\argmin}[1]{\underset{#1}{\operatorname{arg}\,\operatorname{min}}\;}

\title{Multi-level Convolutional Autoencoder Networks for Parametric Prediction of Spatio-temporal Dynamics}
\author{Jiayang Xu\footnote{PhD Candidate, Dept. of Aerospace Engineering, davidxu@umich.edu.},  
Karthik Duraisamy\footnote{Associate Professor, Dept. of Aerospace Engineering, kdur@umich.edu.}} 

\address{University of Michigan, Ann Arbor, MI, 48109}

\begin{document}

\begin{abstract}
    A data-driven framework is proposed towards the end of predictive modeling of complex spatio-temporal dynamics, leveraging nested non-linear manifolds. Three levels of  neural networks are used, with the goal of predicting the future state of a system of interest in a parametric setting. A convolutional autoencoder is used as the top level to encode the high dimensional input data  along spatial dimensions into a sequence of  latent variables. A temporal convolutional autoencoder (TCAE) serves as the second level, which further encodes the output sequence from the first level along the temporal dimension, and outputs a set of latent variables that encapsulate the spatio-temporal evolution of the dynamics. {\color {red}The use of dilated temporal convolutions  grows the receptive field  exponentially with network depth, allowing for efficient processing of long temporal sequences typical of  scientific computations.} A fully-connected network is used as the third level to learn the mapping between these latent variables and the global parameters from training data, and predict them for new parameters. For future state predictions, the second level uses a temporal convolutional network to predict subsequent steps of the output sequence from the top level. Latent variables at the bottom-most level are decoded to obtain the dynamics in physical space at new global parameters and/or at a future time. Predictive capabilities are evaluated on a range of problems  involving discontinuities, wave propagation, strong transients, and coherent structures. The sensitivity of the results to different modeling choices is assessed. The results suggest that given adequate data and careful training, effective data-driven predictive models can be constructed. Perspectives are provided on the present approach and its place in the  landscape of model reduction.
\end{abstract}

\maketitle

\section{Introduction}
Efficient and accurate prediction of complex spatio-temporal processes remains a problem of significant scientific and industrial value. Numerically well-resolved solutions of partial differential equations offer considerable insight into  underlying physical processes, but continue to be prohibitively expensive in many query scenarios such as design, optimization and uncertainty quantification. As a consequence, reduced order and surrogate modeling approaches have emerged as a important avenue for research. In these approaches, an off-line stage aims to extract problem-specific low-dimensional information (such as a projection basis or a latent space) such that relatively inexpensive computations can be performed during the on-line predictive stage. 

In projection-based Reduced Order Models (ROMs)~\cite{benner2015survey}, the most popular method to construct a low-dimensional {\em linear} subspace is the truncated proper orthogonal decomposition (POD)~\cite{berkooz1993proper, rowley2004model}, typically applied to a collection of solution snapshots. The governing equations of the full order model are then projected on to the lower dimensional subspace by choosing an appropriate test basis. Other linear basis construction methods include balanced truncation~\cite{moore1981principal, safonov1989schur}, reduced basis methods~\cite{peterson1989reduced, prud2001reliable, rozza2007reduced}, rational interpolation~\cite{baur2011interpolatory}, and proper generalized decomposition~\cite{de2013basis, berger2016estimation}.   Linear basis ROMs have achieved considerable success in complex problems such as turbulent flows~\cite{rowley2005model, carlberg2011efficient,hijazi2019data} and combustion instabilities~\cite{xu2019reduced, huang2018exploration}.  However, despite the choice of optimal test spaces afforded by Petrov–Galerkin methods~\cite{carlberg2017galerkin}, and closure modeling~\cite{wang2011two, wang2012proper, parish2018adjoint, gouasmi2017priori}, the linear trial space becomes ineffective in advection-dominated problems and many multiscale problems in general. While the associated challenges can be addressed to a certain degree by using adaptive basis~\cite{peherstorfer2015online, peherstorfer2018model, amsallem2011online}, some of the fundamental challenges persist.

To overcome some of the limitations of the choice of a linear trial space, researchers have pursued the extraction of nonlinear trial manifolds. In Ref.~\cite{hartman2017deep},  neural network-based compression in the form of an autoencoder~\cite{demers1993non} instead of POD in the development of a ROM for dynamical systems. Similar approaches have been applied to a range of fluid dynamic problems, including flow over airfoil~\cite{omata2019novel}, reacting flow~\cite{lee2018model}, and improvement in accuracy has been reported, when compared to the use of linear bases. It should also be noted that POD methods do not provide basis functions that are compact spatially and/or temporally, and as a consequence, non-ideal for convection and transport-dominated problems. Neural networks with convolutional layers are capable of representing local features, offering promise in more efficiently representing multi-scale dynamics and convection-dominated problems. Indeed, the advantages of using nonlinear and local manifolds have been demonstrated~\cite{guo2016convolutional,puligilla2018deep,carlberg2019recovering}. 

Independent of the type of basis that is employed, ROM approaches can be broadly categorized into \textit{intrusive} and \textit{non-intrusive} methods. In intrusive ROMs, the full order governing equations are projected onto the reduced dimensional manifold using Galerkin~\cite{rowley2004model, couplet2005calibrated} and  Petrov Galerkin~\cite{carlberg2011efficient, parish2018adjoint} formulations. To achieve computational efficiency in intrusive ROMs of complex non-linear PDEs, additional approximations are required.    Sampling approaches such as  missing point estimation (MPE)~\cite{astrid2004reduction, astrid2008missing}, and empirical interpolation methods~\cite{barrault2004empirical, chaturantabut2009discrete, drmac2016new} have been developed to to restricting the computation of nonlinear terms to a subset of the state variables. Such methods introduce additional complexity and require careful treatment such as adaptive sampling and basis~\cite{peherstorfer2015online, peherstorfer2018model} and oversampling~\cite{peherstorfer2018stabilizing}. It has to be mentioned that effective sampling approaches have not been developed for non-linear manifolds.

As an alternate to intrusive  ROMs, latent data structures  have been exploited to directly infer ROM equations from data. In Ref.~\cite{brunton2016discovering}, sparse system identification  is performed based on a pre-collected dictionary of nonlinear equations. In Ref.~\cite{gu2011qlmor}, the idea of reducing arbitrary nonlinearity to quadratic-bilinearity via lifting is introduced, and applied to reacting flows in Refs.~\cite{kramer2019nonlinear, kramer2019balanced}. Techniques for automated refinement and inference have also been proposed~\cite{schmidt2011automated, daniels2015automated}.

Non-intrusive methods bypass the governing equations and process full order model solutions to develop  data-driven surrogate models. Such methods rely on interpolation~\cite{barthelmann2000high} or regression~\cite{guo2018reduced} operations. In this vein, neural networks have been applied to problems with arbitrary non-linearity~\cite{hesthaven2018non, wang2019non}. With recent advances in time-series processing techniques, the future state of reduced order variables~\cite{mohan2019compressed} and the full field~\cite{lee2018data} have also been directly predicted. Time-series prediction has been popularly addressed using recurrent neural networks (RNN)~\cite{rumelhart1988learning} or long-short term memory networks (LSTM)~\cite{hochreiter1997lstm}.  In Ref.~\cite{gonzalez2018deep}, state variables are  compressed using an autoencoder and the resulting dynamics learned and predicted using a RNN. A similar approach is taken in Ref.~\cite{mohan2019compressed}, with a LSTM network replacing the RNN.  Besides direct prediction of state variables, such techniques have also been applied in ROM closures~\cite{maulik2020time}.

In this work, we leverage neural networks for compression, convolution and regression towards the end of non-intrusive model reduction.  An autoencoder~\cite{demers1993non} consists of an encoder part that  compresses the high dimensional input into low dimensional latent variables, and a decoder part that  reconstructs the original high dimensional input  from the encoded latent variables. The encoder and decoder are trained jointly, yet can be used separately. By using nonlinear layers in an autoencoder, nonlinear model reduction can be performed. Convolutional neural networks (CNN) have been widely applied to image processing and achieved remarkable success.  In the context of scientific computing, the localized nature of the kernel-based convolutional operations enables CNNs to identify and process coherent dynamics in arbitrary parts of the computational domain, which is difficult to achieve using global bases. \textcolor{red}{Convolutional autoencoders have been demonstrated an effective approach for spatial field data compression~\cite{hartman2017deep, guo2016convolutional, lee2018model, carlberg2019recovering}. It should be mentioned that there are also convolutional models that are able to process spatial and temporal dimensions simultaneously, such as the ConvLSTM~\cite{xingjian2015convolutional} and the spatio-temporal convolution~\cite{yu2017spatio}, and have been successfully applied to popular deep learning tasks such as natural language processing and video generation. However, complex spatio-temporal systems in scientific computing often exhibit a much larger data size per frame, therefore such models cannot be easily applied.}

\textcolor{red}{For time series modeling tasks, the family of recurrent networks, e.g. the basic RNN~\cite{elman1990finding}, LSTM~\cite{hochreiter1997long}, GRU~\cite{cho2014properties}, are currently the most popular choice. Such networks process the time series data in a sequential manner, and recurrently updates a vector of hidden states at every input step of the time series data. Tremendous successes have been achieved with recurrent networks, applications including natural language processing~\cite{wang2016attention, wu2016google}, time series forecasting~\cite{qing2018hourly, rather2015recurrent}, automatic music composition~\cite{eck2002first}, etc. Recently,  temporal convolutional networks (TCN)~\cite{oord2016wavenet} has been proposed for time series modeling. The TCN uses causal convolutions, which only operates on data before the current element, to process data in the temporal order. To handle long sequences, dilated convolution is used such that the receptive field grows exponentially with the depth of the network. This feature is especially helpful in practical engineering computations, which may require processing and predictions of thousands of frames. In Ref.~\cite{bai2018empirical}, a comprehensive evaluation is conducted across a diverse range of sequence modeling tasks, and TCNs are shown to  perform favorably compared to recurrent networks and also demonstrate a longer effective memory length, despite the theoretical advantage of unlimited memory for recurrent models. Other advantages of the TCN demonstrated in Ref.~\cite{bai2018empirical} include parallelism, stable gradients and low memory requirement in training. TCNs have proved to be a promising alternative in multiple canonical applications for recurrent models~\cite{gehring2016convolutional, lea2017temporal, dauphin2017language}. However, to our knowledge, TCNs have not been exploited in scientific computing applications.}

  
 We propose a framework that uses multiple levels of neural networks, namely spatial and temporal convolutional neural networks, autoencoders and multi-layer perceptrons to predict spatio-temporal dynamics in a parametric setting. The rest of this paper is organized as follows: \textcolor{red}{An overview of the framework is provided in Sec.~\ref{sec framework}. Detailed neural network architectures and related neural network techniques are introduced in Sec.~\ref{sec network}.}  Numerical tests are presented in Sec.~\ref{sec tests}. Perspectives on the present approach, and its place  in the larger landscape of model reduction is presented in Sec.~\ref{sec perspectives}.  A summary is given in Sec.~\ref{sec conclusion}.  Sensitivity to various modeling choices is further explored in the appendices.

\section{Goals and Framework Overview}\label{sec framework}
We assume that the spatio-temporal process is represented by successive time snapshots of spatial field variables on a fixed uniform Cartesian grid. Further, the snapshots are evenly spaced in time. In each frame, the spatial field is characterized by several variables of interest, e.g. pressure, density, velocity components, etc. 
The variables at time index $i$ and their depedendence on parameters $\mu \in \re^\mu$ is denoted by $\mathbf{q}(i;\mu)\in\re^{n}$, where $n=(\prod_{j}n_j)n_{var}$ is the total degrees of freedom per time step, $n_j$ is the number of grid points in direction $j\in{1,2,3}$, and $n_{var}$ is the number of variables. The  sequence for a given parameter $\mu$ is denoted by $\mathbf{Q}(\mu)=[\mathbf{q}(1;\mu), \dots, \mathbf{q}(n_t;{\mu})]\in\re^{n\times n_t}$, where $n_t$ is the total number of time steps in the sequence. 

The proposed framework targets three types of tasks, namely 1) prediction at new global parameters,  2) prediction for future time steps, and 3) a combination of the above. In the first task,  sequences at several known parameters in a certain time period are provided as training data, and the time series at an unseen parameter in the same time period is predicted. The required input for this type of task is a new parameter $\mu^*$,  and the target output is the corresponding sequence $\mathbf{Q}(\mu^*)$. In the task of prediction for future time steps, the framework incrementally predicts one step in the future. By performing such predictions iteratively, multiple subsequent steps can be obtained. The required input is a sequence $\mathbf{Q}(\mu)$, and the target output is $\{\mathbf{q}(n_t+1;\mu), \dots, \mathbf{q}(n_t^*;{\mu})\}$\textcolor{red}{, where $n_t^*$ is the target final time step to be predicted.} To achieve this, the number of prediction iterations to be performed is $n_p=n_t^*-n_t$.
By combining the first two types of tasks, the framework can predict a time series at unseen parameters, and beyond the training time period. For simplicity, we will refer to the aforementioned tasks as \textit{new parameter prediction}, \textit{future state prediction} and \textit{combined prediction}, respectively. 

\textcolor{red}{Physical processes are typically associated with coherent structures.  In this setting, convolutional neural networks are appropriate as they have the capability to efficiently process local features. Further, properties such as translational and rotational invariance can be naturally encoded. As the main interest of our work is in spatio-temporal dynamics, we employ convolutional operators in both space and time. The framework consists of three levels of neural networks. The rest of this section will provide an overview of the framework, and a detailed introduction to network components and terms is provided in Sec.~\ref{sec network}.}

\begin{figure}
	\centering
	\subfloat{
		\includegraphics[width=0.9\textwidth]{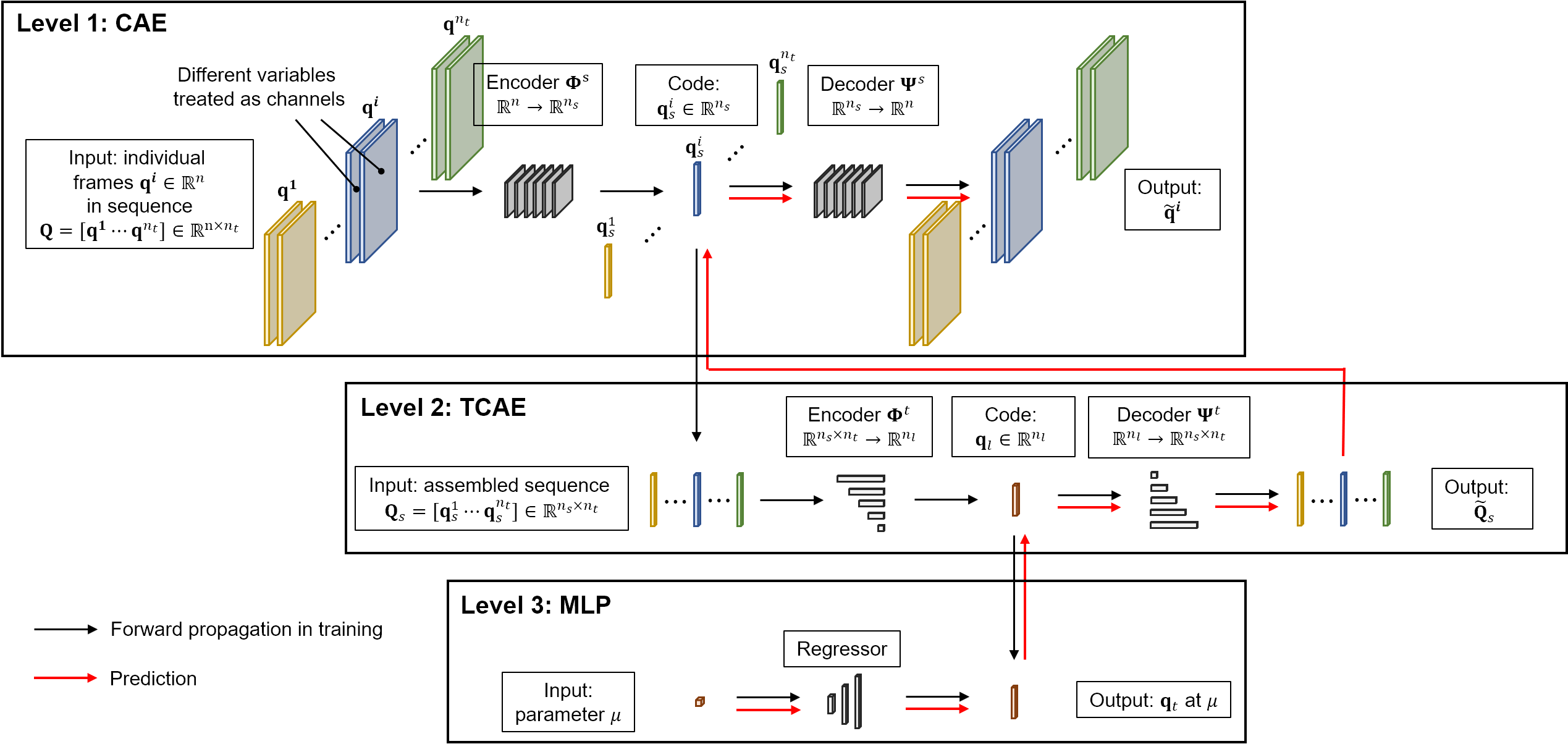}}
	\caption{Schematic of new parameter prediction}\label{fig pipeline1} 
\end{figure}

\begin{figure}
	\centering
	\subfloat{
		\includegraphics[width=0.9\textwidth]{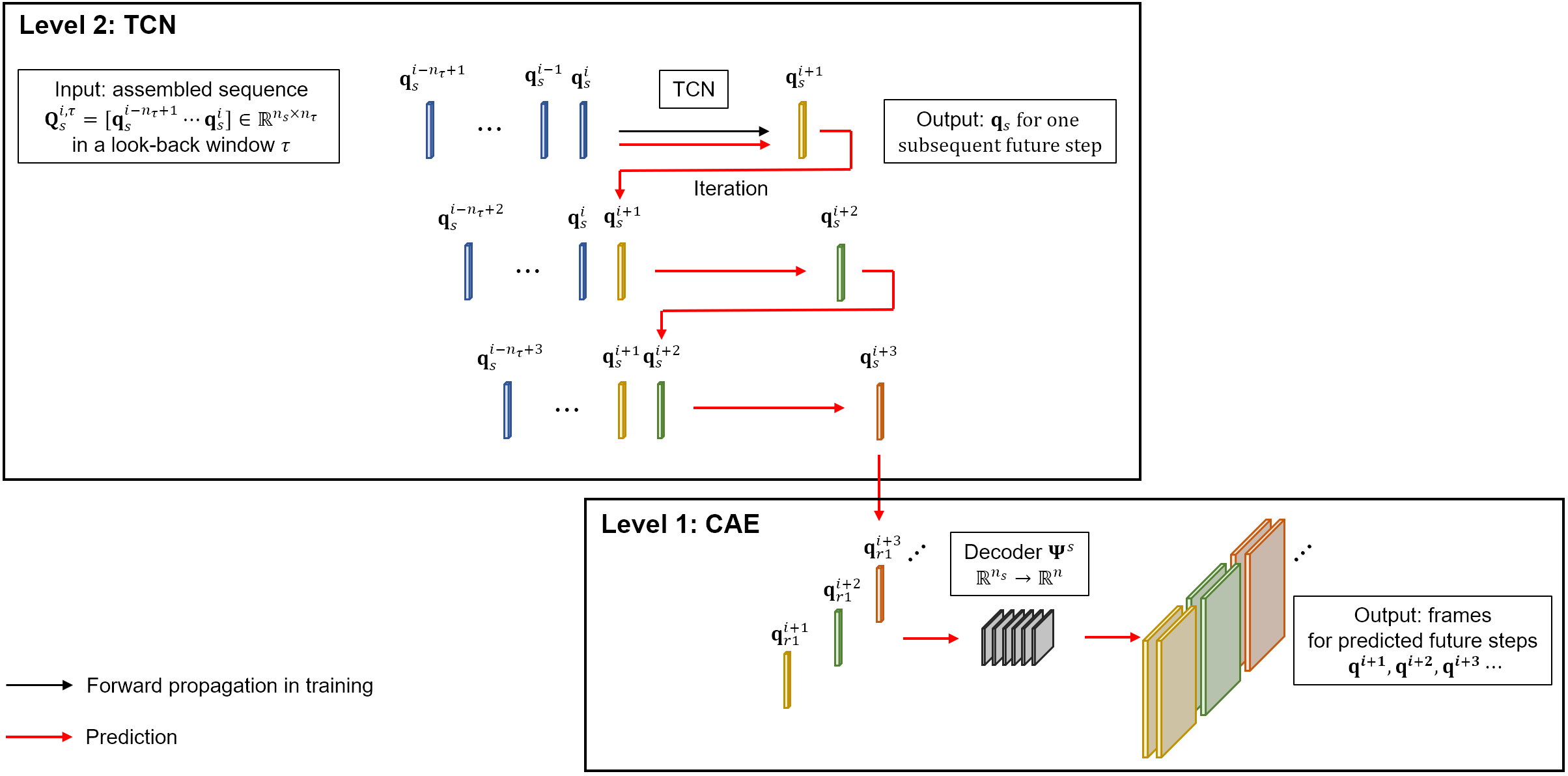}}
	\caption{Schematic for future state prediction}\label{fig pipeline2} 
\end{figure}

\textcolor{red}{Pipeline diagrams for different tasks are provided in Fig.~\ref{fig pipeline1} and ~\ref{fig pipeline2}.} For both tasks, the same top level is shared, which is a convolutional autoencoder (CAE) that encodes the high-dimensional data sequence along spatial dimensions into a sequence of latent variables $\mathbf{Q}_s(\mu)=[\mathbf{q}_s(1;\mu),\dots, \mathbf{q}_s(n_t;\mu)]\in\re^{n_s\times n_t}$, where $n_s$ is the number of  latent variables $\mathbf{q}_s(i;\mu)$ at one time step. Following the CAE, TCNs with different architectures serve as the second level to process the output sequence. \textcolor{red}{As shown in Fig.~\ref{fig pipeline1}, in} new parameter prediction, a temporal convolutional autoencoder (TCAE) is used to  encode $\mathbf{Q}_s$ along the temporal dimension, and outputs a second set of latent variables $\mathbf{q}_l(\mu)\in\re^{n_l}$, which is the encoded spatio-temporal evolution of the flow field. A MLP is used as the third level to learn the mapping between $\mathbf{q}_l$ and $\mu$ from training data, and predict $\mathbf{q}_l(\mu^*)$ for a new parameter $\mu^*$. 
 
\textcolor{red}{For future state predictions, a TCN is used as the second level after the CAE. As shown in Fig.~\ref{fig pipeline2}, TCN is used iteratively to predict future steps $\{\mathbf{q}_s(n_t+1;\mu), \dots, \mathbf{q}_s(n_t^*;{\mu})\}$ subsequent to the output sequence $\mathbf{Q}_s(\mu)$ from the top level.}

In either type of task, outputs at the bottom level are decoded to obtain the high-dimensional sequence at unseen global parameters and/or future states, as indicated by the red arrows in the diagrams.

\section{\textcolor{red}{Network Details}}\label{sec network}
\textcolor{red}{The proposed framework is comprised of multiple levels of neural networks. Following the order of integration, the basic network layers are introduced in Sec.~\ref{sec layer}, baseline network structures are introduced in Sec.~\ref{sec structure}, the constituent sub-level networks of the proposed framework are introduced in Sec.~\ref{sec level}, pipelines for different tasks are described in Sec.~\ref{sec pipeline}, and additional details of the training process are provided in Sec.~\ref{sec training}.}

\subsection{Layer structures}\label{sec layer}
In this work, three types of layers are used, namely \textit{dense}, \textit{convolutional} and \textit{dilated convolutional} layers. 

\subsubsection{Dense layer}
The functional form of a dense layer is given by
\begin{equation}\label{eq dense}
f_{\text{dense}}(\mathbf{h};\mathbf{\Theta})=\sigma(g(\mathbf{h};\mathbf{\Theta})),
\end{equation}
where $g(\mathbf{h}; \mathbf{\Theta})\triangleq\mathbf{Wh+b}$ is a linear function, and $\sigma$ is an element-wise \textit{activation function}, which can be nonlinear. The parameter $\mathbf{\Theta}\in\re^{n_g\times (n_h+1)}$ consists of a bias part $\mathbf{b}\in\re^{n_g}$, and a weight part $\mathbf{W}\in\re^{n_g\times n_h}$.
If all layers in a network are fully connected, the network is called a \textcolor{red}{multilayer perceptron (MLP)}, which is the most basic type of neural network. In our work, nonlinear regression is performed using MLP with nonlinear activation functions.

\subsubsection{Convolutional layers}
A convolutional layer convolves filters with trainable weights with the inputs. Such filters are commonly referred to as \textit{convolutional kernels}. In a convolutional neural network, the inputs and outputs can have multiple  \textit{channels}, i.e. multiple sets of variables defined on the same spatial grid such as the red, green and blue components in a digital colored image. 
For a convolutional layer with $n_{ci}$ input channels and $n_{co}$ output channels, the total number of convolutional kernels is $n_k=n_{ci}\times n_{co}$. Each kernel slides over the one input channel along all spatial directions, and dot products are computed at all sliding steps.
The functional form of a convolutional layer is given by 
\begin{equation}
f_{\text{conv}}(\mathbf{h};\mathbf{k})= \sigma((\mathbf{h}*\mathbf{k})).\label{eq conv}
\end{equation}

In two-dimensional problems, at a sliding step centered at $(i_x,i_y)$, the 2D convolutional dot product of a kernel $\mathbf{k}\in\re^{(2w_x+1)\times(2w_y+1)}$ on an input $\mathbf{x}$ is given by
\begin{equation}\label{eq conv2D}
(\mathbf{x}*\mathbf{k})_{i_x,i_y}\triangleq\sum_{p=w_x}^{-w_x}{\sum_{q=w_y}^{-w_y}{\mathbf{x}_{i_x-p,i_y-q}\mathbf{k}_{p,q}}}.
\end{equation}

\textcolor{red}{\eqr{eq conv2D} can} be easily generalized for 1D or 3D as in \eqr{eq conv1D} and (\ref{eq conv3D})

\begin{eqnarray}
&(\mathbf{x}*\mathbf{k})_{i_x}\triangleq\sum_{p=w_x}^{-w_x}{\mathbf{x}_{i_x-p}\mathbf{k}_{p}},\label{eq conv1D}\\
&(\mathbf{x}*\mathbf{k})_{i_x,i_y,i_z}\triangleq\sum_{p=w_x}^{-w_x}{\sum_{q=w_y}^{-w_y}{\sum_{r=w_z}^{-w_z}{\mathbf{x}_{i_x-p,i_y-q,i_z-r}\mathbf{k}_{p,q,r}}}}.\label{eq conv3D}
\end{eqnarray}

\subsubsection{Dilated convolutional layer}
\textcolor{red}{In standard convolutions, the receptive field, i.e. the range in input that each output element is dependent on, growths linearly with the number of layers and the kernel size. This becomes a major disadvantage when applied to long sequential data, leading to a need for
an extremely deep network or large kernels. As a solution, dilated convolution is performed in TCN.}

The functional form of a dilated convolutional layer is given by 
\begin{equation}
f_{\text{conv},d}(\mathbf{h};\mathbf{k})= \sigma((\mathbf{h}*_d\mathbf{k})).\label{eq dilated conv}
\end{equation}

In one dimension, the dilated convolution is a modified convolution operation with dilated connectivity between the input and the kernel, given by 
\begin{equation}
(\mathbf{x}*_d\mathbf{k})_{i}\triangleq \sum_{p=0}^{w}{\mathbf{x}_{i-dp}\mathbf{k}_{p}},
\end{equation}
where $d$ is called \textit{dilation order}, and $w$ is the 1D kernel size. \textcolor{red}{By increasing $d$ exponentially with the depth of the network, the receptive field also grows exponentially thus long sequences can be efficiently processed.} A visualization of a TCN with multiple dilated convolutional layers of different dilation orders can be found in Ref.~\cite{oord2016wavenet}.

It should be realized that the dilation only affects the kernel connectivity and does not change the output shape, and hence, a sequence processed through it will remain a sequence. To compress the sequence dimension using TCN, the dilated convolution will  be performed in a \textit{strided} manner. In the strided dilated convolution, a stride of $wd$ is used, which means that the kernel will only slide through every $wd$ elements. Thus, any element between the first and last element convolved by the kernel will not be convolved with any other kernel. In contrast, the non-strided dilated convolution will slide through all elements. By using a strided convolution, the output size shrinks after being processed by every layer, and finally only one number is  output for each channel.

\subsection{Network structures}\label{sec structure}
\subsubsection{Feed-forward network}
\textcolor{red}{The feed-forward network is the simplest neural network structure, in which layers are interconnected in a feed-forward, i.e. sequential way. The computation in the $l$-th layer of a network takes the form}
\begin{equation}\label{eq hidden}
\mathbf{h}^{(l)}\triangleq f^{(l)}(\mathbf{h}^{(l-1)};\mathbf{\Theta}^{(l)}),
\end{equation}
where $\mathbf{h}^{(0)}=\mathbf{x}\in\re^{n_x}$ is the input layer, $\mathbf{h}^{(L)}=\tilde{\mathbf{y}}\in\re^{n_y}$ is the output layer, and $\mathbf{\Theta}^{(l)}$ is a set of trainable parameters. Thus the $L$-layer feed-forward neural network can be expressed using a nonlinear function $\mathcal{F}:\re^{n_x}\rightarrow\re^{n_y}$ as
\begin{equation}\label{eq feedforward}
\tilde{\mathbf{y}}=\mathcal{F}(\mathbf{x};\mathbf{\Theta})\triangleq f^{(L)}\circ f^{(L-1)}\circ\cdots\circ f^{(2)}\circ f^{(1)}(\mathbf{x};\mathbf{\Theta}^{(1)}).
\end{equation}

\eqr{eq feedforward} is a serial process, thus if the feed-forward network is cut after the $l$-th layer $\mathbf{h}^{(l)}$ into two parts, with a leading part $$\mathbf{\Phi}(\mathbf{x};\mathbf{\Theta}_{{\Phi}})\triangleq f^{(l)}\circ f^{(l-1)}\circ\cdots\circ f^{(2)}\circ f^{(1)}(\mathbf{x};\mathbf{\Theta}^{(1)}),$$ and a following part $$\mathbf{\Psi}(\mathbf{h}^{(l)};\mathbf{\Theta}_{{\Psi}})\triangleq f^{(L)}\circ f^{(L-1)}\circ\cdots\circ f^{(l+2)}\circ f^{(l+1)}(\mathbf{h}^{(l)};\mathbf{\Theta}^{(l+1)}).$$ 

\textcolor{red}{Then} the output of the full network can be computed from two steps as
\begin{eqnarray}
&\mathbf{y}_l=\mathbf{\Phi}(\mathbf{x}; \mathbf{\Theta}_{\phi}),\label{eq encoder}\\
&\mathbf{\tilde{y}}=\mathbf{\Psi}(\mathbf{y}_l; \mathbf{\Theta}_{\psi}),\label{eq decoder}
\end{eqnarray}

\noindent with $\mathbf{\Phi}:\re^{n_y}\rightarrow\re^{n^{(l)}}$ and $\mathbf{\Psi}:\re^{n^{(l)}}\rightarrow\re^{n_y}$.

\subsubsection{Autoencoder}
The intermediate variable $\mathbf{y}_l$ can be viewed as a set of latent representations for the full variable $\mathbf{y}$. When $\mathbf{y}_l$ is computed from the original input $\mathbf{x}$ using $\mathbf{\Phi}$, it is equal to $\mathbf{h}^{(l)}$ in the uncut network. Alternatively, if $\mathbf{y}_l$ can be obtained from other methods, the output $\mathbf{\tilde{y}}$ can be directly computed from it from the second step, \eqr{eq decoder} without the first step, \eqr{eq encoder}.

An autoencoder is a type of feedforward network with two main characteristics: 1) The output is a reconstruction of the input, i.e. $\mathbf{x}\approx \mathbf{y}$; 2) Autoencoders - in general - assume a converging-diverging shape, i.e. the size of output first reduces then increases along the hidden layers. By cutting an autoencoder after its ``bottleneck'', i.e. the hidden layer with the smallest size, the leading part $\mathbf{\Phi}$ will compress $\mathbf{y}$ into $\mathbf{y}_l$ with the dimension reduced from $n_y$ to $n^{(l)}$, and the following part $\mathbf{\Psi}$ will try to recover $\mathbf{y}$ from $\mathbf{y}_l$.  

In an autoencoder, $\mathbf{\Phi}$ is called a \textit{encoder}, and the transformation to the latent space is called \textit{encoding}. $\mathbf{\Psi}$ is referred to as a \textit{decoder}, and the 
reconstruction process is called \textit{decoding}. The latent variable $\mathbf{y}_l$ is commonly referred to as \textit{code}, and $n^{(l)}$ is called \text{latent dimension}. 


\subsection{Constituent levels}\label{sec level}
\subsubsection{\textcolor{red}{CAE}}
The CAE performs encoding-decoding along the spatial dimensions of the individual frames $\mathbf{q}(i;\mu)$. The encoding and decoding operations are expressed as
\begin{eqnarray}
&\mathbf{q}_s(i;\mu)=\mathbf{\Phi}_s(\mathbf{q}(i;\mu); \mathbf{\Theta}_{\phi s}),\label{eq convencoder}\\
&\mathbf{\tilde{q}}(i;\mu)=\mathbf{\Psi}_s(\mathbf{q}_s(i;\mu); \mathbf{\Theta}_{\psi s}),\label{eq convdecoder}
\end{eqnarray}
 where the encoder  $\mathbf{\Phi}_s:\re^{n}\rightarrow\re^{n_s}$ is  parameterized by $\mathbf{\Theta}_{\phi s}$ and the decoder $\mathbf{\Psi}_s:\re^{n_s}\rightarrow\re^{n}$ is parameterized by $\mathbf{\Theta}_{\psi s}$.

The encoder $\mathbf{\Phi}_s$ begins with several convolution-pooling (Conv-pool) blocks. Each block consists of a 2D/3D convolutional layer with PReLU~\cite{he2015delving} activation, followed by a max-pooling layer that down-samples and reduces the size of the inputs. Zero-padding is applied to the convolutional layers to ensure consistent dimensionality between the inputs and outputs of the convolutional operation. The different variables in the flow field data are treated as different channels in the input layer (i.e. the convolutional layer in the first block). This treatment enables the network to process an arbitrary number of variables of interest without substantial change in structure. Following the Conv-pool operation, blocks are fully-connected (dense) layers with PReLU activation. Before the output layer (i.e. the last dense layer) of $\mathbf{\Phi}_s$, a flattening layer is added, which makes the output encoded latent variable $\mathbf{q}_s$ to be a flattened 1D variable. 

\begin{figure}
	\centering
	\subfloat{
		\includegraphics[width=0.9\textwidth]{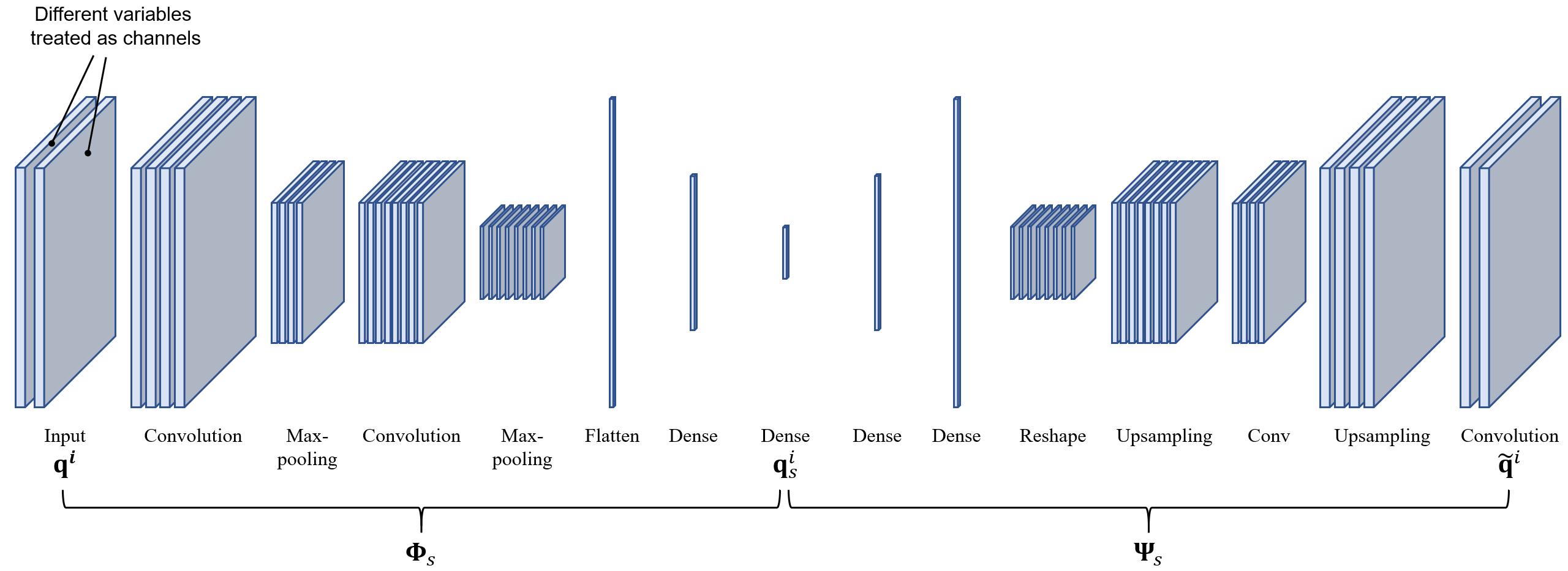}}
	\caption{\textcolor{red}{Sample CAE architecture}. Leftmost and rightmost slabs represent a spatial field.}\label{figl1} 
\end{figure}

The decoder $\mathbf{\Psi}_s$ takes the inverse structure of the encoder, with the Conv-pool blocks replaced by transposed convolution (Conv-trans) layers~\cite{dumoulin2016guide}. In the output layer of $\mathbf{\Psi}_s$, (i.e. the convolutional layer in the last transposed convolution layer), linear activation is used instead of the PReLU activation in other layers. \textcolor{red}{A schematic of a sample CAE architecture with two Conv-pool blocks and two dense layers is given in Fig.~\ref{figl1}}.

\subsubsection{\textcolor{red}{TCAE (for new parameter prediction)}}
In new parameter prediction, a TCAE is used to further compress the temporal dimension, such that each sequence $\mathbf{Q}_s(\mu)=[\mathbf{q}_s(1,\mu), \dots, \mathbf{q}_s({n_t},\mu)]\in\re^{n_s\times n_t}$ is encoded into one set of latent variables $\mathbf{q}_l(\mu)\in\re^{n_l}$. Thus, the total number of training samples for the TCAE equals the number of samples in parameter space  $n_{\mu}$. For the sequence for a single parameter, the encoding and decoding operations are 
\begin{eqnarray}
&\mathbf{q}_l(\mu)=\mathbf{\Phi}_l(\mathbf{Q}_s(\mu);\mathbf{\Theta}_{\phi l}),\label{eq tcnencoder} \\
&\mathbf{\tilde{Q}}_s(\mu)=\mathbf{\Psi}_l(\mathbf{q}_l(\mu);\mathbf{\Theta}_{\psi l}),\label{eq tcndecoder} 
\end{eqnarray}

\begin{figure}
	\centering
	\subfloat{
		\includegraphics[width=0.9\textwidth]{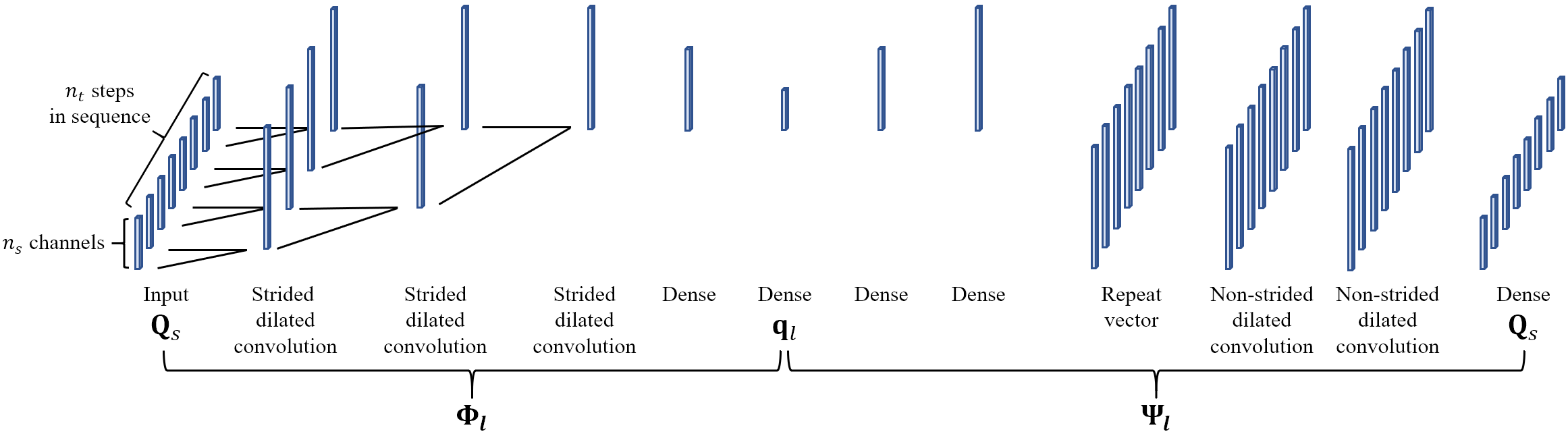}}
	\caption{\textcolor{red}{Sample TCAE architecture}}\label{fig l2} 
\end{figure}

\textcolor{red}{The encoder $\mathbf{\Phi}_l:\re^{n_s\times n_t}\rightarrow\re^{n_l}$ is essentially a TCN by itself, where $\mathbf{Q}_s$ is processed as sequences of length $n_t$ for $n_s$ channels. For each of the channels, strided dilated convolutions are performed along the temporal dimension, and all steps in the sequence are integrated into one number at the last convolution layer.} At this point, the size of the intermediate result is $n_s$ and the temporal dimension is eliminated. After the convolution layers, several dense layers with PReLU activation are used to further compress the intermediate variable into the output code $\mathbf{q}_l$. The decoder $\mathbf{\Psi}_l:\re^{n_l}\rightarrow\re^{n_s\times n_t}$ the inverted structure of $\mathbf{\Phi}_l$, but with non-strided dilated convolution layers and a dense layer with hyperbolic tangent (tanh) activation added to the end as the output layer. \textcolor{red}{A schematic of a sample TCAE architecture with three convolution layers and two dense layers} in $\mathbf{\Phi}_l$ is shown in Fig.~\ref{fig l2}.

\subsubsection{\textcolor{red}{MLP (for new parameter prediction)}}
The first two levels of encoding reduces the dimension of the data  from $n \times n_t$ to $n_{l}$. The third level learns the mapping between the latent variable $\mathbf{q}_l(\mu)$ and the parameters $\mu$, and to predict $\mathbf{q}_l(\mu^*)$ for unseen $\mu^*$. This is achieved by performing regression using a MLP with parameters $\mathbf{\Theta}_{R}$. The regression process is represented as
\begin{equation}\label{eq mlp}
\tilde{\mathbf{q}}_l(\mu)=\mathcal{R}(\mu;\mathbf{\Theta}_{R}).
\end{equation}

The MLP consists of multiple dense layers, where tanh activation is used in the last one and PReLU is used in the rest. 

\subsubsection{\textcolor{red}{TCN} (for future state prediction)}
\textcolor{red}{For future state prediction, the lower levels are different from that for prediction for new parameters. 
The necessity of the change of the lower-level network architecture is determined by the difference in training and prediction scopes for future state prediction from those for the previous task. More specifically, in this task, the goal is to afford flexibility in forecasting arbitrary number of future steps based on the  temporal history patterns. Thus, the training data can be a sequence of arbitrary length instead of multiple fixed-length sequences at different global parameters. As a consequence of such differences, the aforementioned level 3 MLP is not required and the second level is replaced by a TCN serving as a future step predictor $\mathcal{P}:\re^{n_s\times n_{\tau}}\rightarrow\re^{n_s}$.}

\textcolor{red}{
The time-series prediction using the TCN is performed iteratively. For each future step, the prediction is based on the temporal memory via a \textit{look-back window} of $n_\tau$ steps, i.e. $n_{\tau}$ leading steps, $\mathbf{Q}_s(i;\mu;n_\tau)=[\mathbf{q}_s(i-n_\tau+1;\mu),\dots, \mathbf{q}_s(i;\mu)]$, which is represented by
}
\begin{equation}\label{eq tcn predictor} 
    \tilde{\mathbf{q}}_s({i+1};\mu)=\mathcal{P}(\mathbf{Q}_s(i;\mu;\tau);\mathbf{\Theta}_P).
\end{equation}

\textcolor{red}{In general, two types of dynamics may exist in the temporal setting:} 

\textcolor{red}{1. In the first type of dynamics, future states at a spatial point is completely determined by the temporal pattern at that point locally. This is analogous to standing waves. To process such dynamics, the TCN performs strided dilated convolutions along the temporal dimension.}

\textcolor{red}{2. The second type of dynamics involves spatial propagation, thus states of neighboring points are necessary in the prediction. This is analogous to traveling waves. For this type of dynamics, convolutions are performed along the latent dimension of $\mathbf{Q}_s$.}

\begin{figure}
	\centering
	\subfloat{
		\includegraphics[width=0.9\textwidth]{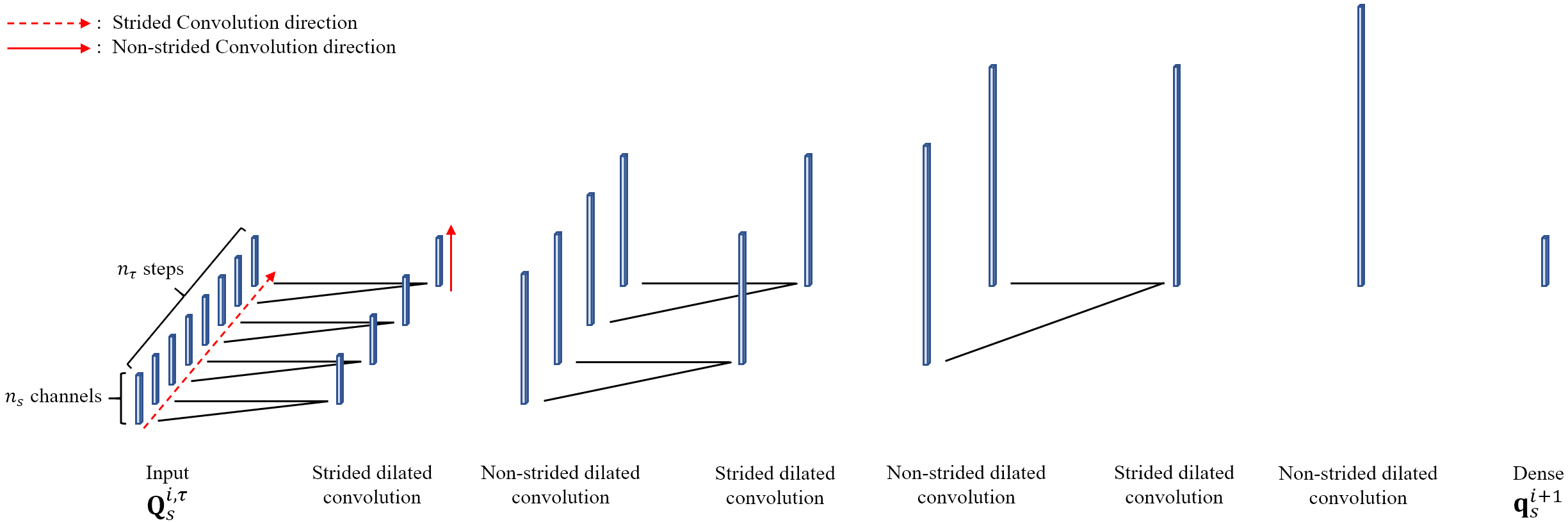}}
	\caption{\textcolor{red}{Sample TCN architecture}}\label{fig l3} 
\end{figure}

\textcolor{red}{A schematic of a sample TCN architecture with convolutions along both spatial and temporal dimensions is shown in Fig.~\ref{fig l3}. It can be seen that the architecture of the predictor $\mathcal{P}$ is similar to the encoder $\mathbf{\Phi}_l$ in the TCAE described previously. The major difference between $\mathcal{P}$ and $\mathbf{\Phi}_l$ is in their outputs and training processes. Instead of being jointly trained with a decoder part and outputting an automatically encoded variable, the output of $\mathcal{P}$ has a more definitive meaning, which is the next latent variable subsequent to the input look-back sequence.}

\textcolor{red}{It should be noted that for simplicity and to demonstrate the primary characteristics in Fig.~\ref{fig l3}, the strided convolutions maintain the number of channels, and the non-strided convolutions do not affect the sequence length, which not necessary in the practical design of the TCN architecture.  As demonstrated in Sec.~\ref{sec tests}, either type of convolution can be used along in the TCN when only one type of dynamics exist in the system.}

\subsection{Framework pipeline}\label{sec pipeline}
\textcolor{red}{Table~\ref{table io} summarizes the inputs and outputs of each component in the framework pipelines. It should be noted that \textit{scaling} ($\mathcal{S}_{\sigma}, \mathcal{S}_{\text{R}}, \mathcal{S}_{\text{SR}}$) is also reflected in the same table, which is described in detail in Sec.~\ref{sec training}}.

\begin{table}[!ht]
	\begin{center}
    \begin{threeparttable}
	\caption {\textcolor{red}{Network inputs/outputs}}\label{table io}
		\begin{tabular}{c| c c}
		\hline
		Network & Input & Output\\
		\hline
		$\mathbf{\Phi}_s$\tnote{[a,b]} & $\mathcal{S}_{\sigma}(\mathbf{q}(i;\mu))\in\re^{n}$ & $\mathbf{q}_s(i;\mu)\in\re^{n_s}$\\
		$\mathbf{\Psi}_s$\tnote{[a,b]} & $\mathbf{q}_s(i;\mu)\in\re^{n_s}$ & $\mathcal{S}_{\sigma}(\Tilde{\mathbf{q}}(i;\mu))\in\re^{n}$\\
		$\mathbf{\Phi}_l$\tnote{[a]} & $\mathcal{S}_{\text{SR}}(\mathbf{Q}_s(\mu))\in\re^{n_s\times n_t}$ & $\mathbf{q}_l(\mu)\in\re^{n_l}$\\
        $\mathbf{\Psi}_l$\tnote{[a]} & $\mathbf{q}_l(\mu)\in\re^{n_l}$ & $\mathcal{S}_{\text{SR}}(\Tilde{\mathbf{Q}}_s(\mu))\in\re^{n_s\times n_t}$\\
		$\mathcal{R}$\tnote{[a]} & $\mathcal{S}_{\text{R}}(\mu)\in\re$ & $\mathcal{S}_{\text{SR}}(\Tilde{\mathbf{q}}_l(\mu))\in\re^{n_l}$\\
		$\mathcal{P}$\tnote{[b]} & $\mathcal{S}_{\text{SR}}(\mathbf{Q}_s(i;\mu;n_{\tau}))\in\re^{n_s\times n_\tau}$ & $\mathcal{S}_{\text{SR}}(\tilde{\mathbf{q}}_s(i+1;\mu))\in\re^{n_s}$\\
		\hline
		\end{tabular}
    \begin{tablenotes}
    \item[{[a]}] For new parameter prediction
    \item[{[b]}] For future state prediction
    \end{tablenotes}
    \end{threeparttable}
	\end{center}  
\end{table}

\subsubsection{New parameter prediction} 
\textcolor{red}{As shown in the pipeline diagram, Fig.~\ref{fig pipeline1},} the framework consists of three levels for this type of task, namely a CAE, a TCAE and a MLP. In the training process, $\mathbf{\Phi}_s$ and  $\mathbf{\Psi}_s$ in the convolutional autoencoder are first trained jointly. Then by encoding the training data using the trained $\mathbf{\Phi}_s$, sequences $\mathbf{Q}_s$ are obtained and used to train  $\mathbf{\Phi}_l$ and  $\mathbf{\Psi}_l$. By encoding $\mathbf{Q}_s$ for different $\mu$ using the trained $\mathbf{\Phi}_l$, latent variables $\mathbf{q}_l$ are obtained, and used along with $\mu$ as the training data for $\mathcal{R}$.

To predict the dynamics for the desired parameter $\mu^{*}$, the  spatio-temporally encoded latent variable is predicted by $\mathbf{q}_l(\mu^*)=\mathcal{R}(\mu^{*})$, which is then sent into $\mathbf{\Psi}_l$ to obtain the sequence of spatially encoded latent variables $\mathbf{Q}_s(\mu^*)=\mathbf{\Psi}_l(\mathbf{q}_l(\mu^*))$. Finally, the frames in the predicted flow field sequence are decoded as $\mathbf{q}(i;\mu^*)=\mathbf{\Psi}_s(\mathbf{q}_s(i;\mu^*))$. 

\subsubsection{Future state prediction}
For future state prediction, the training process shares the same initial step as in the previous task, which is the training of the level 1 convolutional autoencoder. After the sequence $\mathbf{Q}_s(\mu)$ is obtained, the training of $\mathcal{P}$ is performed by taking each step $\mathbf{q}_s(i;\mu)$ with $i>n_\tau$ and its corresponding look-back window $\mathbf{Q}_s(i-1;n_{\tau};\mu)$ as the training data.

\textcolor{red}{As shown in Fig.~\ref{fig pipeline2}, in} the prediction for future states of a given sequence, the look-back window for the last known time step $\mathbf{Q}_s(n_t;n_{\tau};\mu)$ is used as the input, and the spatially encoded latent variable for the subsequent time step is predicted using $\mathbf{q}_s({n_t+1};\mu)=\mathcal{P}(\mathbf{Q}_s(n_t;n_{\tau};\mu)$). The corresponding future  state is decoded by $\mathbf{q}({n_t+1};\mu)=\mathbf{\Psi}_s(\mathbf{q}_s({n_t+1};\mu))$. By performing such predictions iteratively, multiple future steps can be predicted.

\subsection{Training}\label{sec training}
It is noted  in both of the architectures considered above, the training for each level is performed sequentially. \textcolor{red}{After the training of an autoencoder, the encoded variables generated from the trained encoder part serve as the ground truth for the target output for the following network.}
\textcolor{red}{The goal for each individual training process is to minimize a loss function $\mathcal{L}(\mathbf{\Theta})$ by optimizing the network parameters $\mathbf{\Theta}$.} \textcolor{red}{In this work $\mathcal{L}$ is primarily based on the mean squared error (MSE), which is defined on the target output $\mathbf{y}$ and the prediction $\tilde{\mathbf{y}}(\mathbf{\Theta})$ as}
\begin{equation}\label{eq MSE}
\text{MSE}(\mathbf{y}, \tilde{\mathbf{y}}(\mathbf{\Theta}))=\frac{({\tilde{\mathbf{y}}-\mathbf{y}})^T({\tilde{\mathbf{y}}-\mathbf{y}})}{n_y},
\end{equation}
\textcolor{red}{where $n_y$ is the degrees of freedom in $\mathbf{y}$.}

To alleviate \textit{overfitting}, a penalty on $\mathbf{\Theta}$ is added as a \textit{regularization} term to $\mathcal{L}$. Common choices of regularization include the $\ell_1$-norm and $\ell_2$-norm. In this work, the latter is used and the final expressions for the loss function is given by
\begin{equation}
\mathcal{L(\mathbf{\Theta})}=\text{MSE}(\mathbf{y}, \tilde{\mathbf{y}}(\mathbf{\Theta}))+\lambda\left\lVert\mathbf{\Theta}\right\rVert_2,
\end{equation}
where $\lambda$ is a penalty coefficient. The optimal parameters $\mathbf{\Theta}^*$ to be trained are then
\begin{equation}
\Theta^* = \argmin{\Theta}\mathcal{L(\mathbf{\Theta})}.
\end{equation}

\textit{Back propagation} is used to optimize the network parameters iteratively based on the gradient $\nabla_{\mathbf{\Theta}}\mathcal{L}$. There are numerous types of gradient-based optimization methods, and in this work Adam optimizer~\cite{kingma2014adam} is used.

To accelerate training, three types of feature scaling are applied, namely standard deviation scaling $\mathcal{S}_{\sigma}$, \textcolor{red}{minimum-maximum} range scaling $\mathcal{S}_{\text{R}}$ and shifted \textcolor{red}{minimum-maximum} range scaling $\mathcal{S}_{\text{SR}}$. With the original input or output feature, e.g. $\mathbf{q}$ or $\mu$, denoted by $x$, the scaling operations are given by

\begin{eqnarray}
&\mathcal{S}_{\sigma}(x)=\frac{x-\text{mean}(x_{train})}{\sigma(x_{train})}.\label{eq std}\\
&\mathcal{S}_{\text{R}}(x)=\frac{x-\text{min}(x_{train})}{\text{max}(x_{train})-\text{min}(x_{train})}.\label{eq ran}\\
&\mathcal{S}_{\text{SR}}(x)=\frac{x-\text{min}(x_{train})}{\text{max}(x_{train})-\text{min}(x_{train})}-0.5.\label{eq sran}
\end{eqnarray}

\textcolor{red}{The use of the scaling methods in different parts of the framework is summarized in Table \ref{table io}.}

\section{Numerical tests}\label{sec tests}
In this section, numerical tests are performed on three representative  problems of different dimensions and characteristics of dynamics.  These test cases have proven to be challenging for traditional POD-based model reduction techniques. The advection of a half cosine wave is used as a simplified illustrative problem in \ref{sec wave}, to aid the understanding of a few concepts of the framework. In \ref{sec pod},  numerical tests are performed with the POD at the top-level to provide a comparison between POD- and CAE-based methods for reference. A sensitivity study of latent dimensions in the autoencoders is provided in  \ref{sec sensitivity}.

Besides MSE (\eqr{eq MSE}), the relative absolute error (RAE) is taken as another criterion to assess the accuracy of the predictions. For simplicity, we will use $\epsilon_{\text{MSE}}^{y}=\text{MSE}(\mathbf{y},\tilde{\mathbf{y}})$ and $\epsilon_{\text{RAE}}^{y}=\text{RAE}(\mathbf{y},\tilde{\mathbf{y}})$ to denote the two types of errors for variable $y$. The latter is given by
\begin{equation}
    \text{RAE}(\mathbf{y},\tilde{\mathbf{y}})=\frac{\overline{| \tilde{\mathbf{y}}-\mathbf{y}|}}{\text{max}(|\mathbf{y}|)},
\end{equation}
where $\overline{|\tilde{\mathbf{y}}-\mathbf{y}|}$ is the mean of the absolute value of $\tilde{\mathbf{y}}-\mathbf{y}$, and $\text{max}(|\mathbf{y}|)$ is the maximum of the absolute value of $\mathbf{y}$ over the entire computational domain. \textcolor{red}{It should be noted that in the evaluation of subsequent levels, the encoded variables from the preceding levels are taken as the truth.}
Training and testing convergence is reported in ~\ref{sec convergence}.

For a reduced order prediction method aimed at many-query applications, computational efficiency is another important aspect in the framework performance metrics. In this work, all numerical tests are performed using one NVIDIA Tesla P100 GPU and timing results are provided accordingly. 

\subsection{Linear advection}\label{sec wave}
\begin{figure}
    \centering
    \begin{minipage}{0.5\textwidth}
        \centering
    	\includegraphics[width=1\textwidth]{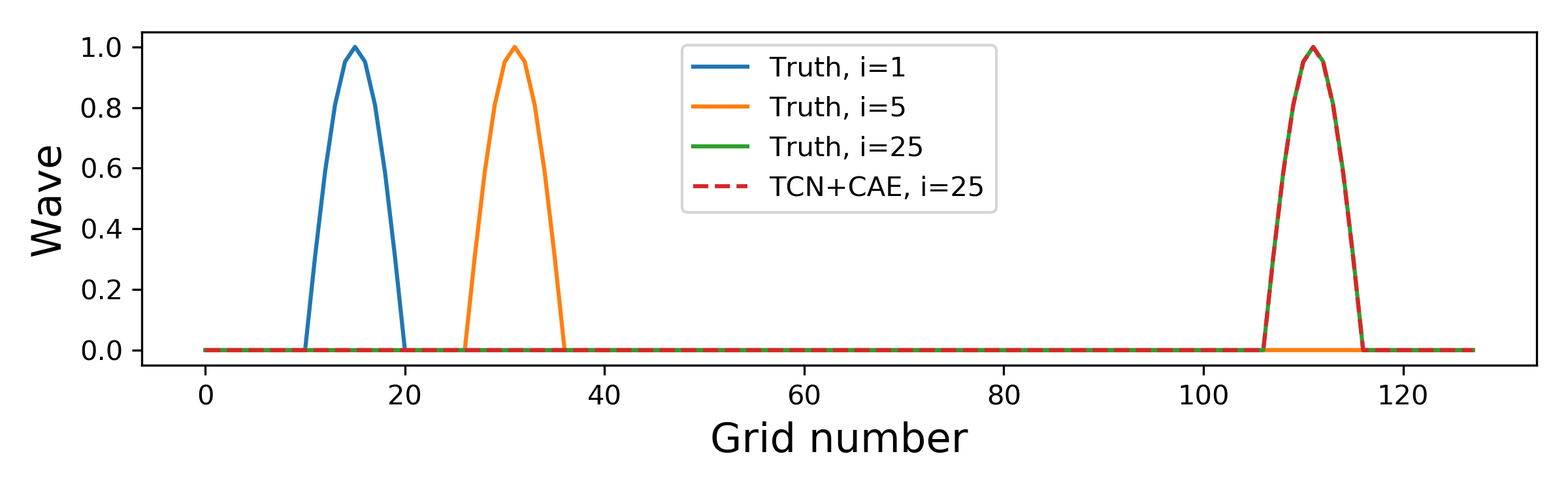}
        \caption{Linear advection: true and predicted profiles}\label{fig wave}
    \end{minipage}
    \begin{minipage}{0.5\textwidth}
        \centering
    	\includegraphics[width=1\textwidth]{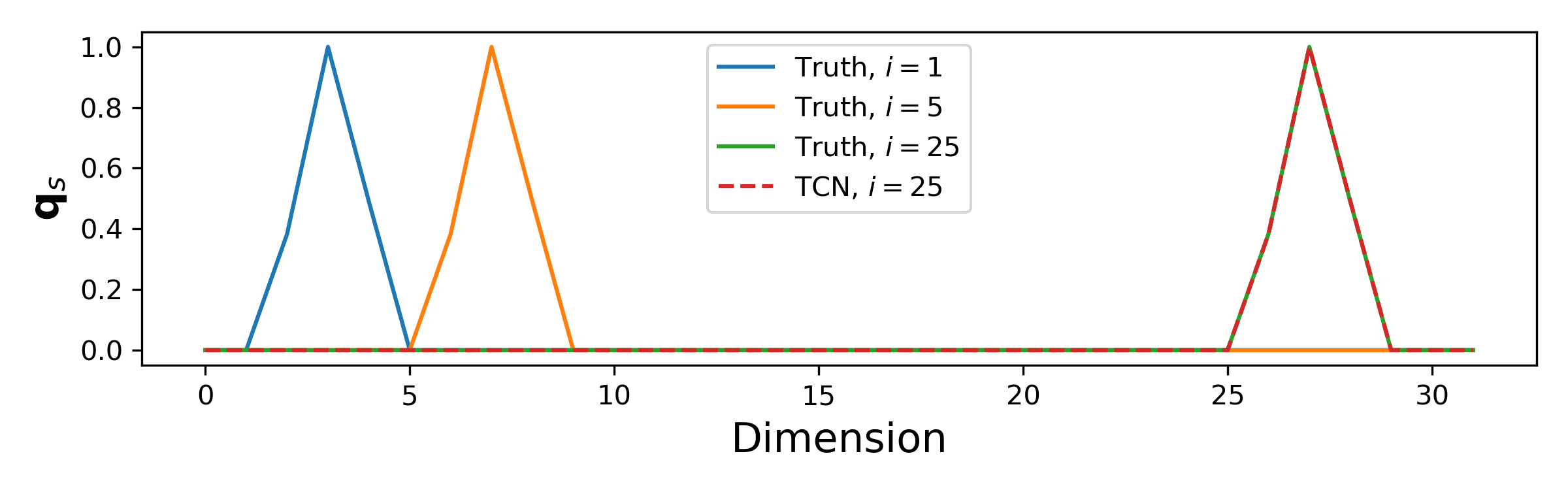}
        \caption{Linear advection: $\mathbf{q}_s$ profiles}\label{fig wave qs}
    \end{minipage}
\end{figure}

The advection of a half cosine wave is used to illustrate  encoding-decoding and future state prediction processes in a simplified setting. In this  problem, a one-dimensional spatial discretization of 128 grid nodes is used. Initially the half cosine wave, spanning 11 grid nodes, is centered at the 15th grid node, and translates 100 grid nodes in 25 frames at a constant advection speed of 4 nodes/frame without any deformation. The initial frame, the last frame in training, $i=5$, and the final frame to be predicted, $i=25$, are shown in Fig.~\ref{fig wave}. In this test, it is assumed that only the first 5 frames of data are available in the training stage, and prediction is performed for the next 20 frames. \textcolor{red}{Errors for the final prediction and different stages of output are summarized in Table~\ref{table wave error}.}

\begin{table}[!ht]
	\centering
	\caption {\textcolor{red}{Linear advection: network architectures}}
	\subfloat[CAE\label{table wave CAE}]{
        \centering
        \small
		\begin{tabular}{c| c c c c c}
    		\hline
    		\multicolumn{6}{c}{$\mathbf{\Phi}_s$}\\
    		\hline
    		Layer & Input & Convolution & Convolution & Convolution & Flatten (Output)\\
    		\hline
    		Output shape & $128\times1$ & $64\times32$ & $32\times64$ & $32\times1$ & 32\\
    		\hline
    		\hline
    		\multicolumn{6}{c}{$\mathbf{\Psi}_s$}\\
    		\hline
    		Layer & Input & Reshape & Conv-trans & Conv-trans & Conv-trans (Output)\\
    		\hline
    		Output shape & 32 & $32\times1$ & $32\times64$  & $64\times32$ & $128\times1$\\
    		\hline
		\end{tabular}
	}
	
	\subfloat[TCN\label{table wave many2many}]{
        \centering
        \small
		\begin{tabular}{c| c c c c}
    		\hline		Layer & Input & 2-layer TCN block & Convolution & Flatten (Output)\\
    		\hline
    		Output shape & $32\times2$  & $32\times32$ & $32\times1$ & 32\\
    		\hline
		\end{tabular}
	}
\end{table}

A simple CAE with three convolutional layers is first used for spatial compression, so that the prediction is performed for 32 latent degrees of freedom instead of the original 128 degrees of freedom for the spatially discretized variable. The detailed CAE architecture is provided in Table~\ref{table wave CAE}. The latent variable \textcolor{red}{$\mathbf{q}_s=\mathbf{\Phi}_s(\mathbf{q})$} is shown in Fig.~\ref{fig wave qs} for the same time instances as in Fig.~\ref{fig wave}. It can be seen that the original half cosine wave is transformed into a triangular wave consisting of two linear halves. 
\textcolor{red}{From Table~\ref{table wave error} it can be seen that the CAE reconstruction shows the same error numbers in both the training and testing stages, which shows that the CAE is able to represent the cosine wave to the same precision regardless of its position in the domain due to its conservation in shape.}

\textcolor{red}{A TCN future step predictor $\mathcal{P}$ is used as the second level to predict $\mathbf{q}_s$ for future steps.} As listed in Table~\ref{table wave many2many}, $\mathcal{P}$ consists of a TCN block with 32 channels and a convolutional layer that reduces the 32 channels into a single output channel which would be the future step of $\mathbf{q}_s$. The TCN block consists of two dilated 1D convolutional layers with dilation orders $d=1,2$. \textcolor{red}{As a propagation problem, both convolution layers are non-strided and operates along the latent dimension.} While a single look-back step is sufficient, to make the case more representative for more complex problems, a look-back window of $n_{\tau}=2$ is used. Thus the input dimension to the TCN is $n\times n_{\tau}=32\times2$. 

With $n_{\tau}=2$, the 5 training frames are grouped into 3 training samples for TCN, i.e. the 1st and 2nd frames as the input, the 3rd frame as the target output; the 2nd and 3rd frames as the input, the 4th frame as the target output, etc. 
After the prediction for a new frame $i$, the second frame in the original look-back window, i.e. frame $i-1$, is used as the first frame in the new window for the prediction of frame $i+1$, and the predicted frame $i$ is used as the second frame in the new window. After 20 iterations, the predicted final frame of $\mathbf{q}_s$ for $i=25$ is obtained. \textcolor{red}{As listed in Table~\ref{table wave error}, the error introduced by the TCN (i.e. in the prediction for $\mathbf{q}_s$) is negligible compared with that by the CAE. Thus the predicted future states $\mathbf{q}^*=\mathbf{\Psi}_s(\Tilde{\mathbf{q}}_s)$ shows an RAE of 0.04\%, which is dominated by the CAE reconstruction error. The predicted final step $i=25$ is shown in Fig.~\ref{fig wave}.}

\begin{table}[!ht]
	\begin{center}
	\caption {\textcolor{red}{Linear advection: errors for different stages of output}}\label{table wave error}
    \small
{\color{red}
		\begin{tabular}{c | c | c| c | c}
		\hline
		Stage & Truth & Output & RAE & MSE\\
		\hline
		\multirow{2}{*}{\makecell{Training\\(frames 1 to 5)}} & $\mathbf{q}$ & $\mathbf{\tilde{q}}=\mathbf{\Psi}_s(\mathbf{\Phi}_s(\mathbf{q}))$ & 0.04\% & \num{1.9e-5}\\
		&$\mathbf{q}_s=\mathbf{\Phi}_s(\mathbf{q})$ & $\tilde{\mathbf{q}}_s=\mathcal{P}(\mathbf{Q}_s)$ & \num{3.0e-8} & \num{1.4e-14}\\
		\hline
		\multirow{3}{*}{\makecell{Testing\\(frames 6 to 25)}} & $\mathbf{q}$ & $\mathbf{\tilde{q}}=\mathbf{\Psi}_s(\mathbf{\Phi}_s(\mathbf{q}))$ & 0.04\% & \num{1.9e-5}\\
		&$\mathbf{q}_s=\mathbf{\Phi}_s(\mathbf{q})$ & $\tilde{\mathbf{q}}_s=\mathcal{P}(\mathbf{Q}_s)$ & \num{1.6e-6} & \num{9.9e-11}\\
		&$\mathbf{q}$ & $\tilde{\mathbf{q}}^*=\mathbf{\Psi}_s(\tilde{\mathbf{q}}_s)$ & 0.04\% & \num{1.9e-5}\\
		\hline
		\end{tabular}
}
	\end{center}  
\end{table}




\subsection{Discontinuous compressible flow}
The Sod shock tube~\cite{sod1978survey} is a classical test and studies the propagation of a rarefaction wave, a shock wave and a discontinuous contact surface initiated by a discontinuity in pressure at the middle of a one dimensional tube filled with ideal gas. This flow is governed by the one dimensional Euler equations of gas dynamics~\cite{danaila2007gas}. The standard initial left and right states are $P_L=1,\rho_L=1,U_L=0,P_R=0.1,\rho_R=0.125,U_R=0$, where $P,\rho,U$ are the density, pressure and velocity respectively. A detailed description of the exact solution for the problem can be found in \cite{danaila2007gas}. 

\begin{figure}
    \centering
    	\includegraphics[width=0.5\textwidth]{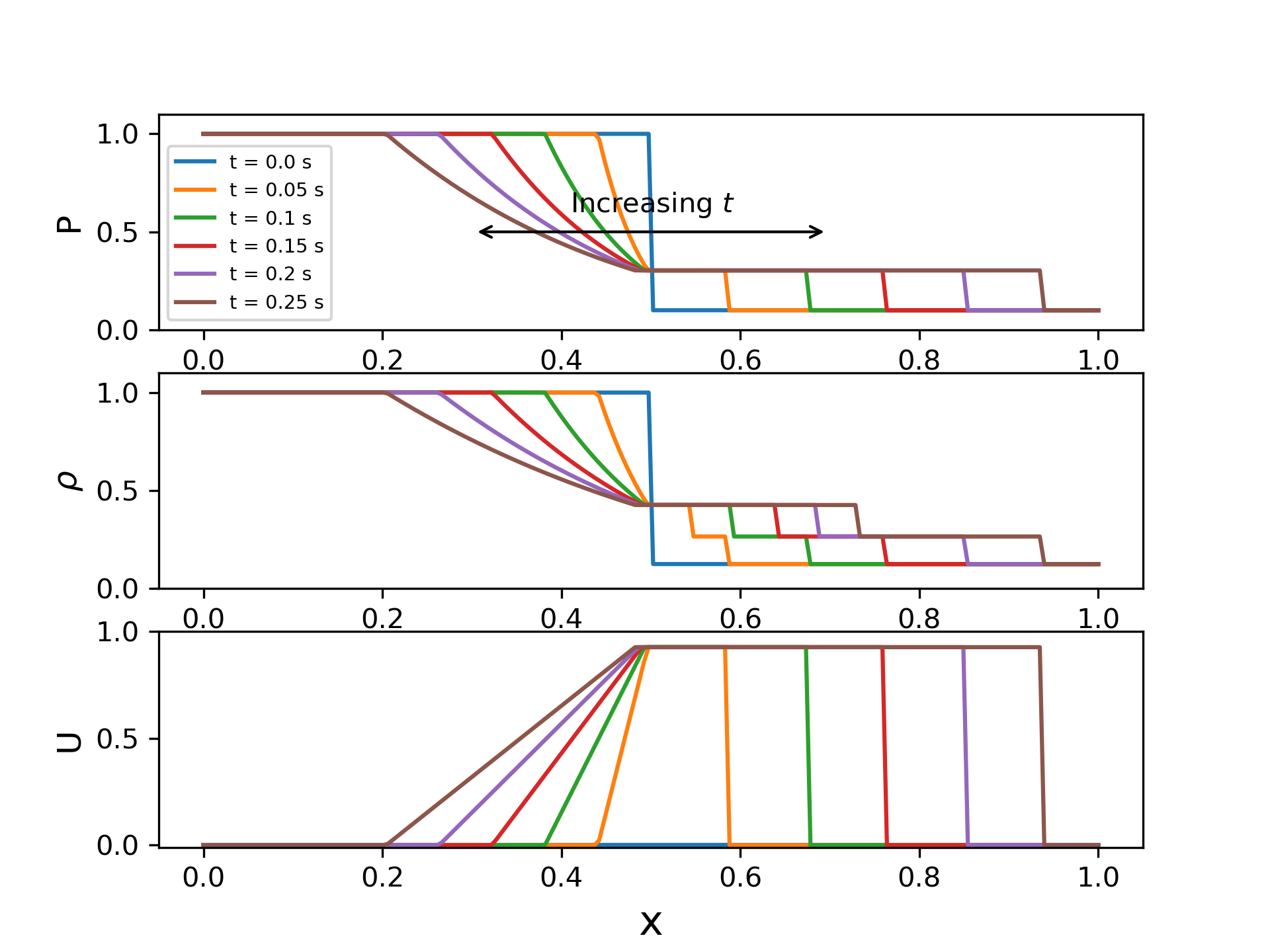}
        \caption{Discontinuous compressible flow: Profiles of pressure, density and velocity.}\label{fig sod}
\end{figure}

In this numerical test, the evolution of $P,\rho,U$ will be predicted in the the time interval 0.1 to 0.25 s based on the dynamics before 0.1 s. The profiles are plotted in Fig.~\ref{fig sod} for several time instances for $0 \leq t \leq 0.25$ s. \textcolor{red}{The ground truth are obtained by mapping the exact solution onto a 200-point 1D grid with a temporal discretization of 35 steps per 0.1 s}. 
Such propagation problems are considered challenging from the perspective of model reduction~\cite{parish2019time} because the dynamics in prediction stage is  not covered in training, leading to failure for ROMs based on global bases. \textcolor{red}{Errors for different stages of output in the proposed framework are summarized in Table~\ref{table sod error}.}

\begin{table}[!ht]
	\centering
	\caption {\textcolor{red}{Discontinuous compressible flow: network architectures}}
	\subfloat[CAE\label{table sod CAE}]{
        \centering
        \small
		\begin{tabular}{c| c c c c}
    		\hline
    		\multicolumn{5}{c}{$\mathbf{\Phi}_s$}\\
    		\hline
    		Layer & Input & Convolution & Convolution & Flatten (Output)\\
    		\hline
    		Output shape & $200\times3$ & $200\times128$ & $200\times1$ & 200\\
    		\hline
    		\hline
    		\multicolumn{5}{c}{$\mathbf{\Psi}_s$}\\
    		\hline
    		Layer & Input & Reshape & Conv-trans & Conv-trans (Output)\\
    		\hline
    		Output shape & 200 & $200\times1$ & $200\times128$ & $200\times3$\\
    		\hline
		\end{tabular}
	}
	
	\subfloat[TCN\label{table sod many2many}]{
        \centering
        \small
		\begin{tabular}{c| c c c c c}
    		\hline		Layer & Input & 2-layer TCN block & 2-layer TCN block & Convolution & Flatten (Output)\\
    		\hline
    		Output shape & $200\times2$  & $200\times200$ & $200\times400$ & $200\times1$ & 200\\
    		\hline
		\end{tabular}
	}
\end{table}

\begin{figure}
	\centering
    \begin{minipage}{0.5\textwidth}
        \centering
    	\includegraphics[width=1\textwidth]{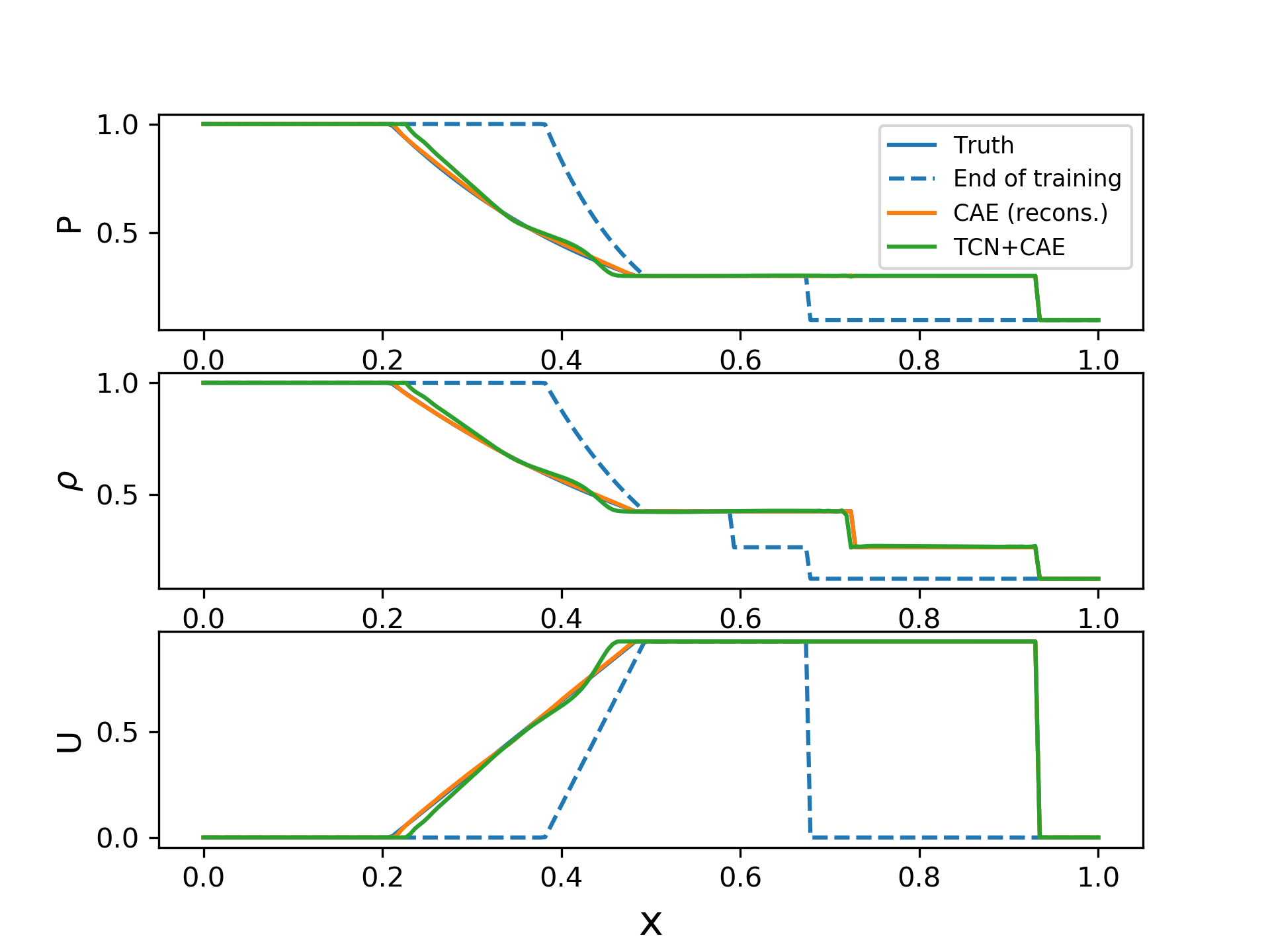}
        \caption{Discontinuous compressible flow: reconstructed (recons.) and predicted profiles for $t=0.25$ s}\label{fig sod decoded}
    \end{minipage}\hfill
    \begin{minipage}{0.5\textwidth}
        \centering
    	\includegraphics[width=1\textwidth]{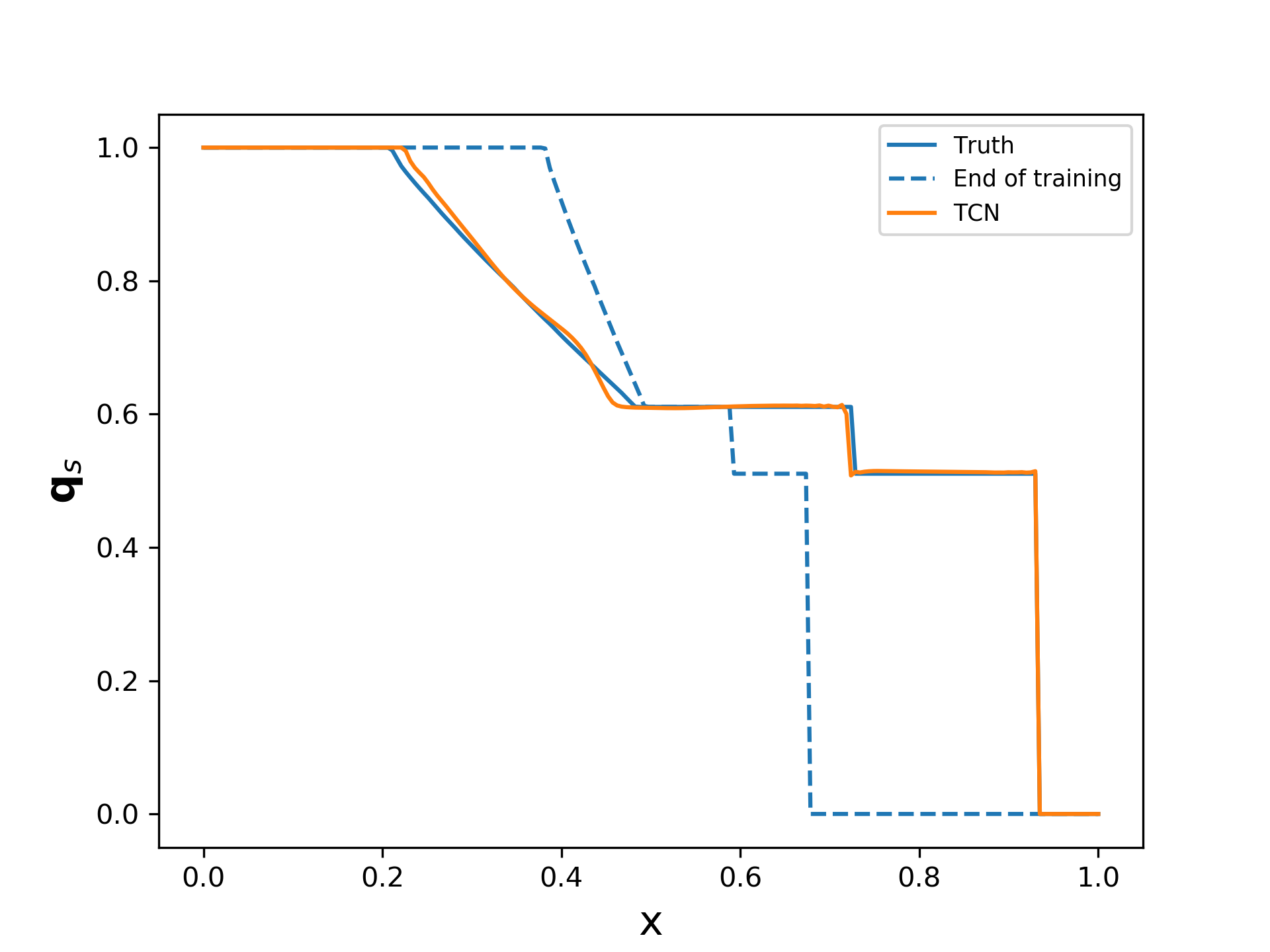}
        \caption{Discontinuous compressible flow: Latent variables $\mathbf{q}_s$ for $t=0.25$ s}\label{fig sod encoded}
    \end{minipage}
\end{figure}

Due to the relatively small grid size an aggressive spatial compression is not necessary, thus no pooling or dense layer is used in the CAE. The detailed CAE architecture is given in Table~\ref{table sod CAE}. \textcolor{red}{The reconstructed variables $\mathbf{\tilde{q}}=\mathbf{\Psi}_s(\mathbf{\Phi}_s(\mathbf{q}))$} is shown at the final time step $t=0.25$ s in Fig.~\ref{fig sod decoded}. This result is effectively an estimate of the lower bound of the error in the prediction case. \textcolor{red}{For the same step, $\mathbf{q}_s$ is present in Fig.~\ref{fig sod encoded}. It can be seen that the encoded variable behaves similarly to a scaled variation of $\rho$, which contains the most complex dynamics with two discontinuities among the three variables studied. This illustrates an efficient dimension reduction by reducing the number of variables in the CAE, even without any pooling operation.}

\textcolor{red}{A short look-back window of $n_{\tau}=2$ is used as in the previous demonstration case, but for this more complicated propagation problem, the size of TCN is significantly increased.} $\mathcal{P}$ consists of two TCN blocks with 200 and 400 channels respectively, followed by one convolutional layer that reduces the 400 channels of the second TCN block into one output channel, which would be the future step of $\mathbf{q}_s$. Each of the TCN blocks consist of two \textcolor{red}{non-strided} dilated 1D convolutional layers with dilation orders $d=1,2$. The detailed architecture of $\mathcal{P}$ is given in Table~\ref{table sod many2many}. \textcolor{red}{53 iterative prediction steps for $\mathbf{q}_s$ are performed for $0 < t \leq 0.25$ s, and the predicted future states are decoded from them. The predicted profiles at $t=0.25$ s for $\mathbf{q}$ and $\mathbf{q}_s$ are present in Fig.~\ref{fig sod decoded} and Fig.~\ref{fig sod encoded}. The RAE in the prediction for the three variables are all below 0.35\%.}

\begin{table}[!ht]
	\begin{center}
	\caption {\textcolor{red}{Discontinuous compressible flow: errors for different stages of output}}\label{table sod error}
    \small
{\color{red}
		\begin{tabular}{c | c | c| c | c}
		\hline
		Stage & Truth & Output & RAE & MSE\\
		\hline
		\multirow{2}{*}{\makecell{Training\\($0\leq t\leq 0.1$ s)}} & $\mathbf{q}\,(P/\rho/U)$ & $\mathbf{\tilde{q}}\,(P/\rho/U)=\mathbf{\Psi}_s(\mathbf{\Phi}_s(\mathbf{q}))$ & 0.03\%/0.01\%/0.03\% & \num{1.2e-6}/\num{2.4e-7}/\num{1.6e-6}\\
		&$\mathbf{q}_s=\mathbf{\Phi}_s(\mathbf{q})$ & $\tilde{\mathbf{q}}_s=\mathcal{P}(\mathbf{Q}_s)$ & 0.008\% & \num{1.2e-6}\\
		\hline
		\multirow{3}{*}{\makecell{Testing\\($0.1<t\leq 0.25$ s)}} & $\mathbf{q}\,(P/\rho/U)$ & $\mathbf{\tilde{q}}\,(P/\rho/U)=\mathbf{\Psi}_s(\mathbf{\Phi}_s(\mathbf{q}))$ & 0.10\%/0.05\%/0.09\% & \num{5.4e-6}/\num{1.5e-6}/\num{5.1e-6}\\
		&$\mathbf{q}_s=\mathbf{\Phi}_s(\mathbf{q})$ & $\tilde{\mathbf{q}}_s=\mathcal{P}(\mathbf{Q}_s)$ & 0.22\% & \num{6.5e-5}\\
		&$\mathbf{q}\,(P/\rho/U)$ & $\tilde{\mathbf{q}}^*\,(P/\rho/U)=\mathbf{\Psi}_s(\tilde{\mathbf{q}}_s)$ & 0.32\%/0.34\%/0.31\% & \num{7.6e-5}/\num{1.6e-4}/\num{8.8e-5}\\
		\hline
		\end{tabular}
}
	\end{center}  
\end{table}

The computing time by different parts of the framework in training and prediction is listed in Table~\ref{table sod time}. The total prediction time for 0.15 s of propagation is 0.6 s. 

\begin{table}[!ht]
	\begin{center}
	\caption {Discontinuous compressible flow: computing time}\label{table sod time}
    \small
		\begin{tabular}{c| c c c}
		\hline
		\multicolumn{4}{c}{Training}\\
		\hline
		Component & CAE & TCN & Total\\
		\hline
		Training epochs & 1000 & 2000 & -\\
		Computing time (s) & 171 & 166 & 337\\
		\hline
		\hline
		\multicolumn{4}{c}{Prediction (53 frames)}\\
		\hline
		Component & $\mathbf{\Psi}_s$ & TCN & Total\\
		\hline
		Computing time (s) & 0.48 & 0.12 & 0.60\\
		\hline
		\end{tabular}
	\end{center}  
\end{table}

 \subsubsection{Impact of training sequence length}
To evaluate the sensitivity of the predictions on the amount of training data, the above TCN is re-trained on training sequences of different length $n_t$, ranging from 20 to 40 frames. 
The comparison is performed on 20 iterative future prediction steps. 

\begin{figure}
	\centering
	\subfloat[TCN training and prediction]{
		\includegraphics[width=0.5\textwidth]{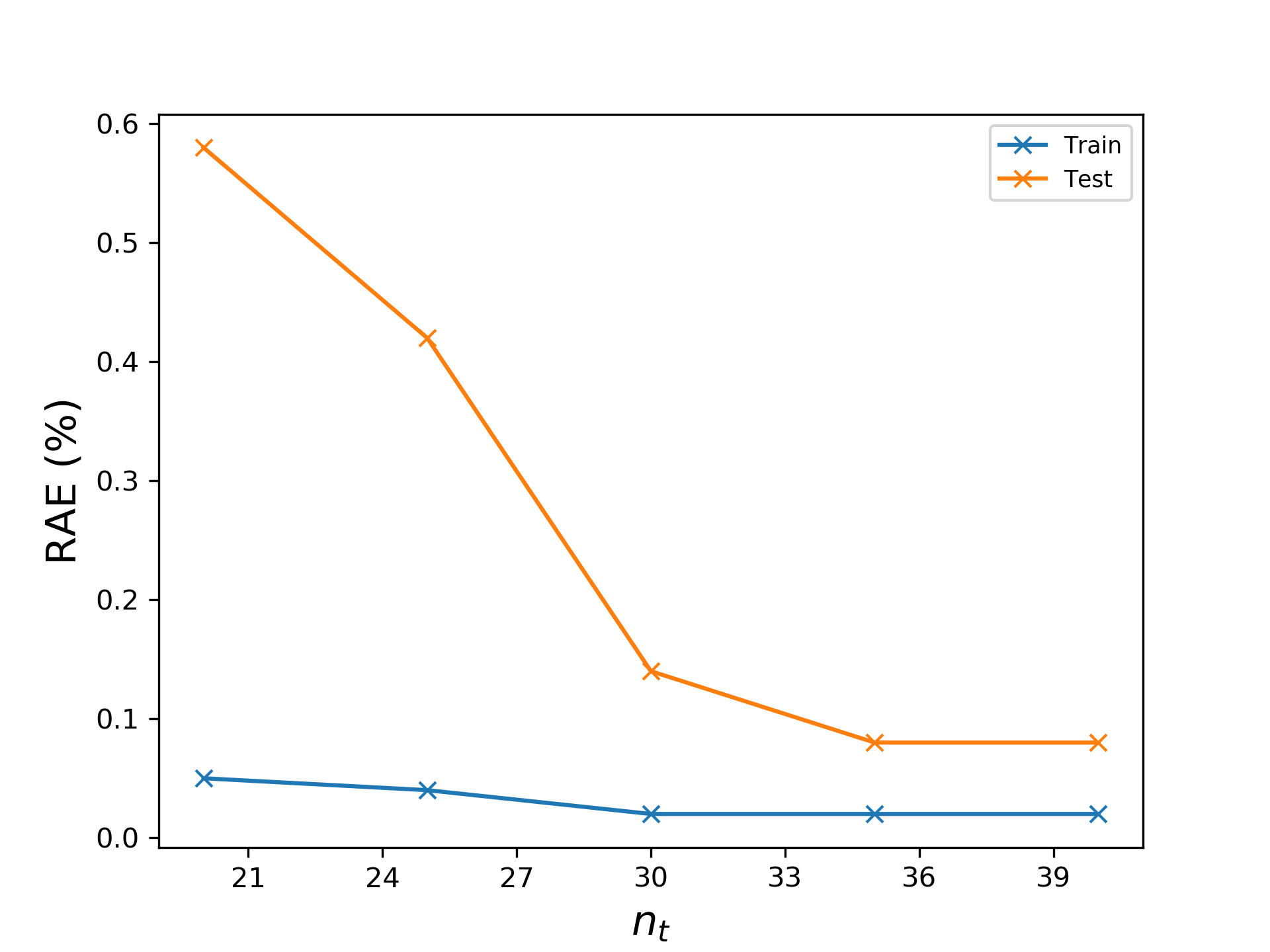}}
	\subfloat[Decoded variables from TCN prediction]{
		\includegraphics[width=0.5\textwidth]{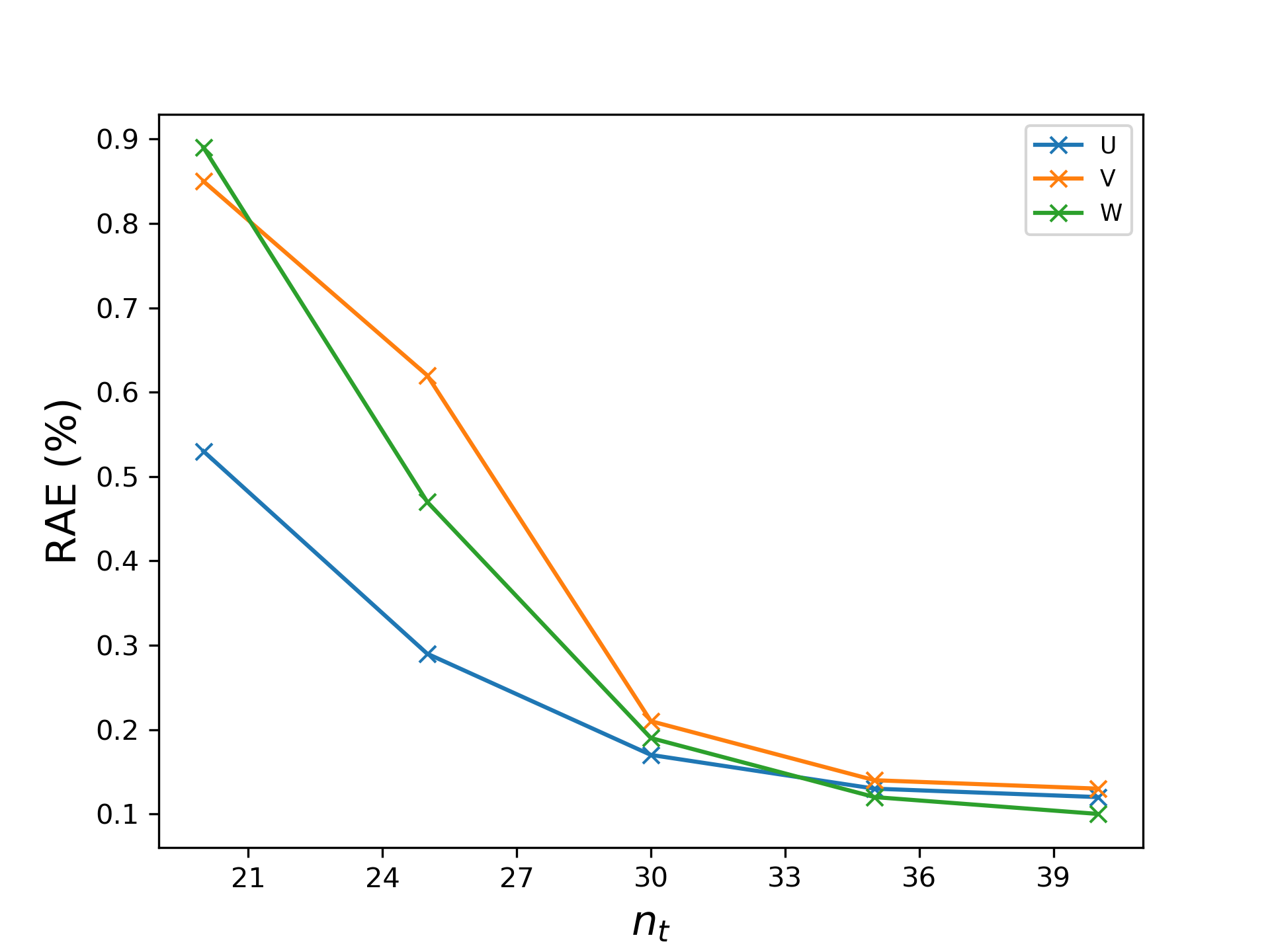}}
	\caption{Discontinuous compressible flow: sensitivity of RAE to number  of training samples $n_t$}\label{fig sod RAE vs nt} 
\end{figure}

\begin{figure}
	\centering
	\subfloat[$n_t=20$]{
		\includegraphics[width=0.33\textwidth]{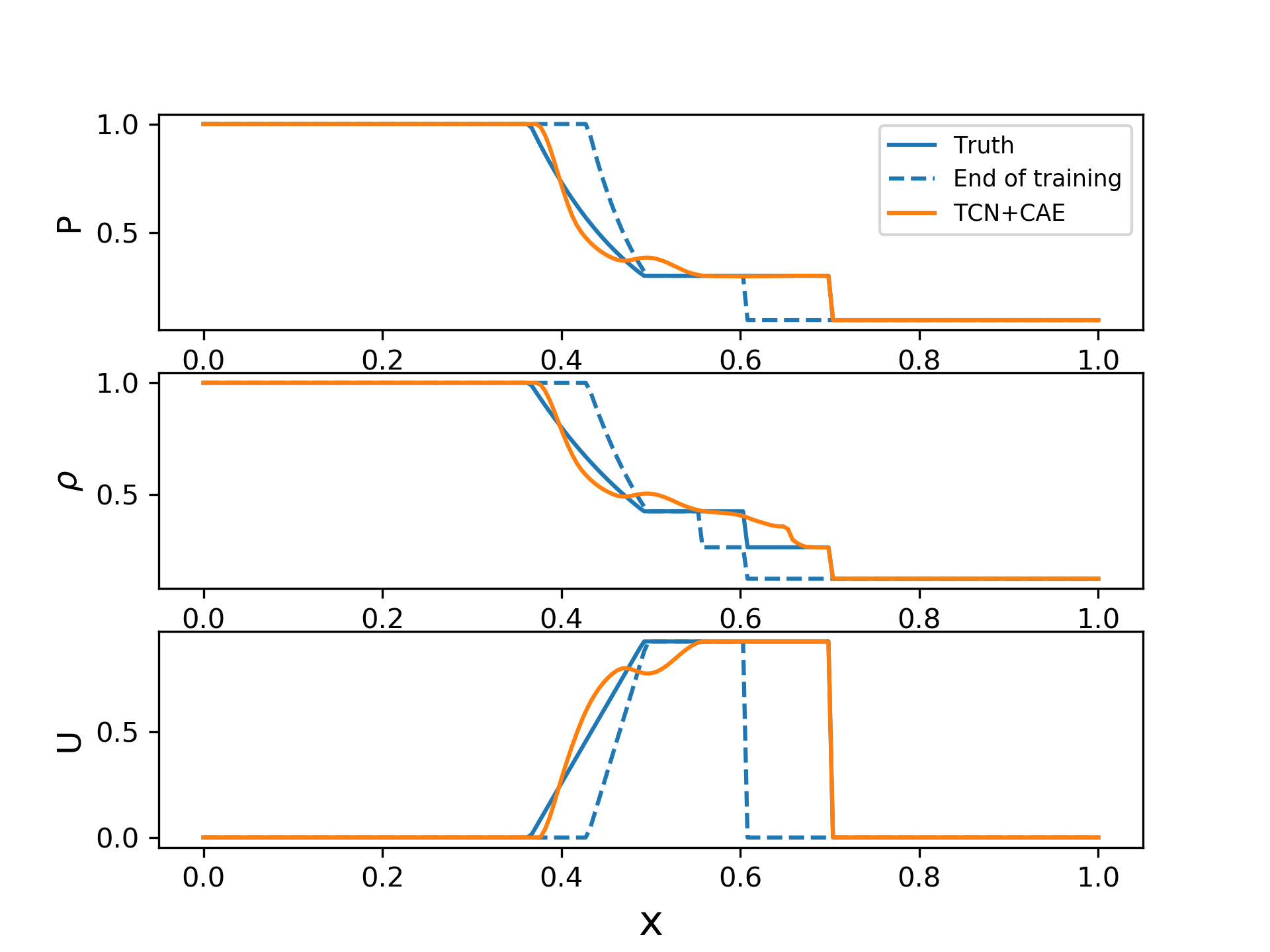}}
	\subfloat[$n_t=30$]{
		\includegraphics[width=0.33\textwidth]{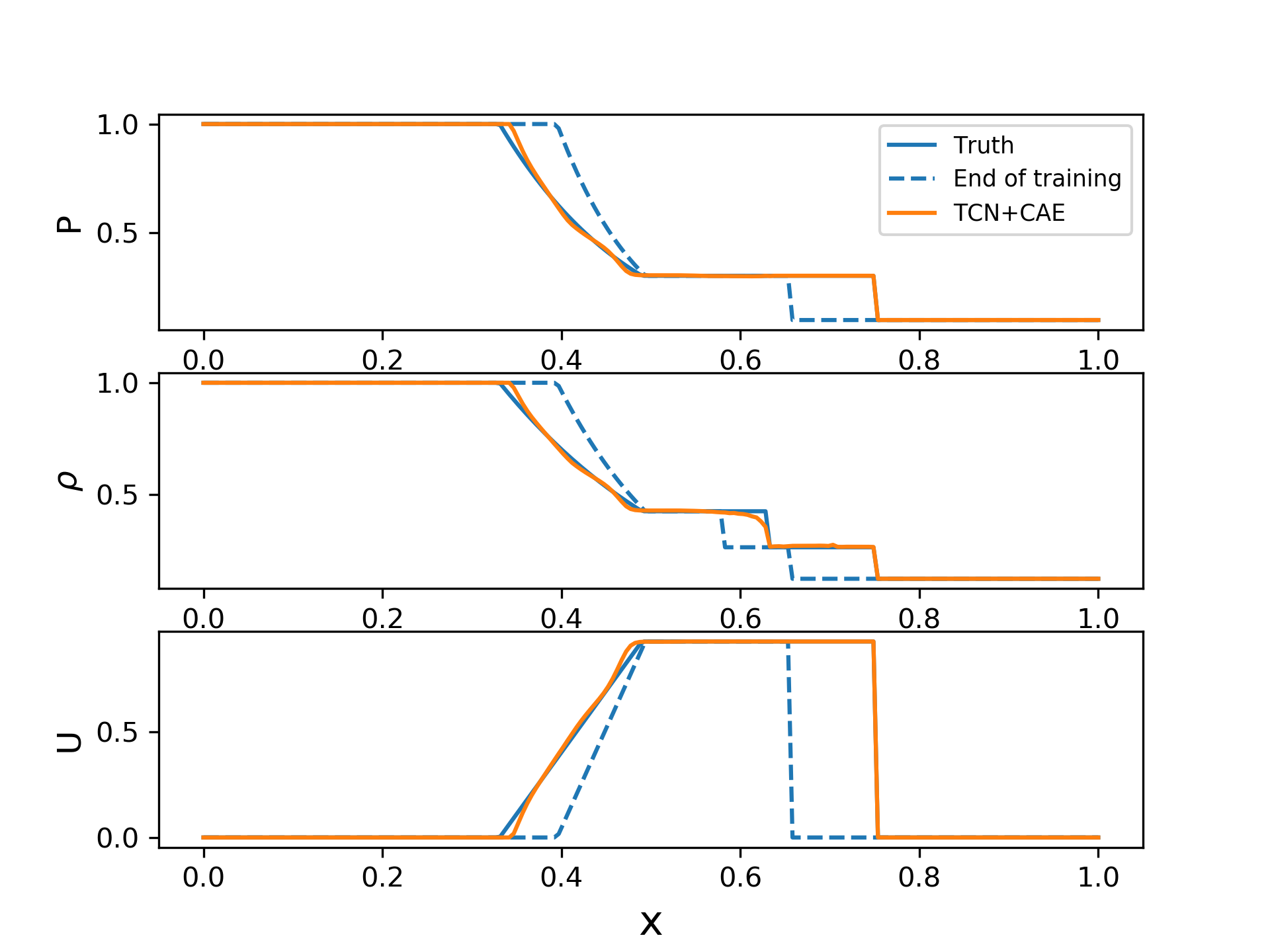}}
	\subfloat[$n_t=40$]{
		\includegraphics[width=0.33\textwidth]{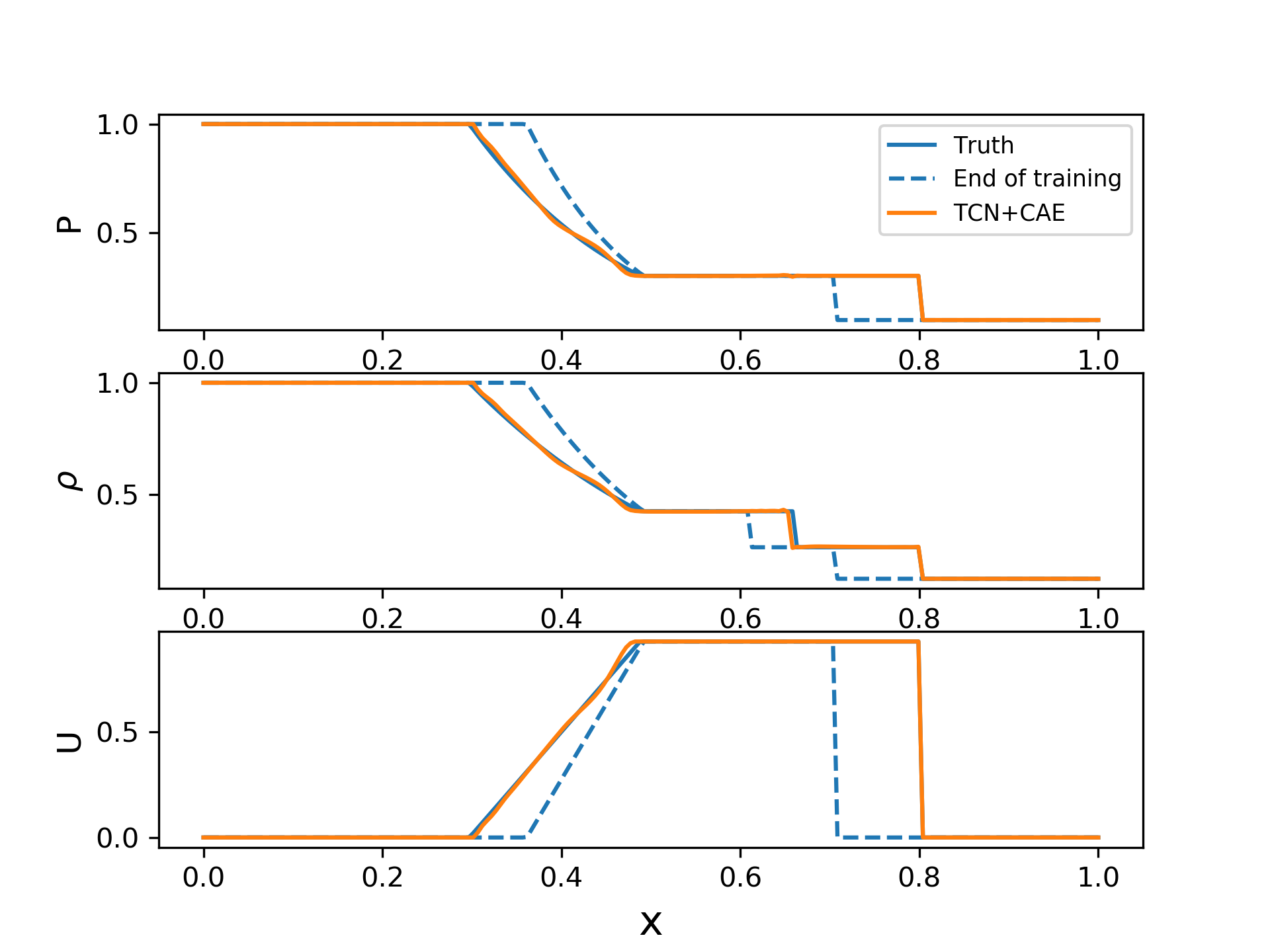}}
	\caption{Discontinuous compressible flow: sensitivity of (20-step) prediction result to number of training samples $n_t$}\label{fig sod nt decoded} 
\end{figure}

The RAE for TCN training and prediction, as well as the decoded result using $\mathbf{\Phi}_s$ is shown in Fig.~\ref{fig sod RAE vs nt}. Sample prediction results are given in Fig.~\ref{fig sod nt decoded}. It is seen that at $n_t=20$, the TCN is unable to provide an accurate future state prediction. The error decreases rapidly with increasing number of training snapshots, and starts to saturate after $n_t=30$. This behavior was observed in other levels of the framework, illustrating the importance of sufficient training data as can be expected of data-driven frameworks. 

\subsection{Transient flow over a cylinder}\label{sec cylinder}
In this test case, transient flow over a cylinder is considered. Despite the simple two-dimensional flow configuration, the complexity is characterized by the evolution of the flow from non-physical, attached flow initial conditions conditions to boundary layer separation and  consequently, vortex shedding. \textcolor{red}{The flow dynamics is determined by a global parameter, the Reynolds number $Re$}, defined as 
\begin{equation}
Re = \frac{U_{\infty}D}{\nu},
\end{equation}
where $U_{\infty}$ is the inflow velocity, $D$ is the diameter of the cylinder and $\nu$ is the viscosity of the fluid. 

The domain of interest spans $20D$ in the x-direction and $10D$ in y-direction. Solution snapshots of the incompressible Navier--Stokes equations are  interpolated onto a $384\times 192$ uniform grid. The $x$-velocity $U$ and $y$-velocity $V$ are chosen as the variables of interest. 
 
\textcolor{red}{For this test case, a parametric prediction is first performed for a new $Re$, and a future state prediction is then appended to the predicted sequence. Errors for different stages of output are summarized in Table~\ref{table cylinder error}.}

\subsubsection{New parameter prediction}
\begin{figure}
	\centering
	\subfloat[Truth\label{fig cylinder contour truth 2500}]{
		\includegraphics[width=0.33\textwidth,trim={3cm 0 2cm 0},clip]{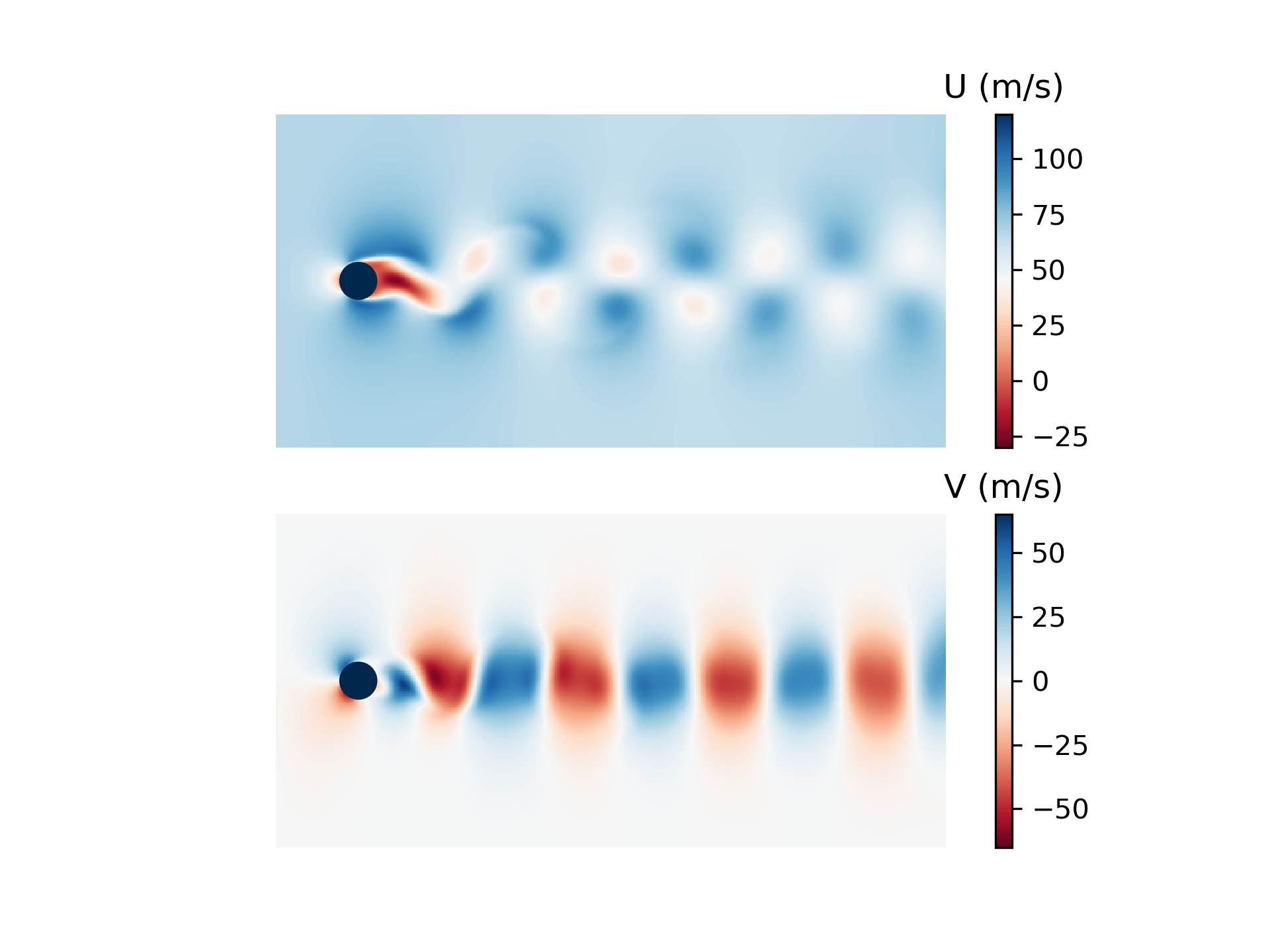}}
	\subfloat[CAE reconstruction\label{fig cylinder contour CAE 2500}]{
		\includegraphics[width=0.33\textwidth,trim={3cm 0 2cm 0},clip]{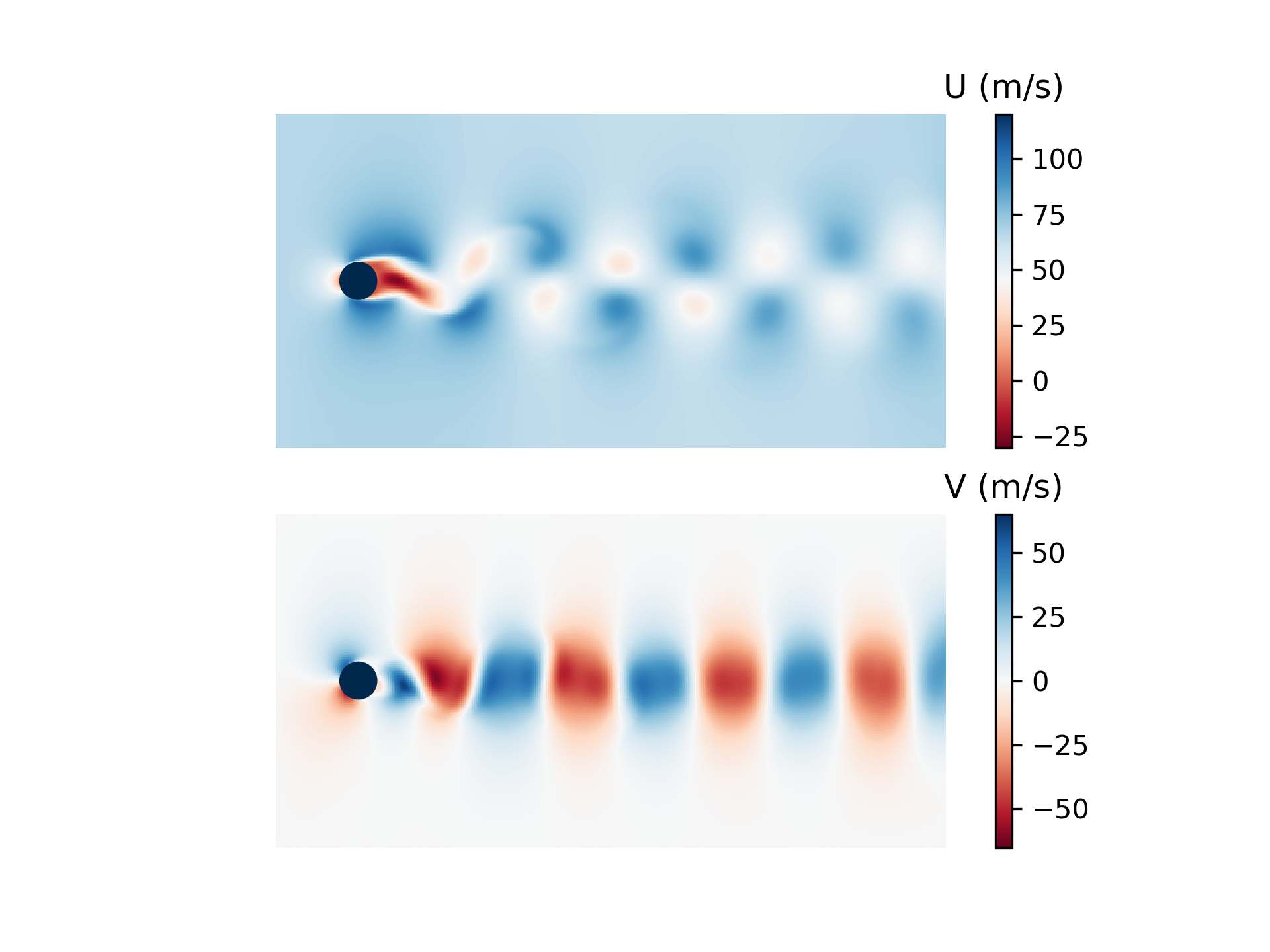}}
	\subfloat[MLP+TCAE+CAE\label{fig cylinder contour feedforward 2500}]{
		\includegraphics[width=0.33\textwidth,trim={3cm 0 2cm 0},clip]{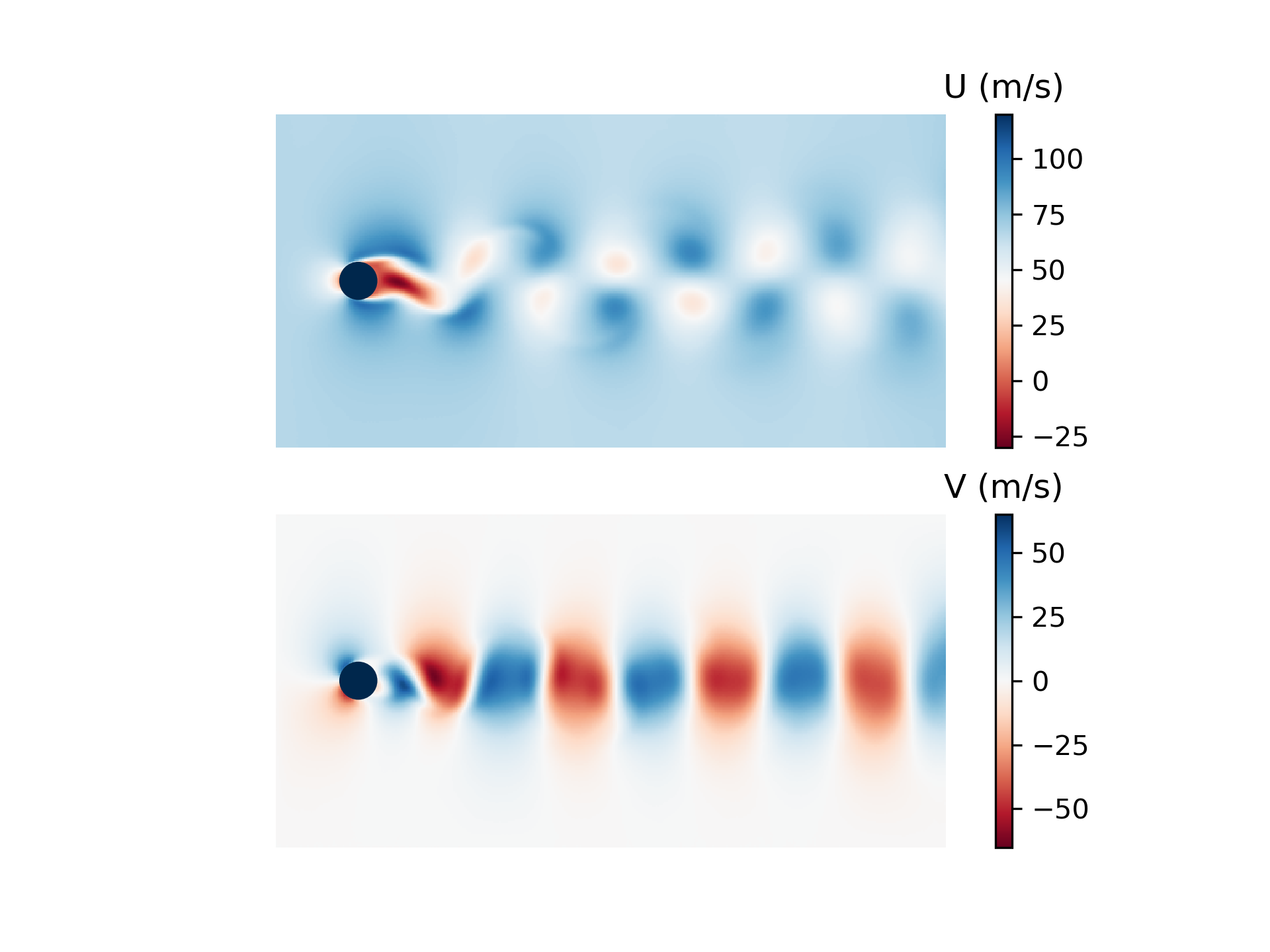}}
	\caption{Transient flow over a cylinder new parameter prediction: flow field contours for $i=2500, Re=200$}\label{fig cylinder contour}
\end{figure}

\begin{figure}
	\centering
    \begin{minipage}{0.5\textwidth}
	\centering
    	\includegraphics[width=1\textwidth]{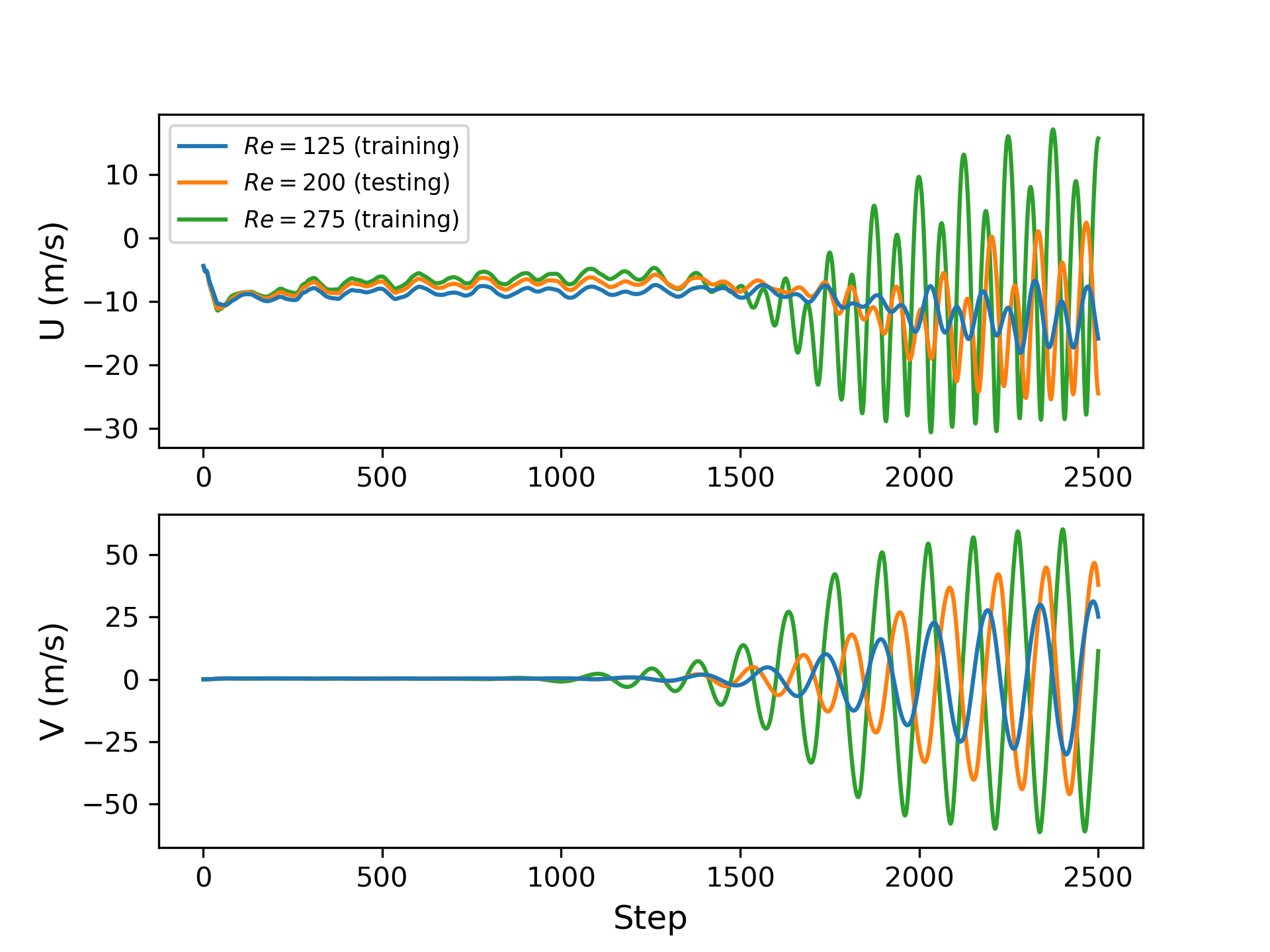}
        \caption{Transient flow over a cylinder: variable history at point monitor}\label{fig cylinder point monitor 2500}
    \end{minipage}\hfill
    \begin{minipage}{0.5\textwidth}
	\centering
    	\includegraphics[width=1\textwidth]{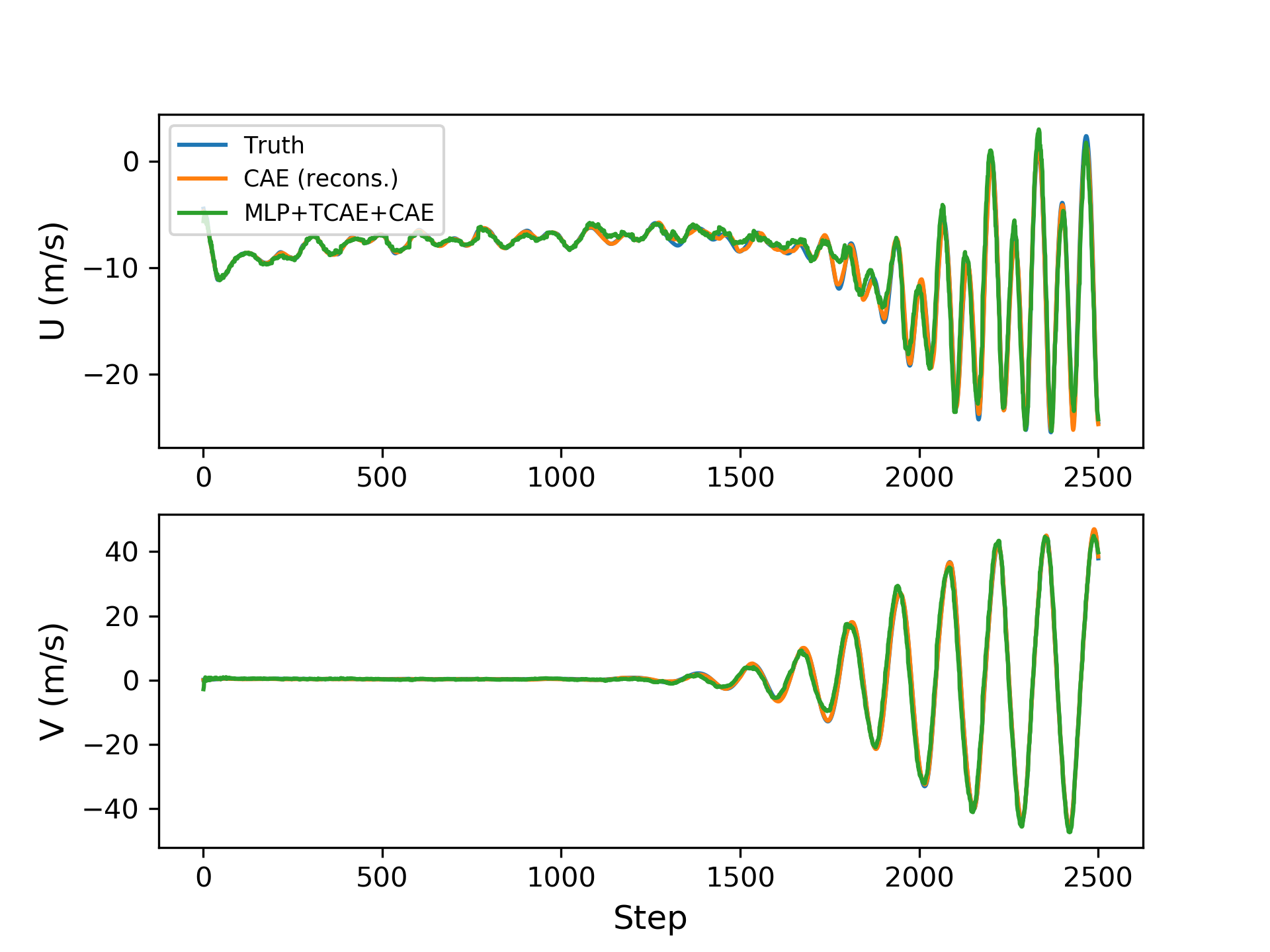}
        \caption{Transient flow over a cylinder new parameter prediction: reconstructed and predicted results at point monitor}\label{fig cylinder point monitor 200 2500}
    \end{minipage}
\end{figure}

In this task, the training data consists of 6 unsteady flow field sequences corresponding to $Re_{\text{train}}=\{125,150,175,225,250,275\}$, and prediction is performed for $Re_{\text{test}}=200$. All sequences are generated from the same steady initial condition corresponding to $Re=20$, each lasting 2500 frames. The sequence length is chosen on purpose to cover the transition to LCO. Depending on $Re$, vortex shedding frequencies are different and the transition to LCO occurs at different rates. A sample contour of flow field is given in Fig.~\ref{fig cylinder contour truth 2500}. To better illustrate the transition process, as well as the difference between different $Re$, a point monitor is placed $1D$ downstream of the center of the cylinder. The velocity at this location is shown for two training parameters $Re=\{125, 275\}$ and the testing parameter $Re=200$ in Fig.~\ref{fig cylinder point monitor 2500}.

\begin{table}[!ht]
	\centering
	\caption {\textcolor{red}{Transient flow over a cylinder: network architectures}}
	\subfloat[CAE\label{table cylinder CAE}]{
        \centering
        \scriptsize
		\begin{tabular}{c| c c c c c c c}
    		\hline
    		\multicolumn{8}{c}{$\mathbf{\Phi}_s$}\\
    		\hline
    		Layer & Input & Conv-pool & Conv-pool & Conv-pool & Conv-pool & Flatten & Dense (Output)\\
    		\hline
    		Output shape & $384\times192\times2$ & $192\times96\times32$ & $96\times48\times64$ & $48\times24\times128$ & $24\times24\times256$ & 147456 & 60\\
    		\hline
    		\hline
    		\multicolumn{8}{c}{$\mathbf{\Psi}_s$}\\
    		\hline
    		Layer & Input & Dense & Reshape & Conv-trans & Conv-trans & Conv-trans & Conv-trans (Output)\\
    		\hline
    		Output shape & 60 & 147456 & $24\times24\times256$ & $48\times24\times128$ & $96\times48\times64$ & $192\times96\times32$ & $384\times192\times2$\\
    		\hline
		\end{tabular}
	}
	
	\subfloat[TCAE\label{table cylinder TCNAE}]{
        \centering
        \small
		\begin{tabular}{c| c c c c c}
    		\hline
    		\multicolumn{6}{c}{$\mathbf{\Phi}_l$}\\
    		\hline
    		Layer & Input & & 12-layer TCN block & Dense & Dense (Output)\\
    		\hline
    		Output shape & $60\times2500$ & & 200 & 100 & 100\\
    		\hline
    		\hline
    		\multicolumn{6}{c}{$\mathbf{\Psi}_l$}\\
    		\hline
    		Layer & Input & Dense & Dense & Repeat vector & 12-layer TCN block (Output)\\
    		\hline
    		Output shape & 100  & 100 & 200 & $200\times2500$ &  $60\times2500$ \\
    		\hline
		\end{tabular}
	}
	
	\subfloat[MLP\label{table cylinder MLP}]{
        \centering
        \small
		\begin{tabular}{c| c c c c}
    		\hline
    		Layer & Input & Dense & Dense & Dense (Output)\\
    		\hline
    		Output shape & 1  & 32 & 256 & 100\\
    		\hline
		\end{tabular}
	}
	
	\subfloat[TCN\label{table cylinder many2one}]{
        \centering
        \small
		\begin{tabular}{c| c c c}
    		\hline
    		Layer & Input & 5-layer TCN block & Dense (Output)\\
    		\hline
    		Output shape & $60\times150$  & $200$ & $60$\\
    		\hline
		\end{tabular}
	}
\end{table}

The top level CAE contains 4 Conv-pool blocks and 1 dense layer in the encoder, with encoded latent dimension $n_s=60$. The decoder assumes a symmetric shape, and the detailed CAE architecture is provided in Table~\ref{table cylinder CAE}. \textcolor{red}{A comparison between the input flow field and the reconstructed variables $\mathbf{\tilde{q}}=\mathbf{\Psi}_s(\mathbf{\Phi}_s(\mathbf{q}))$ for the last frame is provided in Fig.~\ref{fig cylinder contour truth 2500} and \ref{fig cylinder contour CAE 2500}. The point monitored variable history for the reconstructed variable is plotted in Fig.~\ref{fig cylinder point monitor 200 2500}. The maximum reconstruction RAE is 0.19\%.} 

\begin{figure}
	\centering
    \begin{minipage}{0.5\textwidth}
	\centering
    	\includegraphics[width=1\textwidth]{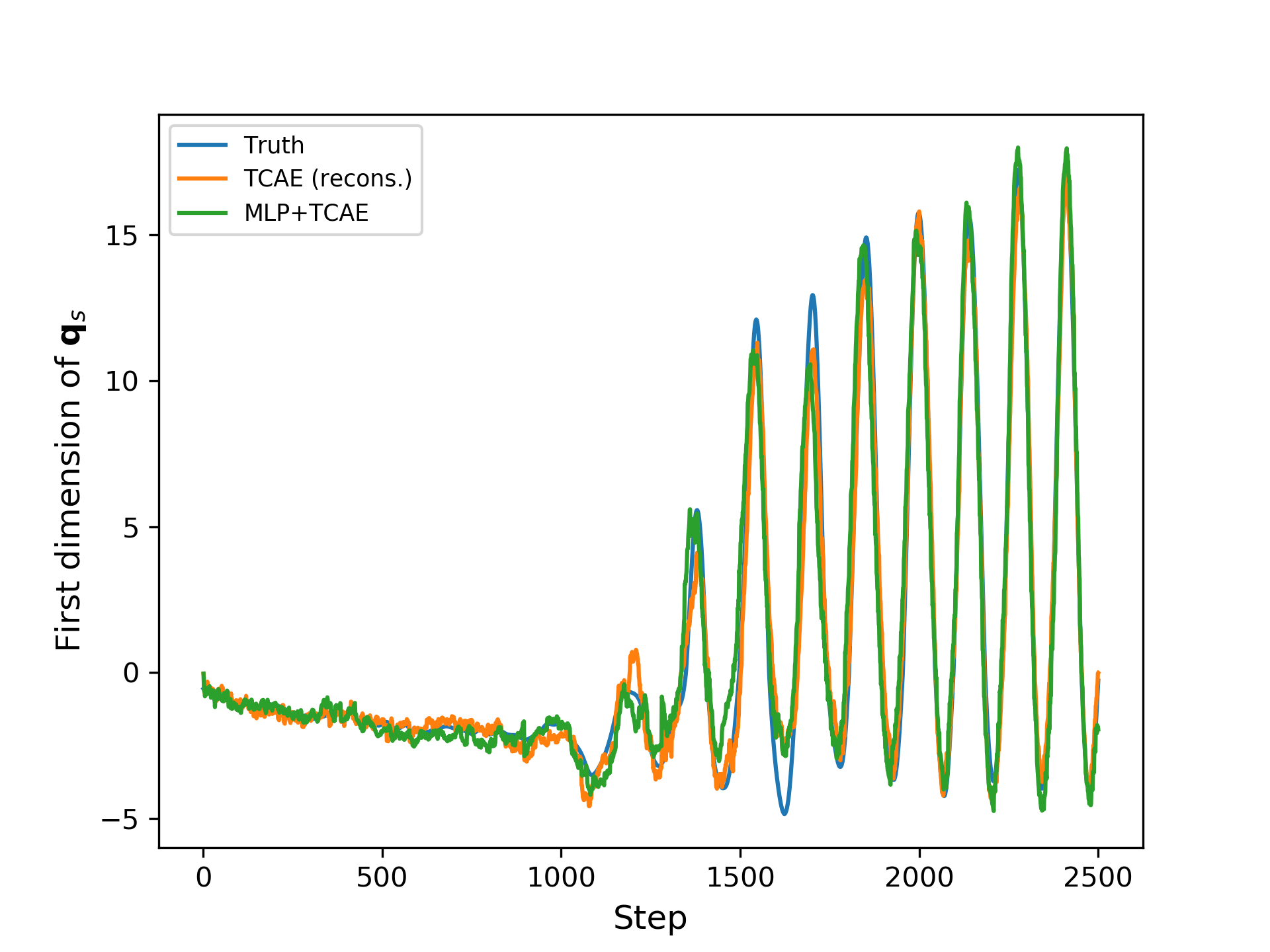}
        \caption{Transient flow over a cylinder new parameter prediction: first dimension of $\mathbf{q}_s$}\label{fig cylinder qs history}
    \end{minipage}\hfill
    \begin{minipage}{0.5\textwidth}
	\centering
    	\includegraphics[width=1\textwidth]{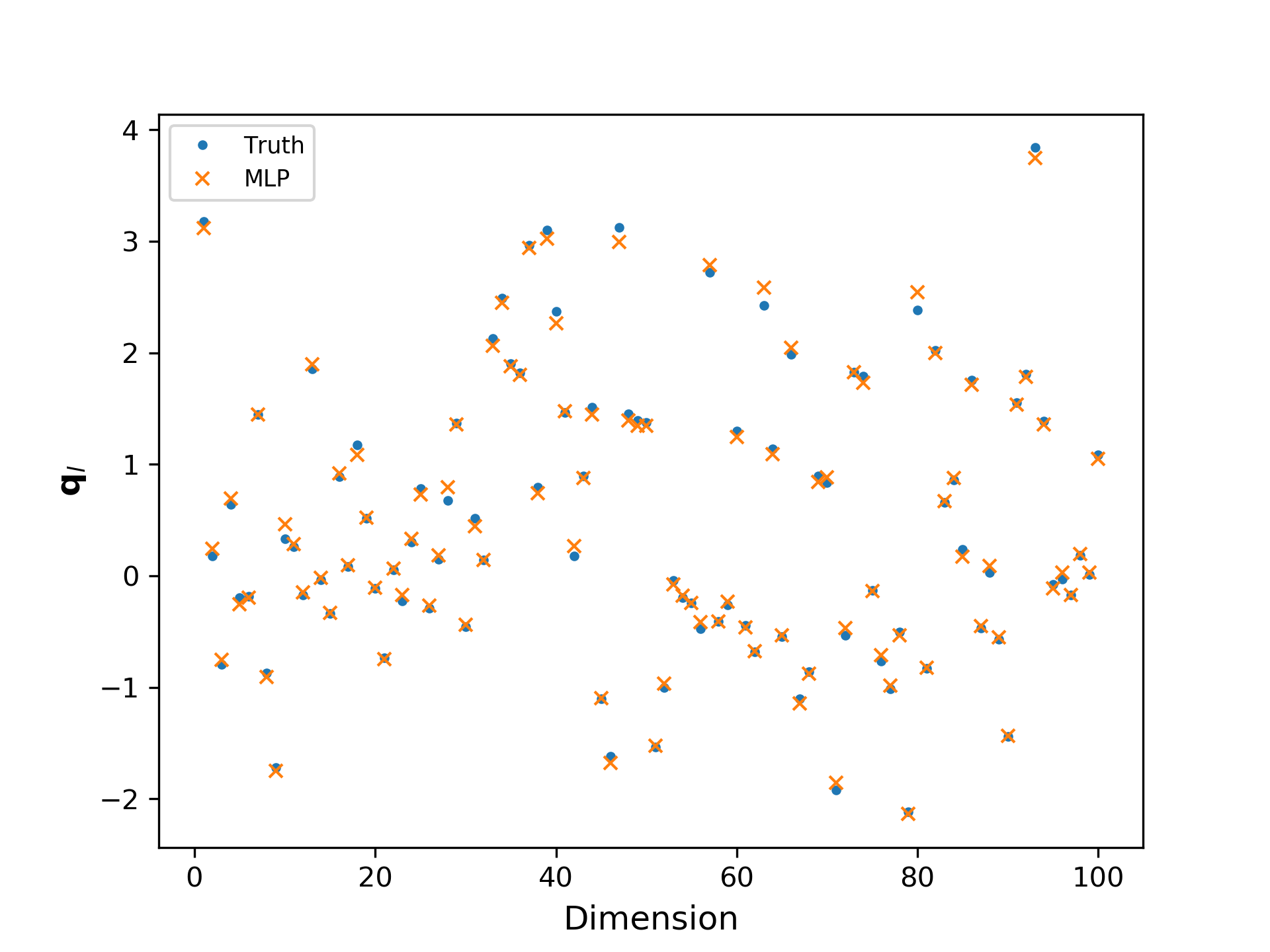}
        \caption{Transient flow over a cylinder new parameter prediction: $\mathbf{q}_l$}\label{fig cylinder ql}
    \end{minipage}
\end{figure}

The spatially encoded variable for the training sequences $\mathbf{q}_s(Re_{train})=\mathbf{\Phi}_s(\mathbf{Q}(Re_{train}))$ is used to train the TCAE. The TCN block in $\mathbf{\Phi}_l$ of the TCAE contains 12 dilated convolutional layers, with uniform kernel size $k=10$ and sequential dilation orders $d=1,2,2^2,\dots,2^{10}, 2^{11}$. Two dense layers are appended after the TCN block, and the encoded latent dimension is $n_l=100$. $l_2$ regularization with a penalty factor $\lambda=\num{1e-8}$ is used in the loss function. The history of the first channel of the true and TCAE reconstructed $\mathbf{Q}_s$ is plotted in Fig.~\ref{fig cylinder qs history}. 

To learn the mapping between $Re$ and $\mathbf{q}_l$, a MLP of 3 dense layers is used. The detailed architecture of the MLP is given in Table~\ref{table cylinder MLP}. The predicted $\mathbf{q}_l^*=\mathcal{R}(Re=200)$ is compared with the truth in Fig.~\ref{fig cylinder ql}. \textcolor{red}{The error in $\mathbf{q}_l^*$ is around 1\%. $\mathbf{\Psi}_l$ and $\mathbf{\Psi}_s$ are then used to decode the final prediction for the flow field. The first channel of the intermediate output $\mathbf{Q}_s^*=\mathbf{\Psi}_l(\mathbf{q}_l^*)$ is compared to the truth in Fig.~\ref{fig cylinder qs history}. The RAE in the velocity components for the final output $\mathbf{Q}^*=\mathbf{\Psi}_s(\mathbf{Q}_s^*)$ is below 1.4\%.} The last frame of the solution is contoured in Fig.~\ref{fig cylinder contour feedforward 2500}, and the point monitor history is shown in Fig.~\ref{fig cylinder point monitor 200 2500}. 

\subsubsection{Combined prediction}
In this task, prediction is performed beyond the training sequences (beyond 2500 frames) at the new parameter $Re=200$ for 500 future frames.

The CAE network is identical to that used in the new parameter prediction task. It is obvious in Fig.~\ref{fig cylinder point monitor 2500} that the unsteady flow field towards the end of the training sequence behaves significantly differently from the first half of the transition process, thus it is redundant to include the latter in the training of the TCN for future state prediction. Instead, only the last 500 frames in $\mathbf{Q}_s(Re_\text{train})$ are used in the training of $\mathcal{P}$. A look-back window of $n_{\tau}=150$ is used which corresponds to approximately one vortex shedding period for $Re=200$. $\mathcal{P}$ consists of a 5-layer \textcolor{red}{strided} TCN with $k=10$ and sequential dilation orders $d=1,2,4,8,16$ with 200 channels followed by a dense layer of output size $n_s=60$. The detailed architecture of $\mathcal{P}$ is given in Table~\ref{table cylinder many2one}.

\begin{figure}
    \begin{minipage}{0.5\textwidth}
	\centering
    	\includegraphics[width=1\textwidth]{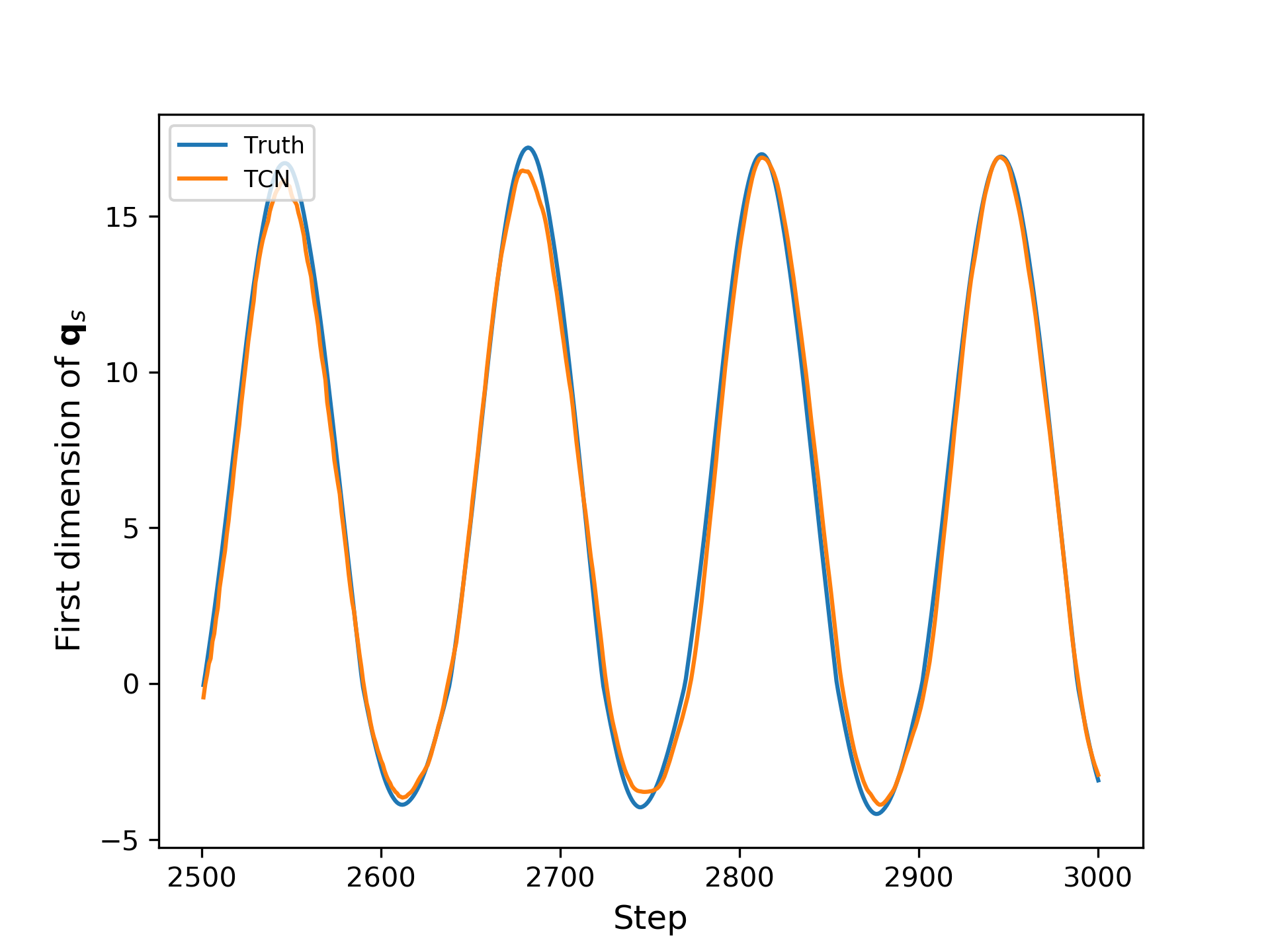}
        \caption{Transient flow over a cylinder combined prediction: first dimension of $\mathbf{q}_s$}\label{fig cylinder qs history 3000_2}
    \end{minipage}
    \begin{minipage}{0.5\textwidth}
	\centering
    	\subfloat{
    		\includegraphics[width=1\textwidth]{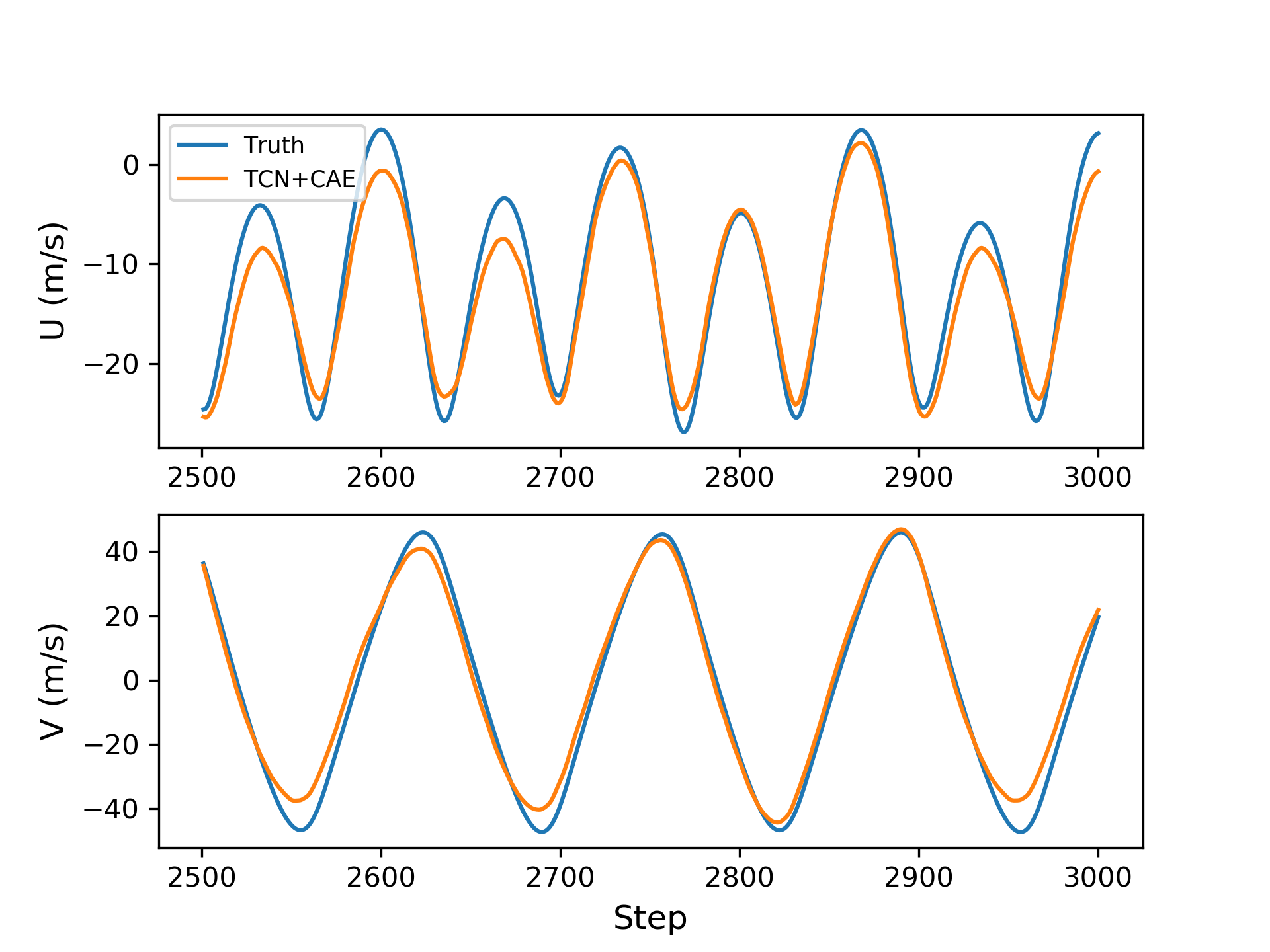}}
    	\caption{Transient flow over a cylinder combined prediction: predicted results at point monitor}\label{fig cylinder point monitor 3000}
    \end{minipage}
\end{figure}

\begin{figure}
	\centering
	\hspace*{\fill}
	\subfloat[Truth\label{fig cylinder contour truth 3000}]{
		\includegraphics[width=0.33\textwidth,trim={3cm 0 2cm 0},clip]{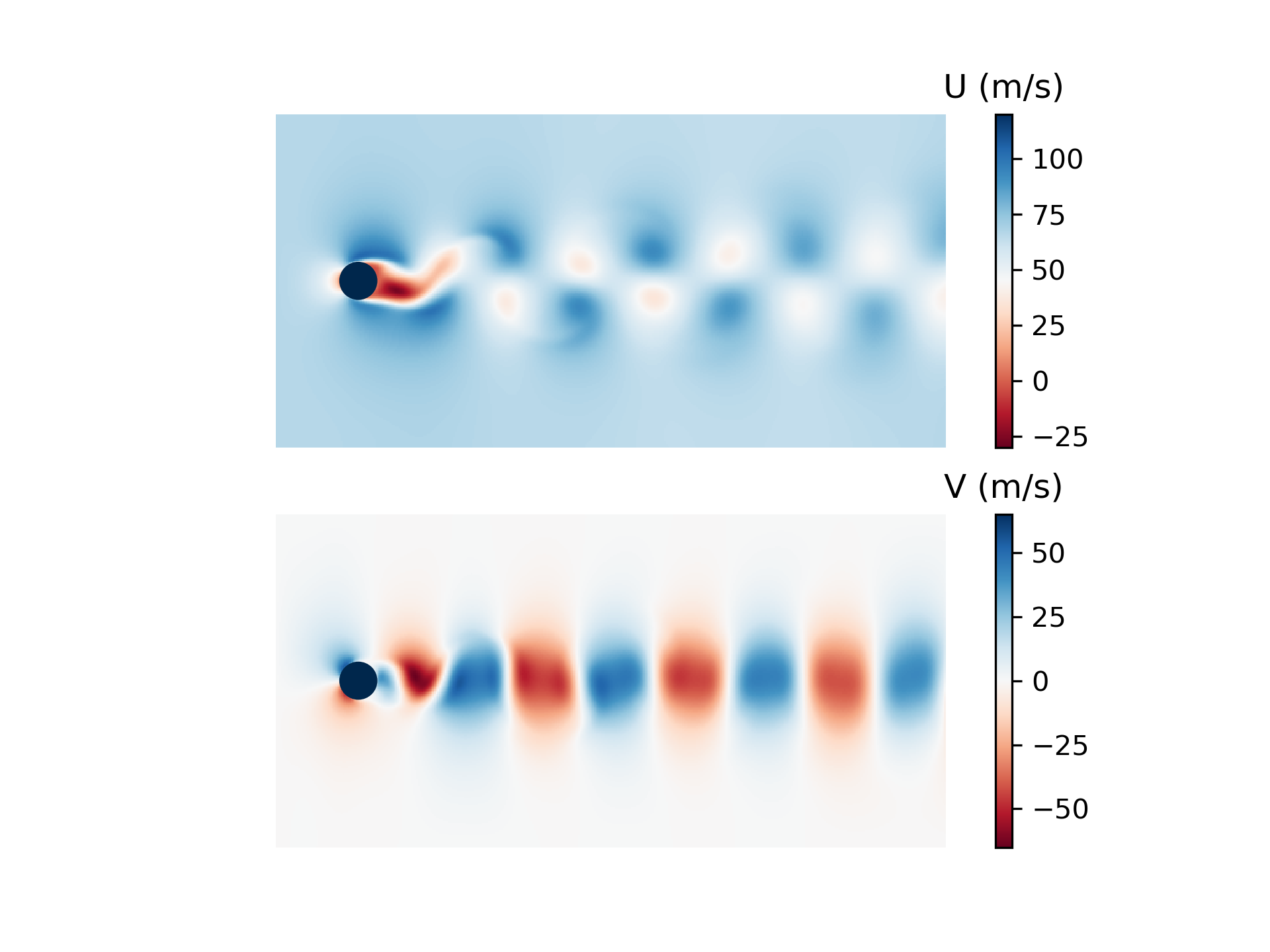}}
    \hfill
	\subfloat[Combined prediction\label{fig cylinder contour combined 3000}]{
		\includegraphics[width=0.33\textwidth,trim={3cm 0 2cm 0},clip]{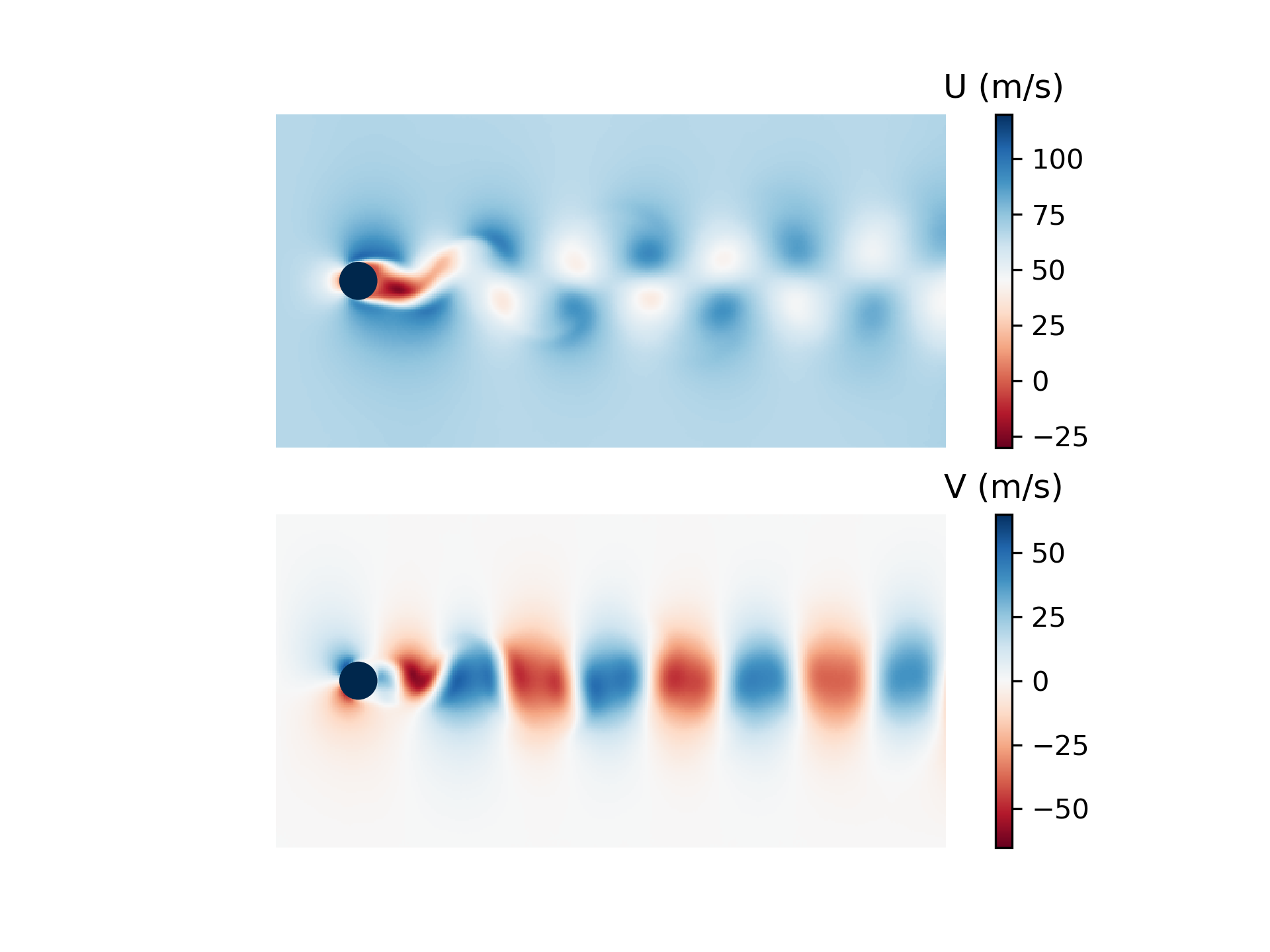}}
	\hspace*{\fill}
	\caption{Transient flow over a cylinder combined prediction: flow field contours for $i=3000, Re=200$}\label{fig cylinder contour future} 
\end{figure}

\textcolor{red}{The predicted evolution of the first latent variable is shown in Fig.~\ref{fig cylinder qs history 3000_2}. The point monitor results of the decoded variables are shown in Fig.~\ref{fig cylinder point monitor 3000}, and the contours for the last predicted step at $i=3000$ are provided in Fig.~\ref{fig cylinder contour combined 3000}. The RAE for the predicted velocity components is below 2\%.}

\begin{table}[!ht]
	\begin{center}
	\caption {\textcolor{red}{Transient flow over a cylinder: errors for different stages of output}}\label{table cylinder error}
    \footnotesize
    
{\color{red}
		\begin{tabular}{c c | c | c| c | c}
		\hline
		\multicolumn{2}{c|}{Stage} & Truth & Output & RAE & MSE\\
		\hline
		\multicolumn{2}{c|}{\multirow{3}{*}{\makecell{Training\\($Re=\{125,150,175,225,250,275\},$\\frames 1 to 2500)}}} & $\mathbf{q}\,(U/V)$ & $\mathbf{\tilde{q}}\,(U/V)=\mathbf{\Psi}_s(\mathbf{\Phi}_s(\mathbf{q}))$ & 0.05\%/0.09\% & 0.010/0.011\\
		& &$\mathbf{Q}_s=\mathbf{\Phi}_s(\mathbf{Q})$ & $\tilde{\mathbf{Q}}_s=\mathbf{\Psi}_l(\mathbf{\Phi}_l(\mathbf{Q}_s))$ & 0.82\% & 0.06\\
		& &$\mathbf{q}_l=\mathbf{\Phi}_(\mathbf{Q}_s)$ & $\mathbf{q}_l^*=\mathcal{R}(Re)$ & 0.56\% & \num{1.2e-3}\\
		\hline
		\multicolumn{1}{c|}{\multirow{6}{*}{\makecell{Testing\\($Re=200$)}}} & \multirow{5}{*}{Frames 1 to 2500} & $\mathbf{q}\,(U/V)$ & $\mathbf{\tilde{q}}\,(U/V)=\mathbf{\Psi}_s(\mathbf{\Phi}_s(\mathbf{q}))$ & 0.11\%/0.19\% & 0.059/0.059\\
		\multicolumn{1}{c|}{}& &$\mathbf{Q}_s=\mathbf{\Phi}_s(\mathbf{Q})$ & $\tilde{\mathbf{Q}}_s=\mathbf{\Psi}_l(\mathbf{\Phi}_l(\mathbf{Q}_s))$ & 1.71\% & 0.31\\
		\multicolumn{1}{c|}{}& &$\mathbf{q}_l=\mathbf{\Phi}_(\mathbf{Q}_s)$ & $\mathbf{q}_l^*=\mathcal{R}(Re)$ & 1.02\% & \num{2.6e-3}\\
		\multicolumn{1}{c|}{}& &$\mathbf{Q}_s=\mathbf{\Phi}_s(\mathbf{Q})$ & $\tilde{\mathbf{Q}}_s^*=\mathbf{\Psi}_l(\mathbf{q}_l^*)$ & 1.99\% & 0.52\\
		\multicolumn{1}{c|}{}& &$\mathbf{Q}\,(U/V)$ & $\tilde{\mathbf{Q}}^*\,(U/V)=\mathbf{\Psi}_s(\tilde{\mathbf{Q}}_s^*)$ & 0.68\%/1.37\% & 2.37/7.02\\
		\hhline{~-----}
		\multicolumn{1}{c|}{} & \multirow{2}{*}{Frames 2501 to 3000} & $\mathbf{q}_s=\mathbf{\Phi}_s(\mathbf{q})$ & $\tilde{\mathbf{q}}_s=\mathcal{P}(\mathbf{Q}_s)$ & 2.94\% & 0.80\\
		\multicolumn{1}{c|}{}& &$\mathbf{q}\,(U/V)$ & $\tilde{\mathbf{q}}^*\,(U/V)=\mathbf{\Psi}_s(\tilde{\mathbf{q}}_s)$ & 1.18\%/1.92\% & 3.72/8.34\\
		\hline
		\end{tabular}
}
	\end{center}  
\end{table}

The computing time by different parts of the framework in training and prediction for the combined prediction task is listed in Table~\ref{table cylinder time}. \textcolor{red}{The total prediction time for 3000 frames for a new Reynolds number is less than 6.2 s. In comparison, the original CFD simulation is conducted with an in-house solver GEMS (General Equation and Mesh Solver) developed by Purdue University~\cite{huang2020investigations}, which uses a implicit time marching method with 10 sub-iterations per time step. The simulation takes 15870.2 s on a 18-core Intel Xeon Gold 6154 CPU running at 3.70GHz. Thus, more than three orders of magnitude acceleration is achieved.} 

\begin{table}[!ht]
	\begin{center}
	\caption {Transient flow over a cylinder: computing time}\label{table cylinder time}
    \footnotesize
		\begin{tabular}{c| c c c c c}
		\hline
		\multicolumn{6}{c}{Training}\\
		\hline
		Component & CAE & TCAE & MLP & TCN & Total \\
		\hline
		Training epochs & 200 & 1500 & 2000 & 300 & -\\
		Computing time (s) & 8735 & 979 & 84 & 343 & 10141\\
		\hline
		\hline
		\multicolumn{6}{c}{Prediction}\\
		\hline
		Component & $\mathbf{\Psi}_s$ (3000 frames) & $\mathbf{\Psi}_l$ (2500 frames) & MLP & TCN (500 frames) & Total (3000 frames)\\
		\hline
		Computing time (s) & 5.46 & 0.13 & 0.011 & 0.57 & 6.171\\
		\hline
		\end{tabular}
	\end{center}  
\end{table}

\subsection{Transient \textcolor{red}{ship} airwake}\label{sec airwake}
In this test, a new parameter prediction is performed on a three-dimensional flow behind a shipstructure. The inflow side-slip angle $\alpha$, i.e. the angle between inflow and ship cruising directions, is taken as the studied global parameter.  Small variations in $\alpha$ introduce significant changes in the flow structures in the wake of the ship. The streamwise ($x$-direction), beamwise ($z$-direction) and vertical ($z$-direction) velocity components, $U,V,W$ are the variables of interest. The ship geometry is based on the Simple Frigate Shape Version 2 (SFS2) model~\cite{lee2005simulation}, which features a double-level ship structure that results in significant flow separation over the deck. More details about the model can be found in Ref.~\cite{lee2005simulation}

\begin{figure}
	\centering
	\subfloat[$\alpha=0\degree$]{
    	\begin{minipage}{0.33\linewidth}
    		\includegraphics[width=0.9\textwidth]{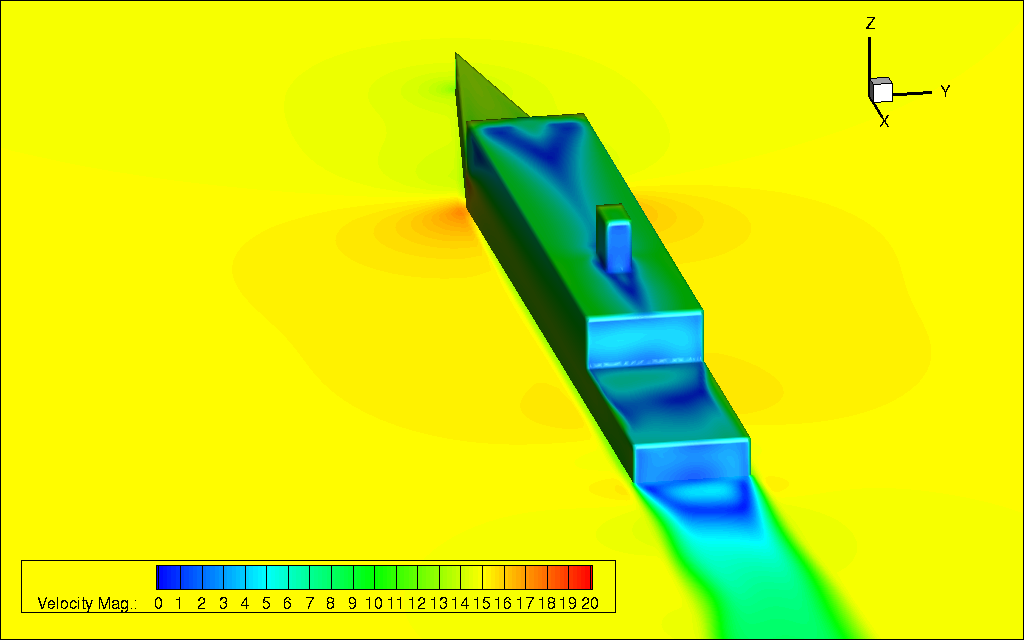}\\
    		\includegraphics[width=1\textwidth,trim={1cm 0 2cm 0},clip]{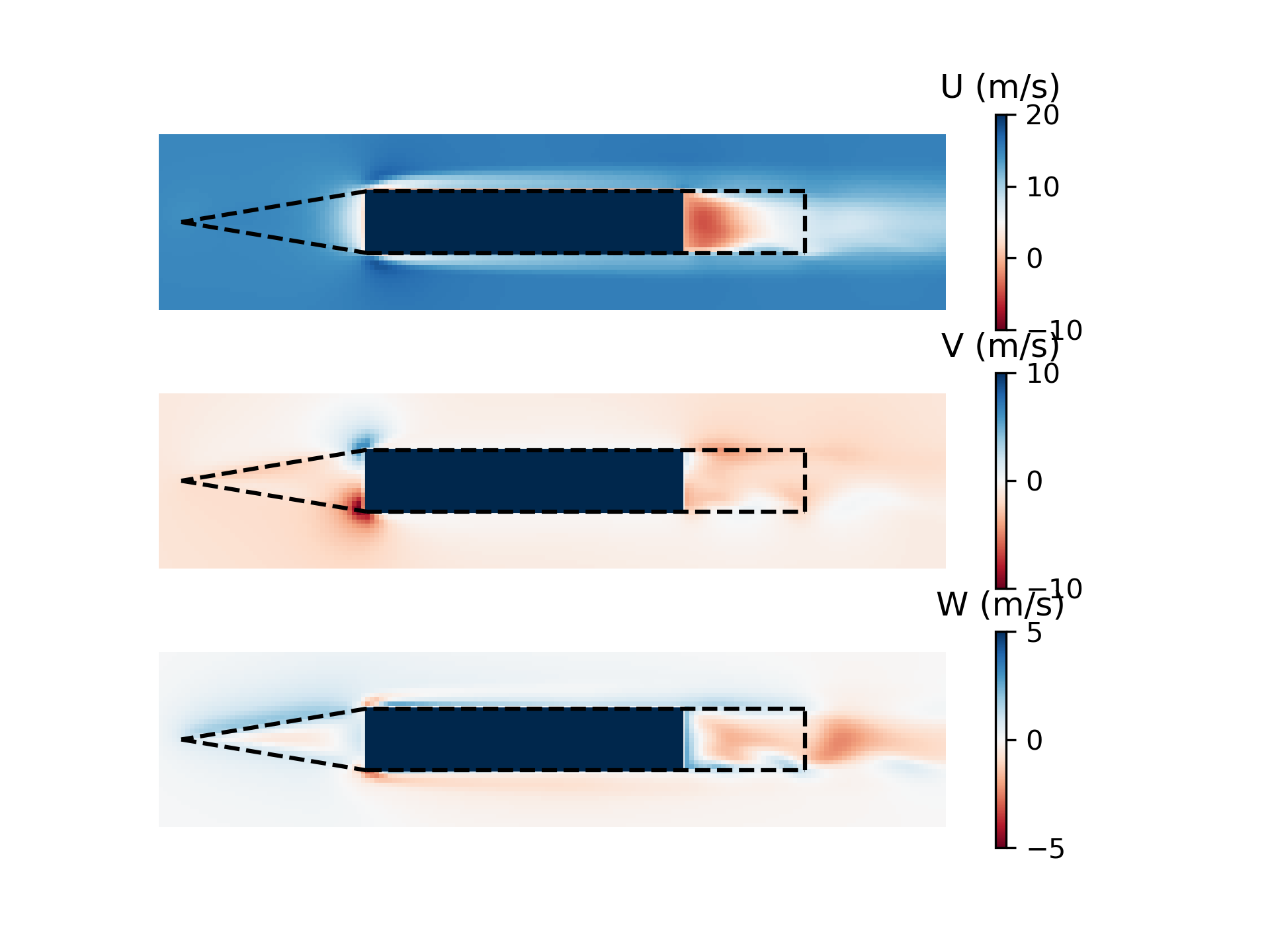}
    	\end{minipage}
    }
	\subfloat[$\alpha=10\degree$]{
    	\begin{minipage}{0.33\linewidth}
        	\includegraphics[width=0.9\textwidth]{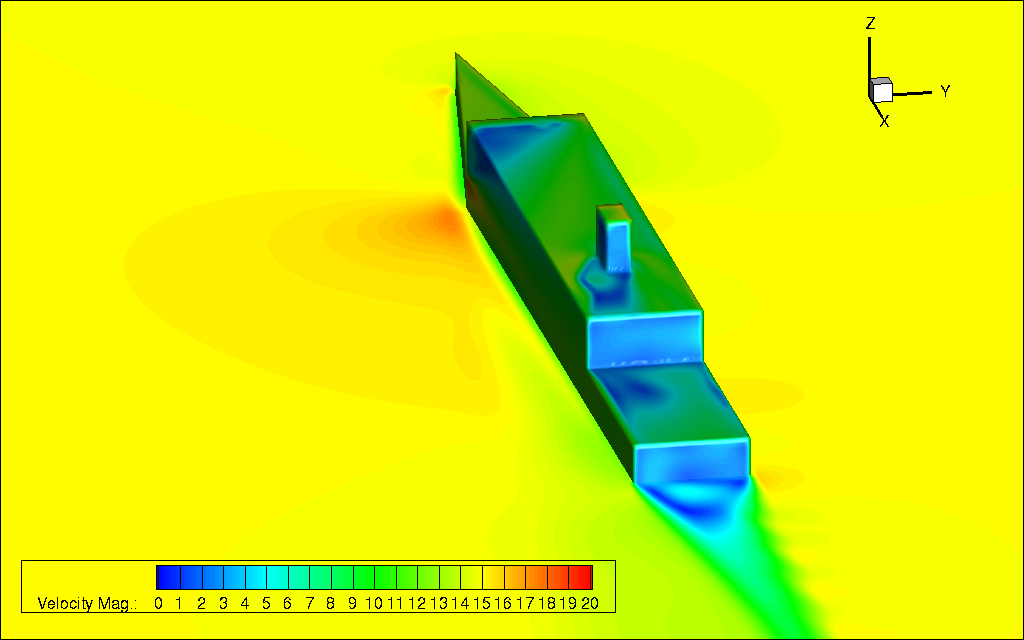}\\
    		\includegraphics[width=1\textwidth,trim={1cm 0 2cm 0},clip]{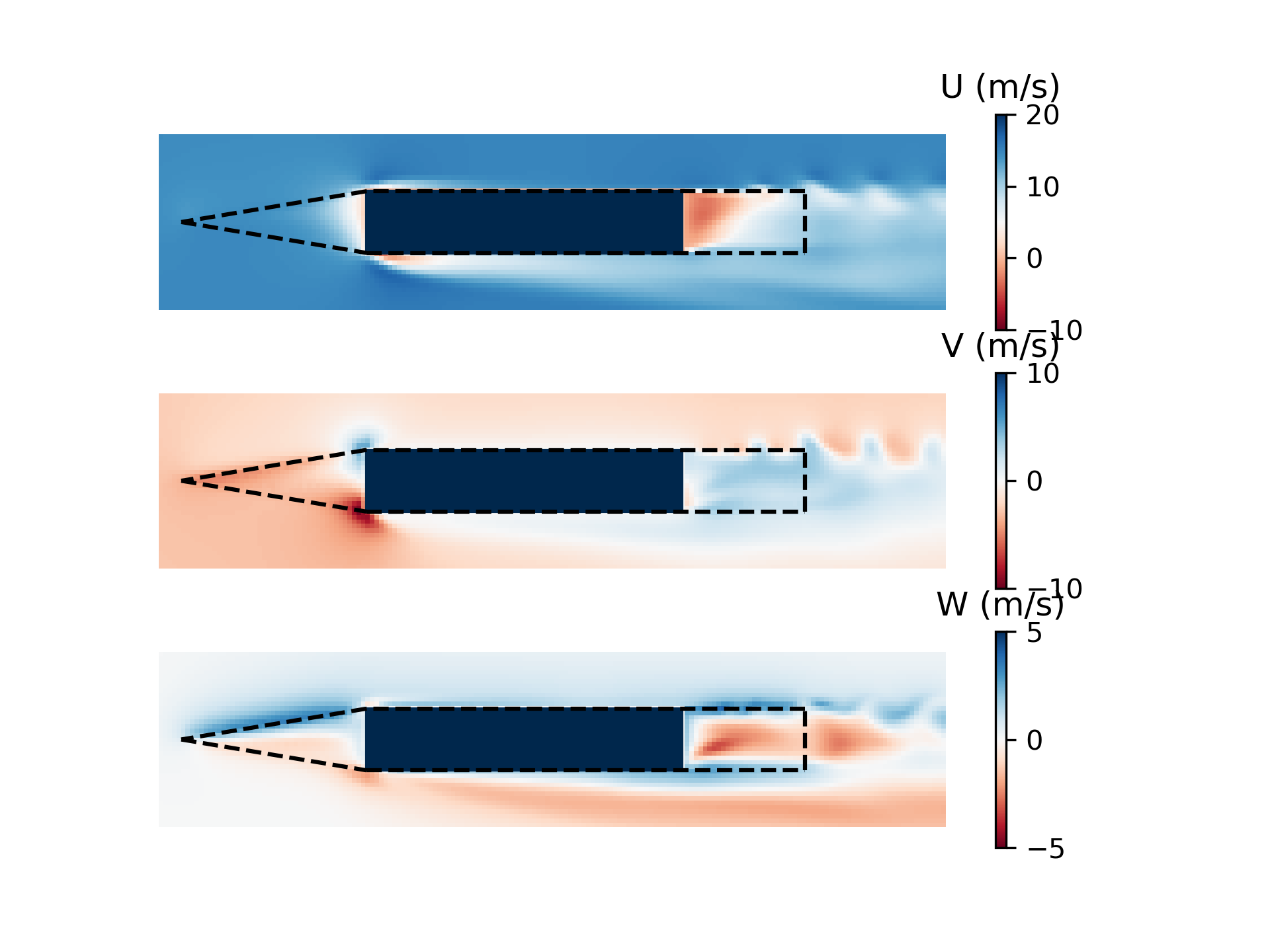}
    	\end{minipage}
    }
	\subfloat[$\alpha=20\degree$]{
    	\begin{minipage}{0.33\linewidth}
        	\includegraphics[width=0.9\textwidth]{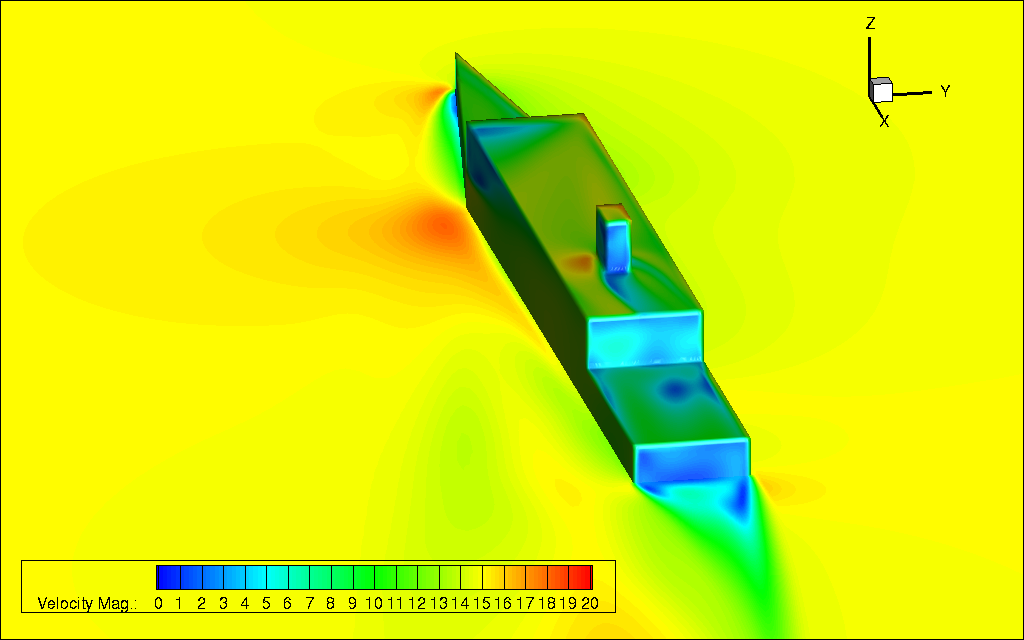}\\
    		\includegraphics[width=1\textwidth,trim={1cm 0 2cm 0},clip]{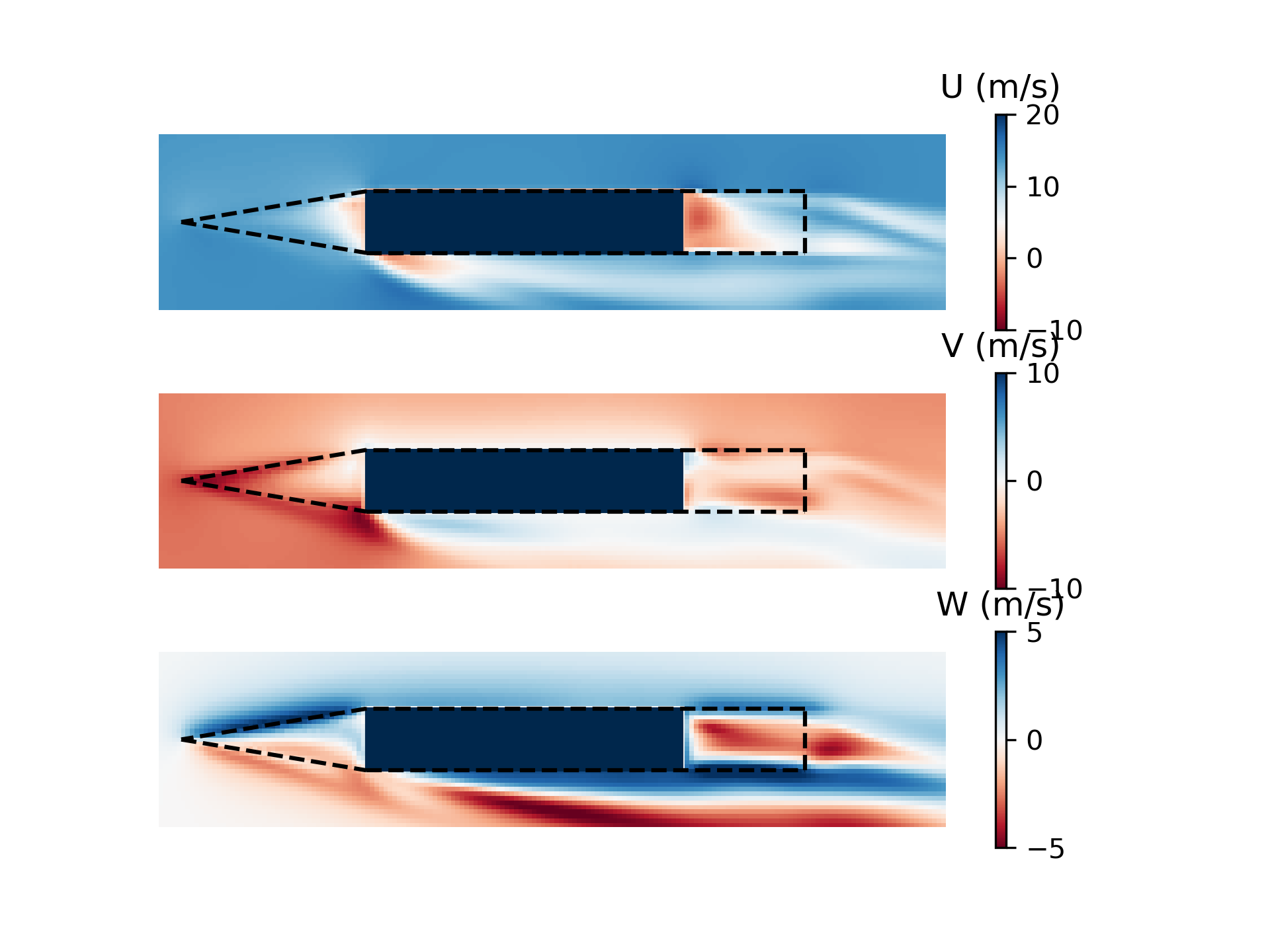}
    	\end{minipage}
    }
	\caption{Transient ship airwake: sample frame of settled unsteady patterns. Top: velocity magnitudes; bottom: velocity components.}\label{fig airwake} 
\end{figure}

The \textcolor{red}{ground truth} flow field is computed using unsteady Reynolds Averaged Navier-Stokes equations with the $k-\omega$ turbulence model~\cite{wilcox1988reassessment}. 4 sets of training data corresponding to the new inflow angles $\alpha=\{5\degree,10\degree,15\degree,20\degree\}$ are provided, and prediction is performed for $\alpha=12.5\degree$.
In all the cases, the ship is originally cruising in a headwind condition, i.e. $\alpha=0\textcolor{red}{\degree}$, then the side-slip angle is impulsively changed, causing the flow to transition to a different unsteady state.  
The velocity magnitude for a sample frame of the settled unsteady pattern for three different $\alpha$ and the corresponding velocity components on a x-y plane 5 meters above the sea level (slightly above the deck) are shown in Fig.~\ref{fig airwake}.

Predictions are focused on a 3D near-field region surrounding the ship of dimensions length $\times$ width $\times$ height = 175 m $\times$ 39 m $\times$ 23 m. The size of the interpolated Cartesian grid is $n_x\times n_y\times n_z=176\times 40\times 24$. Each data set includes 400 frames corresponding to 40 s of physical time.  \textcolor{red}{Errors for different stages of output are summarized in Table~\ref{table airwake error}.}

\begin{table}[!ht]
	\centering
	\caption {\textcolor{red}{Transient ship airwake: network architectures}}
	\subfloat[CAE\label{table airwake CAE}]{
        \centering
        \notsotiny
		\begin{tabular}{c| c c c c c c c}
    		\hline
    		\multicolumn{8}{c}{$\mathbf{\Phi}_s$}\\
    		\hline
    		Layer & Input & Conv-pool & Conv-pool & Conv-pool & Conv-pool & Flatten & Dense (Output)\\
    		\hline
    		Output shape & $400\times40\times24\times3$ & $200\times20\times12\times64$ & $100\times10\times6\times128$ & $50\times5\times6\times256$ & $25\times5\times6\times256$ & 192000 & 20\\
    		\hline
    		\hline
    		\multicolumn{8}{c}{$\mathbf{\Psi}_s$}\\
    		\hline
    		Layer & Input & Dense & Reshape & Conv-trans & Conv-trans & Conv-trans & Conv-trans (Output)\\
    		\hline
    		Output shape & 20 & 192000 & $25\times5\times6\times256$ & $50\times5\times6\times256$ & $100\times10\times6\times128$ & $200\times20\times12\times64$ & $400\times40\times24\times3$\\
    		\hline
		\end{tabular}
	}
	
	\subfloat[TCAE\label{table airwake TCNAE}]{
        \centering
        \small
		\begin{tabular}{c| c c c c}
    		\hline
    		\multicolumn{5}{c}{$\mathbf{\Phi}_l$}\\
    		\hline
    		Layer & Input & & 9-layer TCN block & Dense (Output)\\
    		\hline
    		Output shape & $20\times400$ & & 400 & 20\\
    		\hline
    		\hline
    		\multicolumn{5}{c}{$\mathbf{\Psi}_l$}\\
    		\hline
    		Layer & Input & Dense & Repeat vector & 9-layer TCN block (Output)\\
    		\hline
    		Output shape & 20 & 400 & $400\times400$ &  $20\times400$ \\
    		\hline
		\end{tabular}
	}
	
	\subfloat[MLP\label{table airwake MLP}]{
        \centering
        \small
		\begin{tabular}{c| c c c c}
    		\hline
    		Layer & Input & Dense & Dense & Dense (Output)\\
    		\hline
    		Output shape & 1  & 16 & 50 & 20\\
    		\hline
		\end{tabular}
	}
\end{table}

\begin{figure}
	\centering
	\subfloat[Truth, $\mathbf{q}$\label{fig airwake contour truth}]{
		\includegraphics[width=0.33\textwidth,trim={1cm 0 2cm 0},clip]{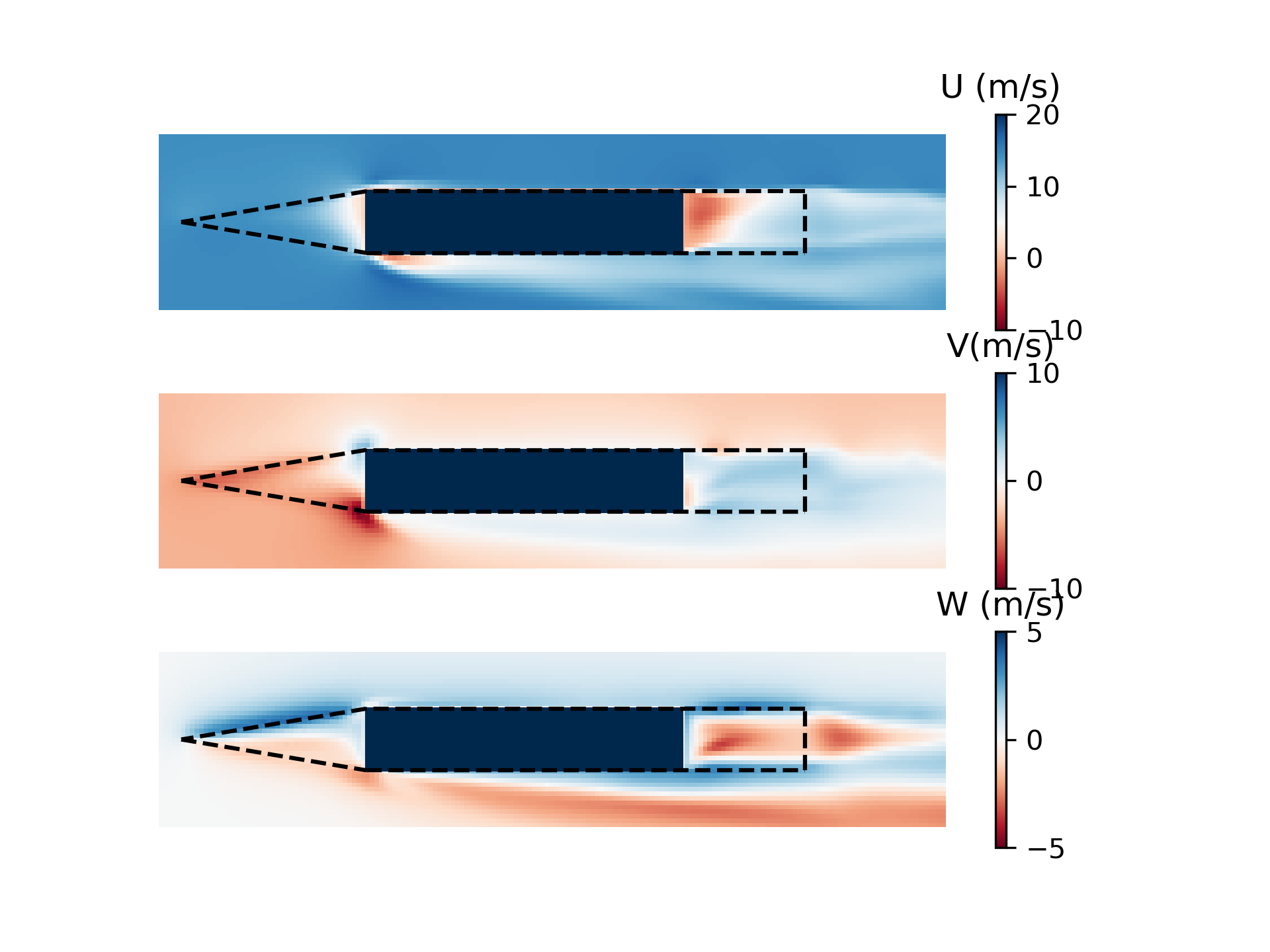}}
	\subfloat[CAE reconstruction\label{fig airwake contour CAE}]{
		\includegraphics[width=0.33\textwidth,trim={1cm 0 2cm 0},clip]{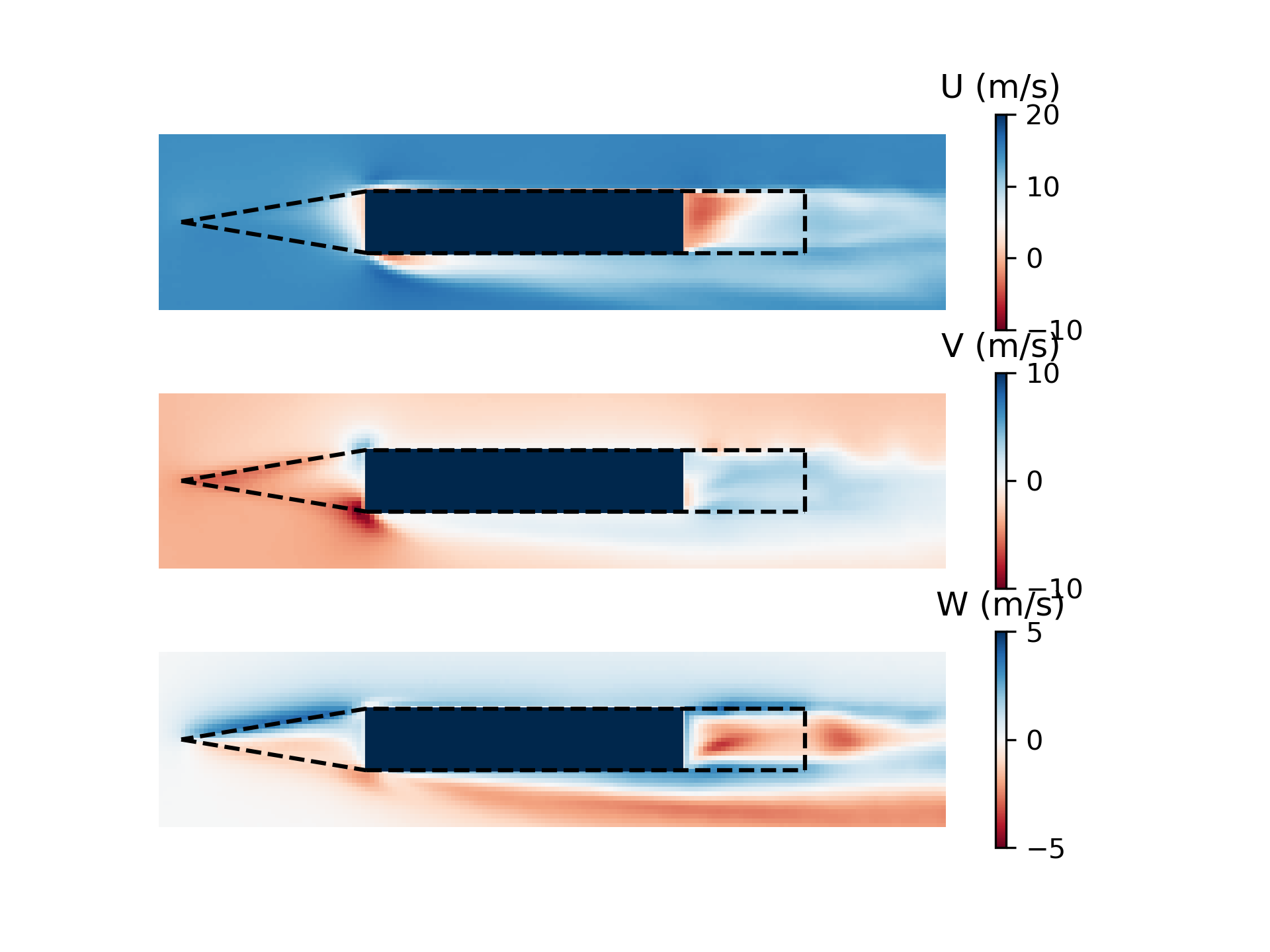}}
	\subfloat[MLP+TCAE+CAE\label{fig airwake contour feedforward}]{
		\includegraphics[width=0.33\textwidth,trim={1cm 0 2cm 0},clip]{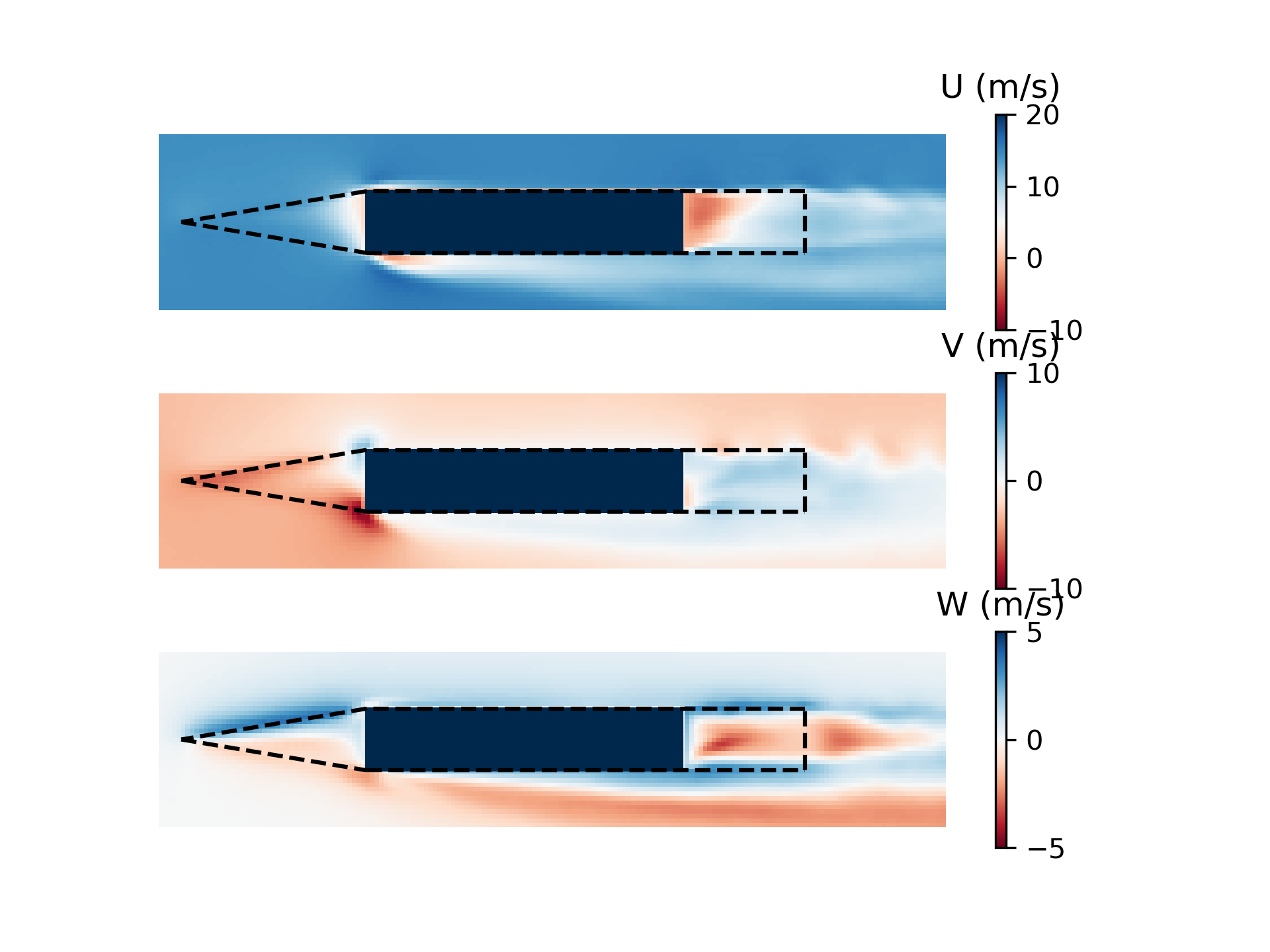}}
	\caption{Transient ship airwake: x-y plane velocity components for $i=400, \alpha=12.5\degree$}\label{fig airwake contour} 
\end{figure}

\begin{figure}
	\centering
    \begin{minipage}{0.5\textwidth}
	\centering
    	\includegraphics[width=1\textwidth]{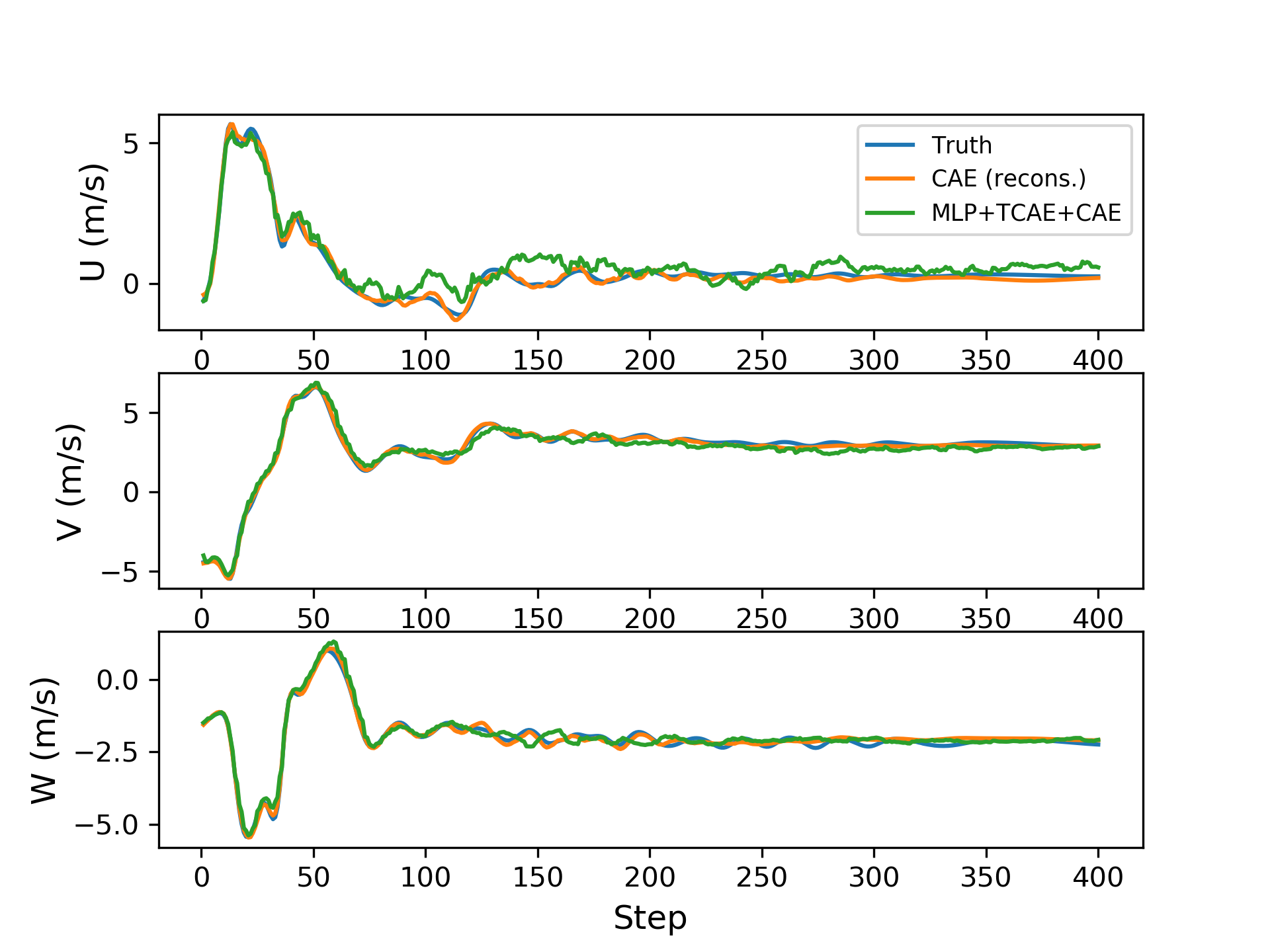}
        \caption{Transient ship airwake: reconstructed and predicted results at point monitor}\label{fig airwake point monitor}
    \end{minipage}\hfill
    \begin{minipage}{0.5\textwidth}
	\centering
    	\includegraphics[width=1\textwidth]{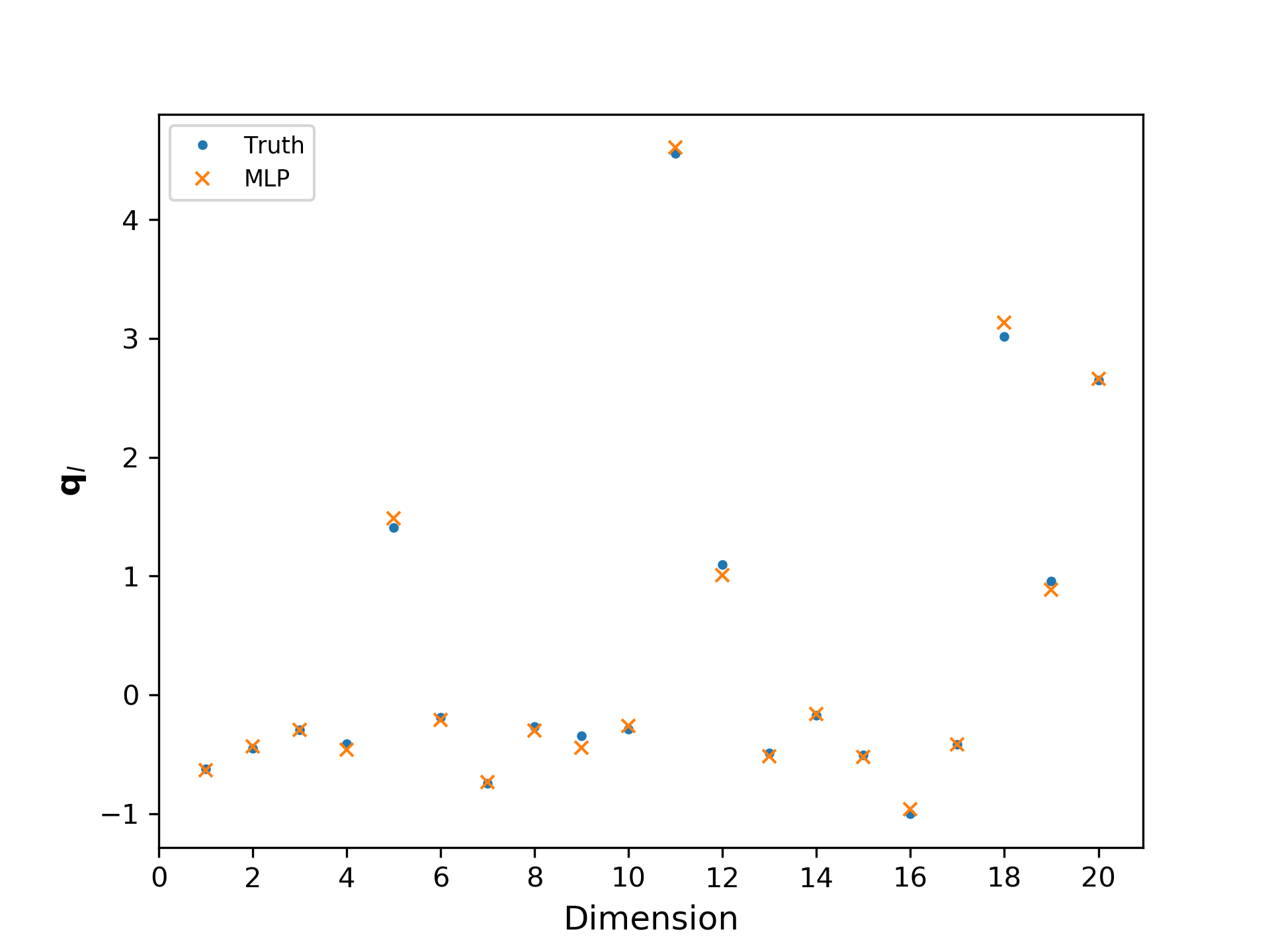}
        \caption{Transient ship airwake: $\mathbf{q}_l$}\label{fig airwake ql}
    \end{minipage}
\end{figure}

The top level CAE contains 4 Conv-pool blocks and 1 dense layer in the encoder with a encoded latent dimension of $n_s=20$. The detailed CAE architecture is provided in Table~\ref{table airwake CAE}. \textcolor{red}{The CAE reconstruction $\mathbf{\tilde{q}}=\mathbf{\Psi}_s(\mathbf{\Phi}_s(\mathbf{q}))$ shows a RAE around 0.1\%. A comparison between the original and reconstructed flow field for $\alpha=12.5\degree$ at the 400-th frame is provided in Fig.~\ref{fig airwake contour truth} and \ref{fig airwake contour CAE}. The monitored history for the reconstructed variables at a point 1 meter above the center of the deck of the SFS2 model is compared with the ground truth in Fig.~\ref{fig airwake point monitor}.} 

The TCN block in $\mathbf{\Phi}_l$ of the TCAE contains 9 dilated convolution layers, with a uniform kernel size $k=10$ and sequential dilation orders $d=1,2,2^2,\dots,2^{7}, 2^{8}$. A dense layer is appended after the TCN block, and the encoded latent dimension is $n_l=20$. \textcolor{red}{The reconstructed sequence  $\tilde{\mathbf{Q}}_s=\mathbf{\Psi}_l(\mathbf{\Phi}_l(\mathbf{Q}_s))$} shows a RAE of 4.47\% in the testing stage.

A MLP of 3 dense layers is used for the third level, the detail\textcolor{red}{s} of which is given in Table~\ref{table airwake MLP}. The accuracy of the \textcolor{red}{MLP} prediction is assessed in Fig.~\ref{fig airwake ql}.
\textcolor{red}{The RAE for final predicted velocity components $\mathbf{Q}^*=\mathbf{\Psi}_s(\mathbf{\Psi}_l(\mathbf{q}_l^*))$ are below 0.9\%.} The last frame of the solution is shown in Fig.~\ref{fig airwake contour feedforward}. The predictions are seen to be accurate, despite significant variations in unsteady patterns between the testing condition and the training ones represented in Fig.~\ref{fig airwake contour}. The point monitor history is shown in Fig.~\ref{fig airwake point monitor} where the trend of dynamics is captured closely.

The computing time by different parts of the framework in training and prediction for the combined prediction task is listed in Table~\ref{table airwake time}. \textcolor{red}{The total prediction time for 400 frames of the 3D unsteady flow field is about 6.6 s. In comparison, the original CFD simulation takes 4011.5 s using a commercial parallel solver Ansys Fluent on a six-core Intel Xeon E5-1650 CPU running at 3.60GHz. Thus, $>600\times$ reduction in computational cost is achieved.}

\begin{table}[!ht]
	\begin{center}
	\caption {\textcolor{red}{Transient ship airwake: errors for different stages of output}}\label{table airwake error}
    \footnotesize
{\color{red}
		\begin{tabular}{c | c | c| c | c}
		\hline
		Stage & Truth & Output & RAE & MSE\\
		\hline
		\multirow{3}{*}{\makecell{Training\\($\alpha=\{5\degree,10\degree,15\degree,20\degree\}$)}} & $\mathbf{q}\,(U/V/W)$ & $\mathbf{\tilde{q}}\,(U/V/W)=\mathbf{\Psi}_s(\mathbf{\Phi}_s(\mathbf{q}))$ & 0.12\%/0.09\%/0.08\% & \num{2.0e-3}/\num{5.9e-4}/\num{2.5e-4}\\
		&$\mathbf{Q}_s=\mathbf{\Phi}_s(\mathbf{Q})$ & $\tilde{\mathbf{Q}}_s=\mathbf{\Psi}_l(\mathbf{\Phi}_l(\mathbf{Q}_s))$ & 0.63\% & 2.18\\
		&$\mathbf{q}_l=\mathbf{\Phi}_(\mathbf{Q}_s)$ & $\mathbf{q}_l^*=\mathcal{R}(\alpha)$ & 0.60\% & \num{1.6e-3}\\
		\hline
		\multirow{5}{*}{\makecell{Testing\\($\alpha=12.5\degree$)}} & $\mathbf{q}\,(U/V/W)$ & $\mathbf{\tilde{q}}\,(U/V/W)=\mathbf{\Psi}_s(\mathbf{\Phi}_s(\mathbf{q}))$ & 0.30\%/0.38\%/0.29\% & \num{1.4e-2}/\num{5.8e-3}/\num{2.7e-3}\\
		&$\mathbf{Q}_s=\mathbf{\Phi}_s(\mathbf{Q})$ & $\tilde{\mathbf{Q}}_s=\mathbf{\Psi}_l(\mathbf{\Phi}_l(\mathbf{Q}_s))$ & 4.47\% & 65.54\\
		&$\mathbf{q}_l=\mathbf{\Phi}_(\mathbf{Q}_s)$ & $\mathbf{q}_l^*=\mathcal{R}(\alpha)$ & 0.85\% & \num{2.6e-3}\\
		&$\mathbf{Q}_s=\mathbf{\Phi}_s(\mathbf{Q})$ & $\tilde{\mathbf{Q}}_s^*=\mathbf{\Psi}_l(\mathbf{q}_l^*)$ & 4.44\% & 65.82\\
		&$\mathbf{Q}\,(U/V/W)$ & $\tilde{\mathbf{Q}}^*\,(U/V/W)=\mathbf{\Psi}_s(\tilde{\mathbf{Q}}_s^*)$ & 0.51\%/0.89\%/0.62\% & 0.045/0.049/0.014\\
		\hline
		\end{tabular}
}
	\end{center}  
\end{table}

\begin{table}[!ht]
	\begin{center}
	\caption {Transient ship airwake: computing time}\label{table airwake time}
    \small
		\begin{tabular}{c| c c c c}
		\hline
		\multicolumn{5}{c}{Training}\\
		\hline
		Component & CAE & TCAE & MLP & Total \\
		\hline
		Training epochs & 500 & 500 & 2500 & -\\
		Computing time (s) & 12705 & 213 & 27 &12945\\
		\hline
		\hline
		\multicolumn{5}{c}{Prediction}\\
		\hline
		Component & $\mathbf{\Psi}_s$ (400 frames) & $\mathbf{\Psi}_l$ (400 frames) & MLP & Total (400 frames)\\
		\hline
		Computing time (s) & 5.71 & 0.89 & 0.005 & 6.605\\
		\hline
		\end{tabular}
	\end{center}  
\end{table}

\section{Perspectives}\label{sec perspectives}
Overall, the results suggest that given adequate data and careful training, effective data-driven models can be constructed using multi-level neural networks.
A pertinent question - and one of the original motivations for the authors to pursue this research - is how the present approach compares to 
classical and emerging intrusive projection-based model reduction techniques. The input data for this framework is similar to that used in projection-based model reduction techniques. The present approach is however completely non-intrusive in the sense that the governing equations and partial differential equation solvers  and codes therein are not used. These techniques, therefore, fall in the class of non-intrusive operator inference methods such as those presented in Refs.~\cite{peherstorfer2015online,kramer2019nonlinear,wang2019non}, which leverage POD rather than neural networks to identify the first level (spatially encoded) of latent variables $\mathbf{Q}_s$. 

The non-intrusive nature of the framework greatly simplifies the development process and time. Further, the proposed techniques are orders of magnitude less expensive in execution time than standard projection-based models. Indeed, model training was performed in a few GPU hours, and all the predictions were executed in a few seconds. The availability of efficient open-source libraries for deep learning is another advantage of this approach.

 To the contrary, it can be argued that the use of governing equations such as in projection-based reduced order models can reduce the amount of data required, and present more opportunities to enforce physical constraints. Although the authors have not seen clear evidence of advantages of intrusive methods based on a few experiments, this aspect nevertheless requires disciplined evaluations on a series of benchmark problems. It has to be recognized, however, that intrusive ROMs are prone to and highly sensitive to numerical oscillations and require special treatment~\cite{huang2020investigations,huang2020data} to ensure robustness in complex flows, whereas non-intrusive ROMs are much more flexible.  
 
 Lee and Carlberg~\cite{lee2018model} use autoencoders to identify the latent space, and solve the governing equations on this manifold using Galerkin and Petrov-Galerkin approaches, instead of TCN-based time steppers as in the present work. Ref.~\cite{lee2018model} is an intrusive approach, and thus requires further development of sparse sampling techniques to be more efficient than the full order model it approximates. Nevertheless, such techniques possess the added appeal of guarantees on consistency and optimality, and would be a natural candidate to evaluate the pro and cons of the present approach. 
 
 A general critique on neural network-based approaches is that they require ``more data'' compared to linear, and more structured compression techniques. Expressivity and generalization has to be balanced with the cost of obtaining data. Bayesian approaches can be used to construct operators and to optimally design data-generating experiments. Data requirements can also be reduced by including the governing equations in the loss function~\cite{raissi2019physics}. Moving beyond the reputation of neural networks being ``black boxes,'' there are  prospects to enforce additional structure. For instance, Ref.~\cite{pan2019physics} imposes structured embedding with guarantees on stability. Transformation of the state variables either by hand~\cite{swischuk2019learning} or using learning algorithms~\cite{gin2019deep,pan2019physics} may also be leveraged.

Finally, it should be recognized that the convolution operations pursued in this work are defined on Cartesian grids. Replacing the Euclidean convolution operations by graph convolutions~\cite{kipf2016semi} would enable application to predictions over generalized unstructured meshes.


\section{Summary}\label{sec conclusion}
A multi-level deep learning approach is proposed for the direct prediction of spatio-temporal dynamics. The framework is designed to address  parametric and future state prediction. The data is processed as a uniformly sampled sequence of time snapshots of the spatial field. A convolutional autoencoder (CAE) serves as the top level, encoding each time snapshot into a set of latent variables. Temporal convolutional networks (TCNs) serve as the second level to process the output sequence. The TCN is a unique feature of the framework. Unlike in standard convolutions in which the receptive field growths linearly with the number of layers and the kernel size, dilated convolutions in the TCN allow for exponential growth of the reception field, leading to more efficient processing of  long temporal sequences.
 For parametric predictions, a TCAE is used to further encode the sequence of spatially encoded variables $\mathbf{Q}_s$ along the temporal dimension into a second set of latent variables $\mathbf{q}_l$, which is the encoded spatio-temporal evolution. A multi-layer perceptron (MLP) is used as the third level to learn the mapping between $\mathbf{q}_l$ and the global parameters $\mu$. For future state predictions, the second level iteratively uses a TCN to predict $\mathbf{Q}_s(\mu)$. In either type of task, outputs at the bottom level are decoded to obtain the predictions for high-dimensional spatio-temporal dynamics for desired conditions. 

Numerical tests in a one-dimensional compressible flow problem demonstrate the capability of the framework to accurately predict the motion of discontinuities and waves beyond the training period. It is notable that this capability is impossible in linear basis techniques (such as POD), even when the governing equations are used intrusively as in projection-based ROMs. 
Evaluations on transient two and three dimensional flows with coherent structures  demonstrates parametric and future state prediction capabilities, even when there are significant changes in the flow topology for the prediction configuration.  The CAE is also shown to perform significantly better than POD to extract the latent variables, especially when future state prediction is involved.  Sensitivity of the results to the amount of input data, the size of the latent space and other modeling choices is explored. 

As such, we do not recommend our approach as a replacement to traditional model reduction methods, or that CAEs should necessarily replace POD for all classes of problems. Parsimoniously parameterized models informed by PDE-based physics constraints have formed the basis of scientific modeling, and will continue to be useful. However, we find the results compelling enough to be worthy of serious debate, especially when adequate data is available.

\section{Acknowledgments}
The authors acknowledge  support from the Air Force under the Center of Excellence grant FA9550-17-1-0195, titled {\em Multi-Fidelity Modeling of Rocket Combustor Dynamics} (Program Managers: Dr. Mitat Birken and Dr. Fariba Fahroo), and the Office of Naval Research grant N00014-16-1-2728 titled {\em A Surrogate Based Framework for Helicopter/Ship Dynamic Interface,} (Program manager: Dr. Brian Holm-Hansen).

\section*{References}
\bibliographystyle{elsarticle-num-names}
\bibliography{ref.bib}

\appendix
\section{Replacement of CAE with POD}\label{sec pod}
In this section, the top level CAE is replaced with POD to provide a direct performance comparison between linear and nonlinear latent spaces. The POD basis is computed in a \textit{coupled} manner, i.e.  all the variables are flattened into a vector $\mathbf{s}\in\re^{n}$ for each frame of data, and these vectors are combined as columns in the snapshot matrix $\mathbf{S}\in\re^{n\times n_t}$. Before the flattening and collection of snapshots, the same standard deviation scaling as for CAE is applied to the variables. In other words, $\mathbf{S}$ is simply a flattened version of the input $\mathbf{Q}$ of CAE. For clarity, we will  use $\mathbf{q}$ and $\mathbf{Q}$ to refer to a single frame and the entire sequence of inputs for both POD and CAE. SVD is performed on the snapshot matrix to obtain the POD basis. Each mode of the basis contains information about all variables, enabling a direct comparison with the CAE. 

Assuming $n_k$ to be the designed number of modes, POD basis $\mathbf{V}\in\re^{n\times n_k}$ is constructed from the first $n_k$ left-singular vectors from the SVD. The corresponding spatially compressed variable  is 
\begin{equation}
    \mathbf{q}_s=\mathbf{V}^T\mathbf{q}.
\end{equation}

For a given $\mathbf{q}_s$, the full order solution is then approximated by 
\begin{equation}
    \mathbf{\tilde{q}}=\mathbf{V}\mathbf{q}_s.
\end{equation}

 We will refer to the compression and reconstruction processes as \textit{encoding} and \textit{decoding}, respectively.

\subsection{Discontinuous compressible flow}
\begin{figure}
	\centering
	\subfloat{
		\includegraphics[width=0.5\textwidth]{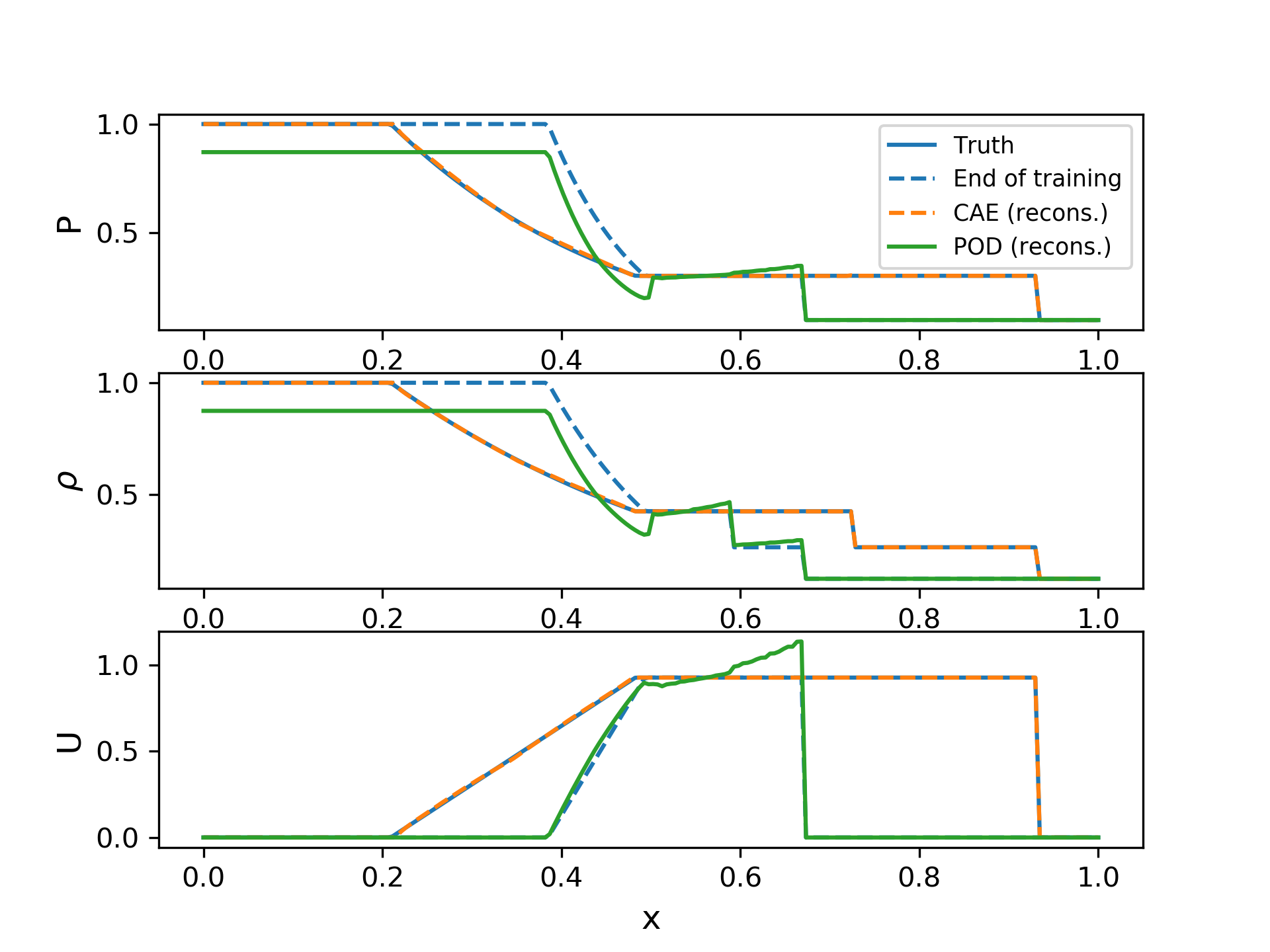}}
	\caption{Discontinuous compressible flow: POD reconstructed variables for $t=0.25$ s}\label{fig sod pod}
\end{figure}

The training data for the  shock tube only contains 35 time steps, and thus the total number of effective POD modes is limited to $n_k=35$. Since all POD modes are used, an exact reconstruction of the training data is achieved. However, in prediction, the waves propagate outside the region in which training features were present, and thus the global basis fails immediately. The reconstruction results for $t=0.25$ s are shown in Fig.~\ref{fig sod pod} along with the truth and the ending frame in the training data, which corresponds to $t=0.1$ s.

\subsection{Transient flow over a cylinder}
In this test, the number of POD modes is set to be the same as the latent dimension of the CAE, i.e. $n_k=n_s=60$. Other than replacing the CAE with POD, the neural network settings remain identical to the CAE-based combined prediction in Sec.~\ref{sec cylinder}. The same training and predictions as in Sec~\ref{sec cylinder} are repeated and the RAE for different stages is summarized in Table~\ref{table cylinder pod}.

\begin{table}[!ht]
	\begin{center}
	\caption {Transient flow over a cylinder: RAE for frameworks with CAE- and POD-based spatial compression}\label{table cylinder pod}
\footnotesize
		\begin{tabular}{c | c| c c | c c}
		\hline
		\multirow{2}{*}{Frame} & \multirow{2}{*}{Variable} & \multicolumn{2}{c|}{CAE-based} & \multicolumn{2}{c}{POD-based}\\
		\hhline{~~----}
		& & Train & Test & Train & Test\\
		\hline
		\multirow{4}{*}{1 to 2500} & CAE/POD reconstruction, $\mathbf{Q}\,(U/V)$ & 0.05\%/0.09\% & 0.11\%/0.19\% & 0.39\%/0.41\% & 0.38\%/0.45\%\\
		& TCAE reconstruction, $\mathbf{Q}_s$ & 0.75\% & 1.71\% & 0.66\% & 1.62\%\\
		& MLP prediction, $\mathbf{q}_l$ & 0.56\% & 1.02\% & 0.8\% & 2.2\%\\
		& Framework prediction, $\mathbf{Q}\,(U/V)$ & - &  0.68\%/1.37\% & - & 1.01\%/1.97\%\\
		\hline
		\multirow{3}{*}{2501 to 3000} & CAE/POD reconstruction, $\mathbf{Q}\,(U/V)$ & - & 0.19\%/0.41\% & - &  0.47\%/0.62\% \\ & TCN prediction, $\mathbf{Q}_s$ & 0.21\% & 2.94\% & 0.15\% &  2.20\% \\ & Framework prediction, $\mathbf{Q}\,(U/V)$ & - &  1.18\%/1.92\% & - & 1.69\%/4.14\%\\
		\hline
		\end{tabular}
	\end{center}  
\end{table}

\begin{figure}
	\centering
    \begin{minipage}{0.5\textwidth}
	\centering
    	\includegraphics[width=1\textwidth]{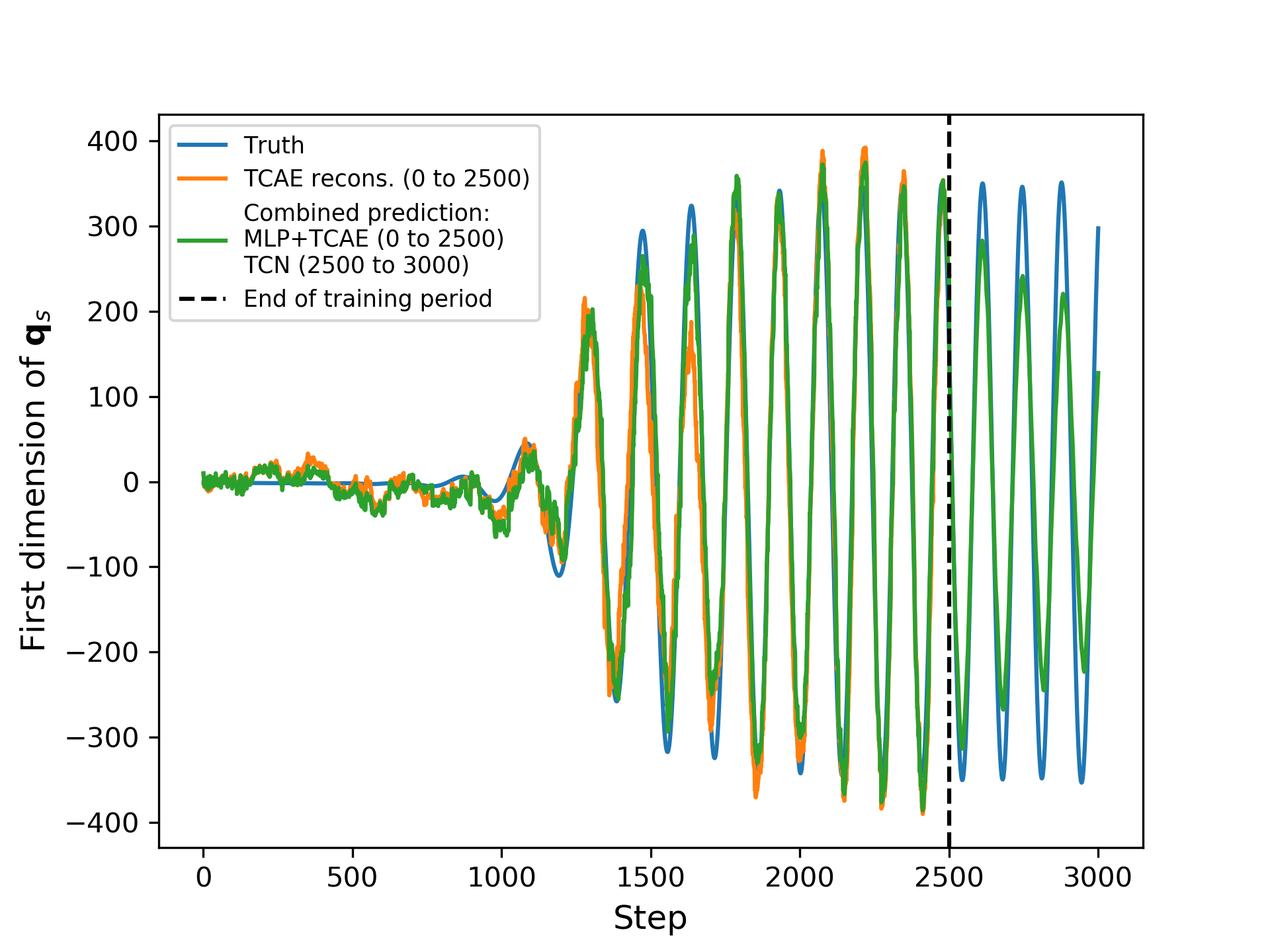}
        \caption{Transient flow over a cylinder: first dimension of $\mathbf{q}_s$ in POD-based framework}\label{fig cylinder pod qs}
    \end{minipage}\hfill
    \begin{minipage}{0.5\textwidth}
	\centering
    	\includegraphics[width=1\textwidth]{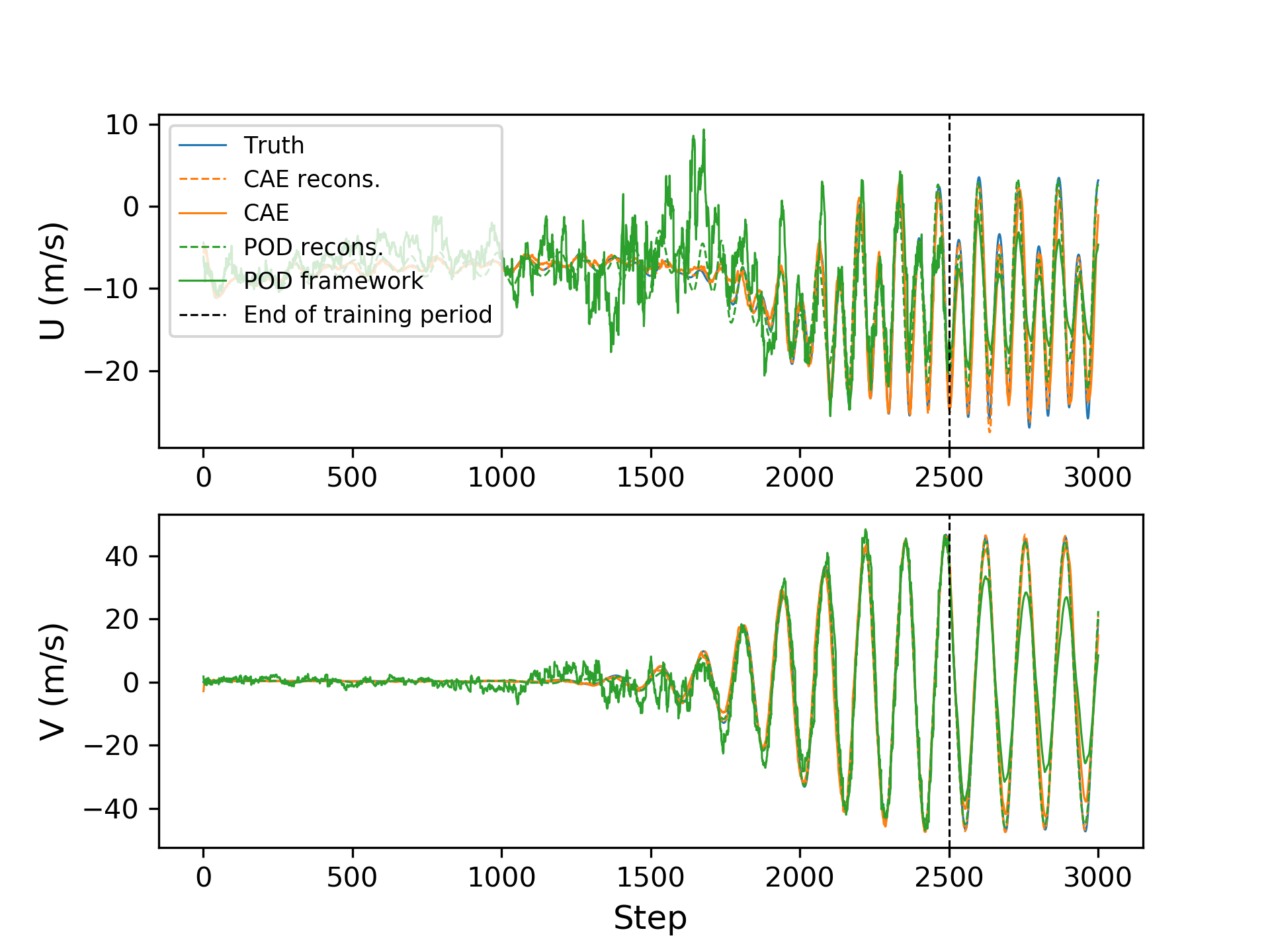}
        \caption{Transient flow over a cylinder: point monitor results for CAE- and POD-based frameworks}\label{fig cylinder pod point monitor}
    \end{minipage}
\end{figure}

It can be seen that the error for spatial reconstruction using POD is significantly larger than that of the CAE in both training and testing. This is  visualized by the dashed lines in Fig.~\ref{fig cylinder pod point monitor}.

The performance of the time-series processing of the encoded sequence $\mathbf{Q}_s$ using TCAE or TCN is independent of the spatial encoding. Indeed, from the plot of the first dimension of $\mathbf{q}_s$ in Fig.~\ref{fig cylinder pod qs}, it can be seen that curve is smoother than the point monitor result for the physical variables in Fig.~\ref{fig cylinder pod point monitor} or the latent variable encoded using CAE in Fig.~\ref{fig cylinder qs history}, thus reducing the error in the TCAE and TCN predictions for the latent variable slightly. 

\begin{figure}
	\centering
	\subfloat[Truth, $i=1650$]{
		\includegraphics[width=0.33\textwidth,trim={3cm 0 2cm 0},clip]{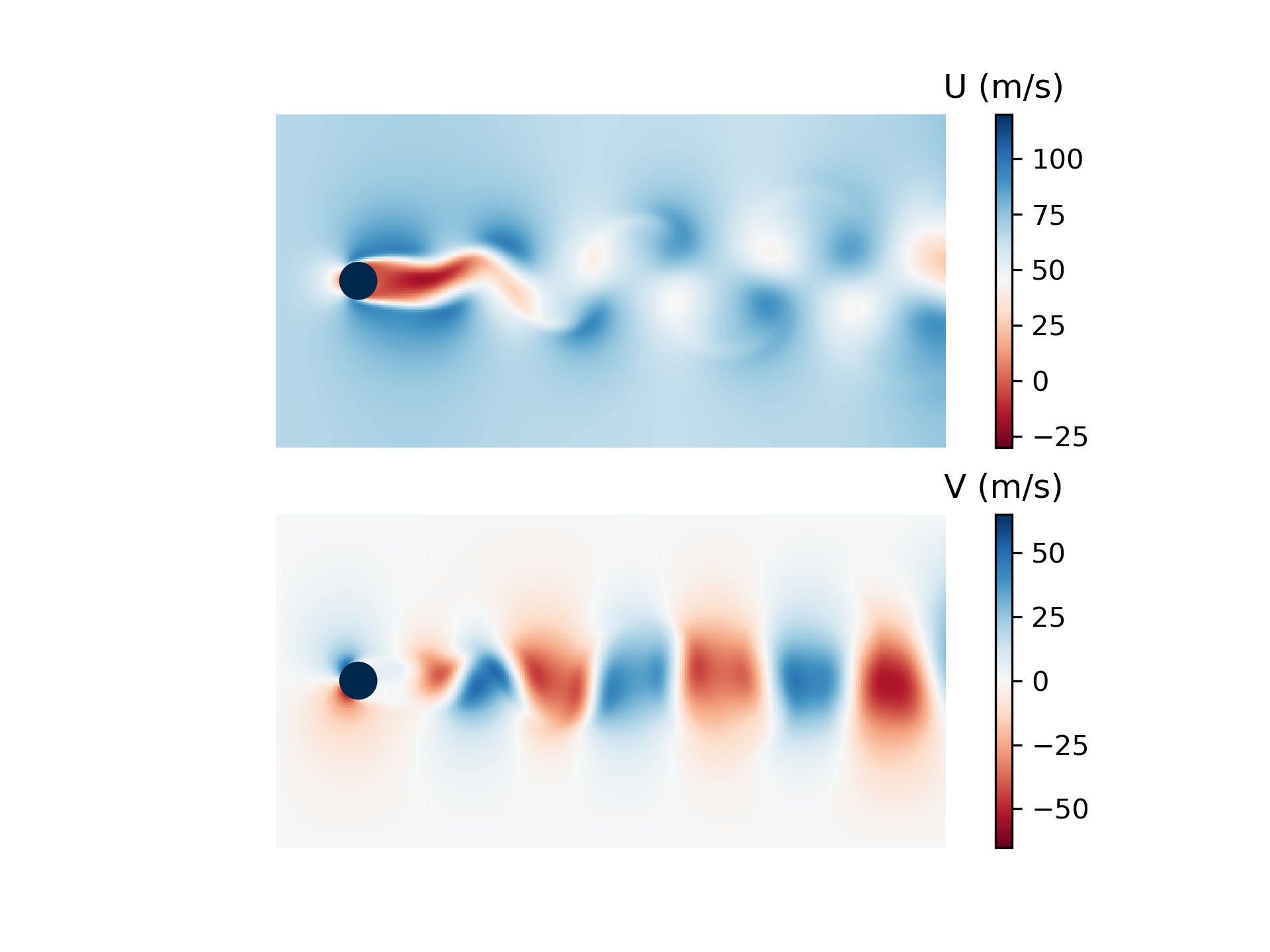}}
	\subfloat[CAE prediction, $i=1650$]{
		\includegraphics[width=0.33\textwidth,trim={3cm 0 2cm 0},clip]{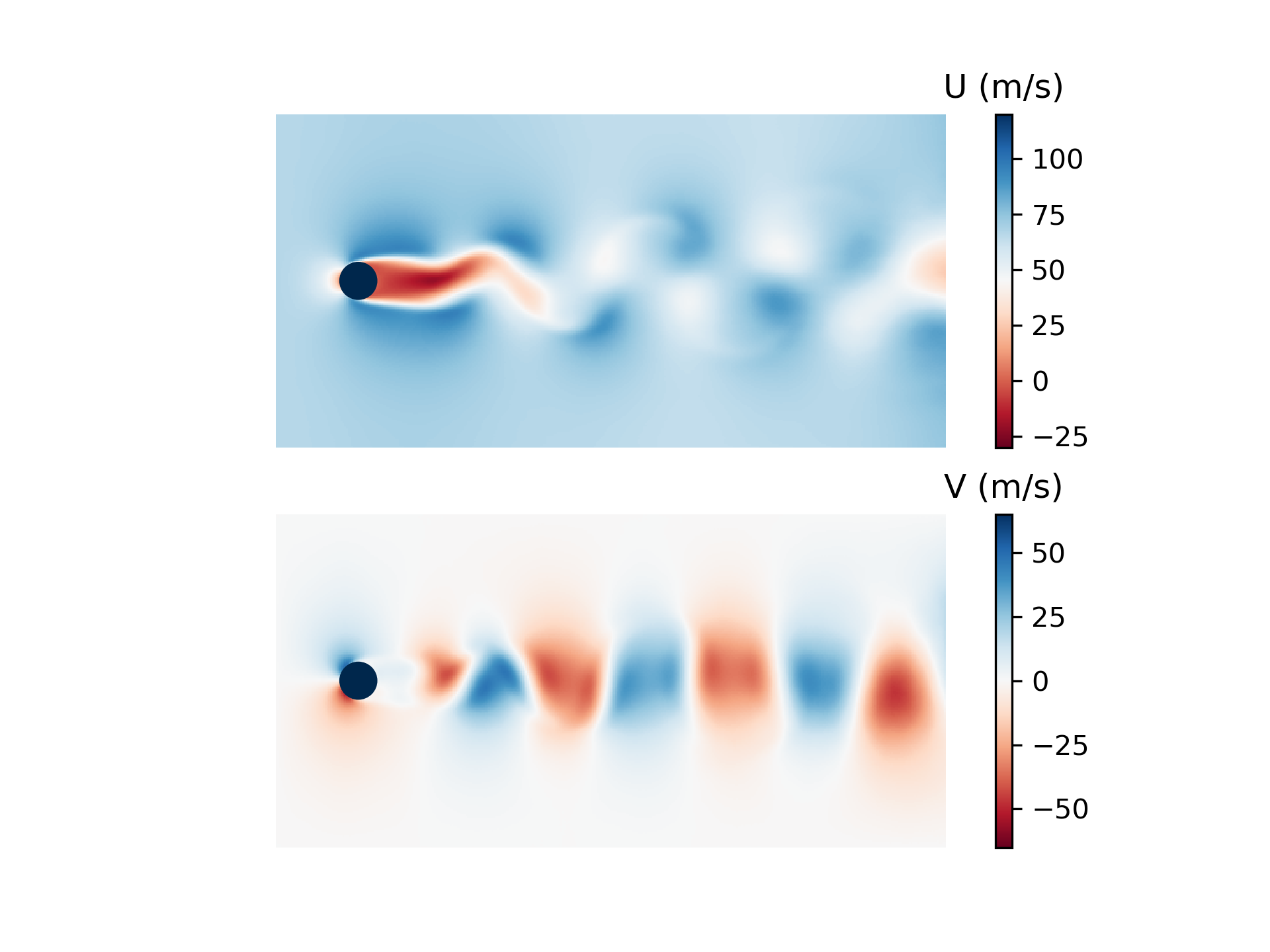}}
	\subfloat[POD prediction, $i=1650$]{
		\includegraphics[width=0.33\textwidth,trim={3cm 0 2cm 0},clip]{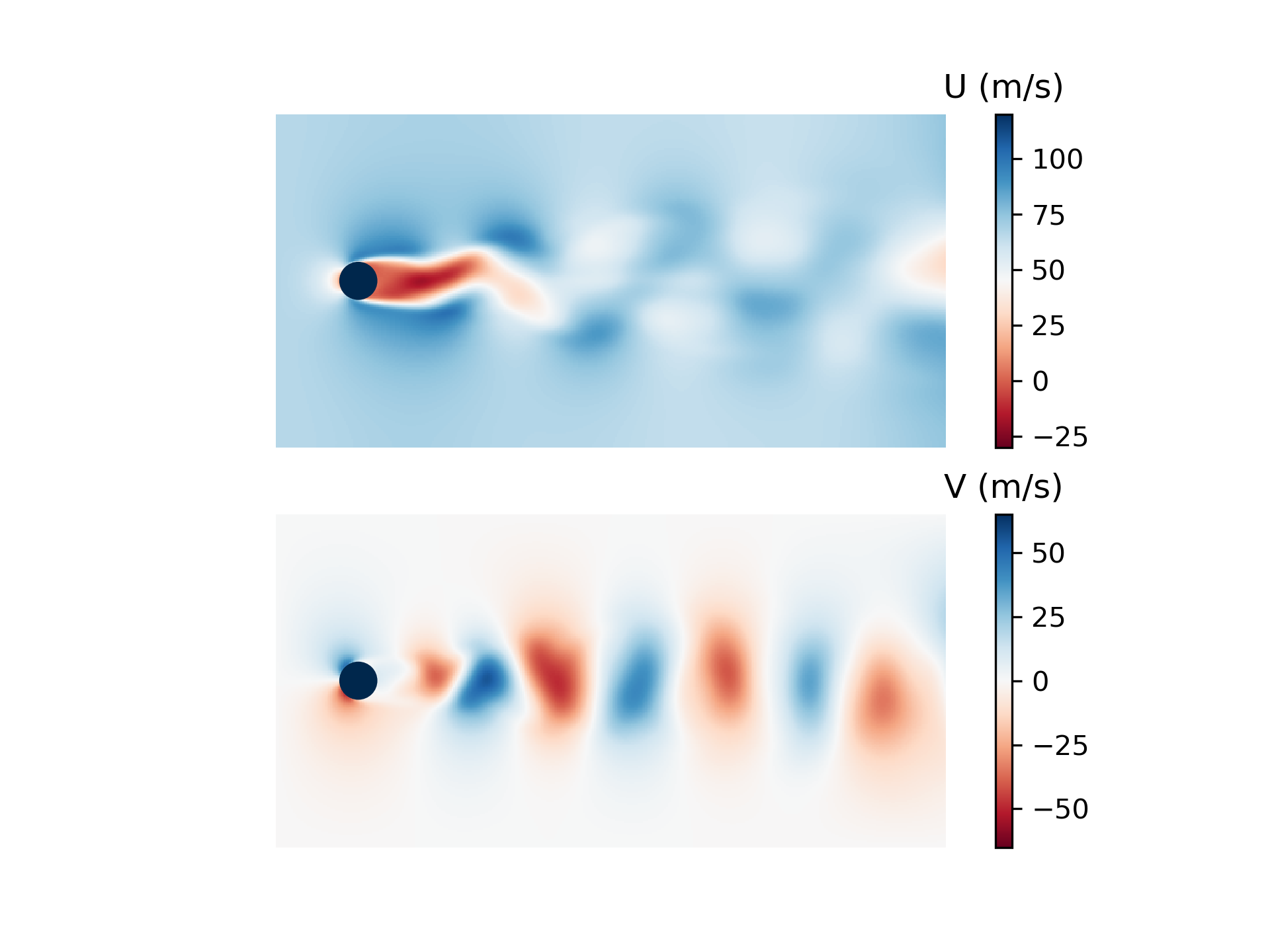}}\\
	\subfloat[Truth, $i=3000$]{
		\includegraphics[width=0.33\textwidth,trim={3cm 0 2cm 0},clip]{images/cylinder/200_3000.png}}
	\subfloat[CAE prediction, $i=3000$]{
		\includegraphics[width=0.33\textwidth,trim={3cm 0 2cm 0},clip]{images/cylinder/decoded_from_tcn_from_feedforward_from_timeseries_200_3000.png}}
	\subfloat[POD prediction, $i=3000$]{
		\includegraphics[width=0.33\textwidth,trim={3cm 0 2cm 0},clip]{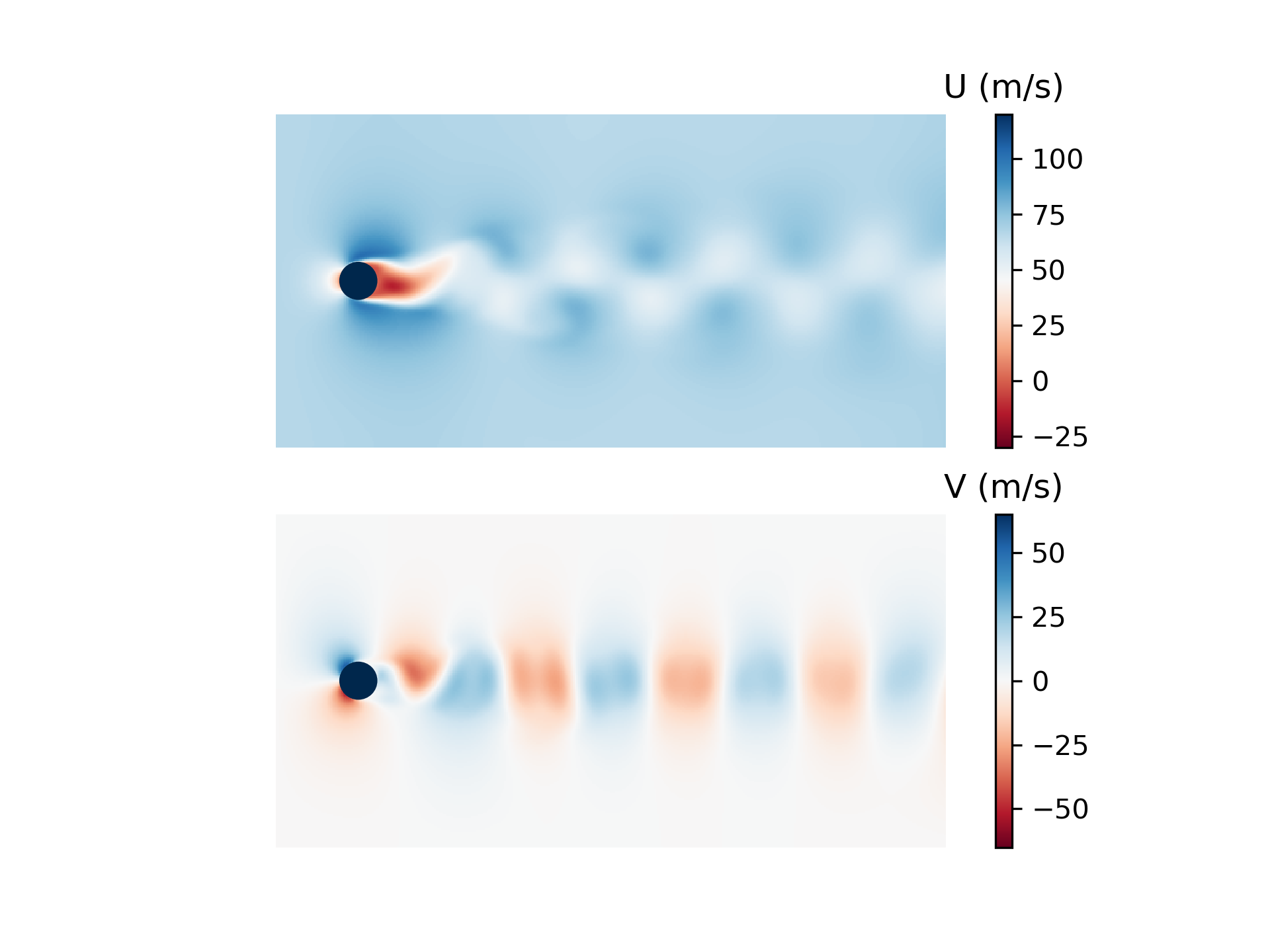}}
	\caption{Transient flow over a cylinder: flow contours from CAE- and POD-based frameworks}\label{fig cylinder contour pod} 
\end{figure}

The final prediction accuracy is largely limited by the POD performance when the predicted encoded variables are decoded. The final POD-based prediction result fails to represent the dynamics, as is clear from the point monitor result in Fig.~\ref{fig cylinder pod point monitor}. It is noted that the relative error listed in Table~\ref{table cylinder pod} is noticeably larger than that for the CAE-based result, though it is not visible in the figure. This is because that the error is averaged over the spatio-temporal span including a large range of low-amplitude unsteady or even steady data. To provide a more direct comparison of the prediction results, the flow field contours for the 1650th frame, where a large deviation from the truth is observed in the POD result, and the last, i.e. the 3000th frame, where the accumulative error in the future state prediction is maximized, are shoown in Fig.~\ref{fig cylinder contour pod}. It can be seen that the POD-based framework fails to capture certain details of the vortex structures whereas the CAE-based prediction is significantly closer to the truth. 

\subsection{Transient ship airwake}
As in the previous example, the encoded dimensions are set to be the same for POD and CAE, i.e. $n_k=n_s=20$. The same second and third level network architectures are used as in Sec.~\ref{sec airwake}. The RAE for different stages is summarized in Table~\ref{table airwake pod}.

\begin{table}[!ht]
	\begin{center}
	\caption {\textcolor{red}{Transient ship airwake: RAE for frameworks with CAE- and POD-based spatial compression}}\label{table airwake pod}
    \scriptsize
		
\color{red}{\begin{tabular}{c| c c | c c}
		\hline
		\multirow{2}{*}{Variable} & \multicolumn{2}{c|}{CAE-based} & \multicolumn{2}{c}{POD-based}\\
		\hhline{~----}
		& Train & Test & Train & Test\\
		\hline
		CAE/POD reconstruction, $\mathbf{Q}\,(U/V/W)$ & 0.12\%/0.09\%/0.08\% & 0.30\%/0.38\%/0.29\% & 0.34\%/0.37\%/0.29\% & 0.42\%/0.44\%/0.41\%\\
		First 100 frames of reconstruction & - & 2.28\%/1.47\%/1.18\% & - & 7.77\%/6.20\%/7.13\%\\
		TCAE reconstruction, $\mathbf{Q}_s$ & 0.63\% & 4.47\% & 0.20\% & 1.91\%\\
		MLP prediction, $\mathbf{q}_l$ & 0.60\% & 0.85\% & 0.96\% & 6.23\%\\
		Framework prediction, $\mathbf{Q}\,(U/V/W)$ & - & 0.51\%/0.89\%/0.62\% & - & 0.61\%/0.91\%/0.70\%\\
		First 100 frames of prediction & - & 4.52\%/3.94\%/2.74\% & - & 9.02\%/6.49\%/7.91\%\\
		\hline
		\end{tabular}
}
	\end{center}  
\end{table}

\begin{figure}
	\centering
	\subfloat{
		\includegraphics[width=0.5\textwidth]{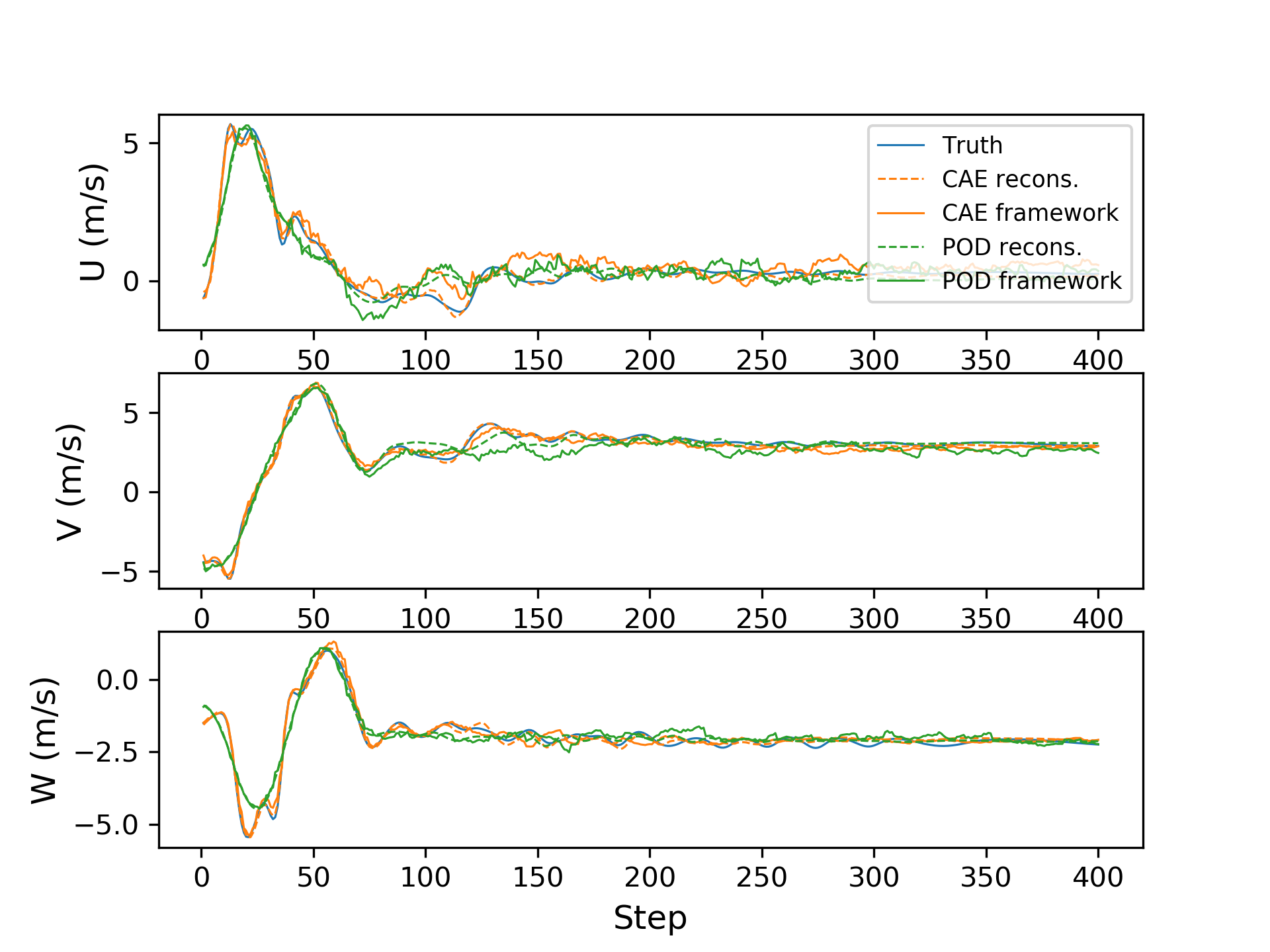}}
	\caption{Transient ship airwake: point monitor results for CAE- and POD-based frameworks}\label{fig airwake pod point monitor}
\end{figure}

\begin{figure}
	\centering
	\subfloat[Truth]{
		\includegraphics[width=0.33\textwidth,trim={3cm 0 2cm 0},clip]{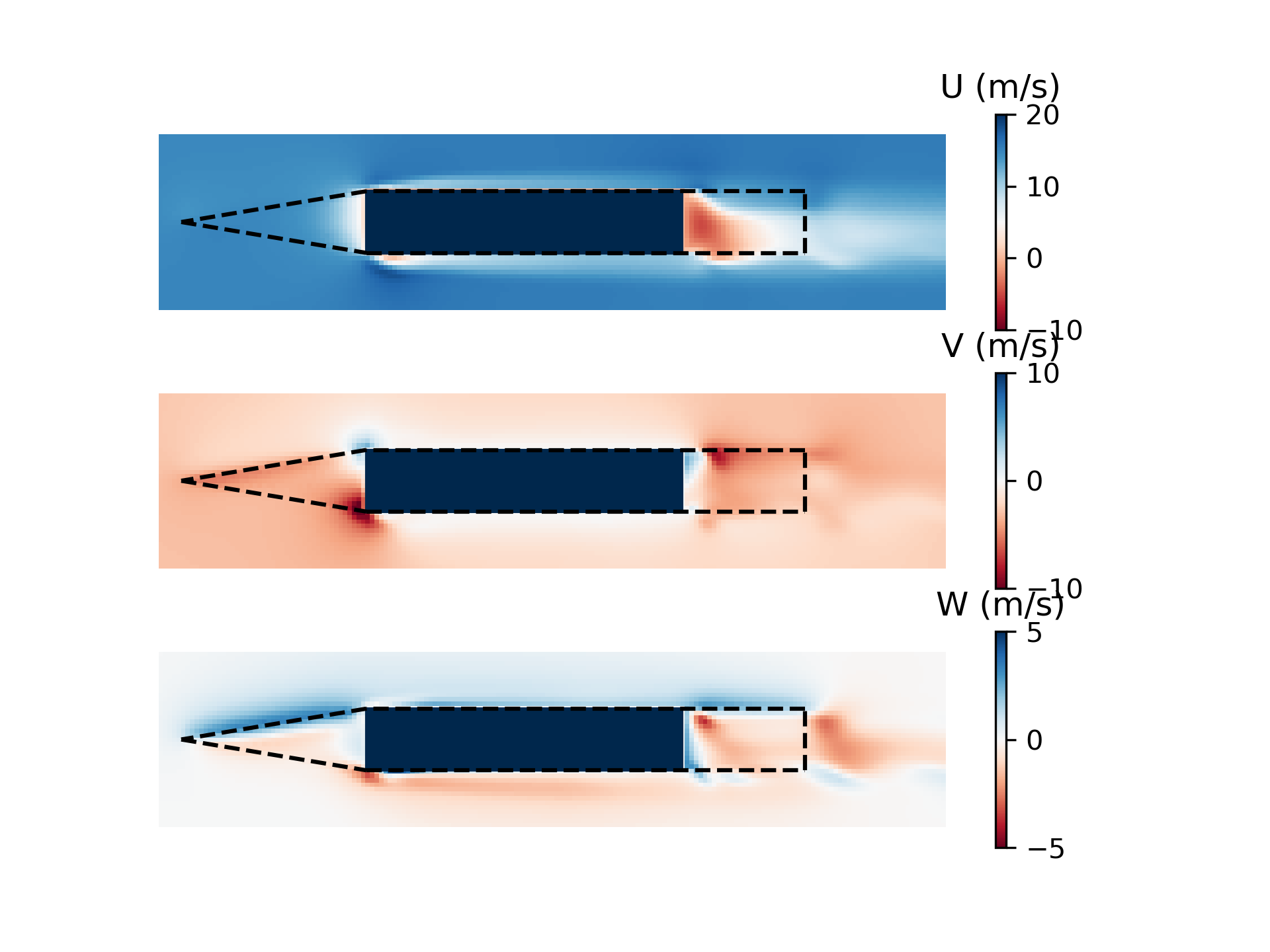}}
	\subfloat[CAE prediction]{
		\includegraphics[width=0.33\textwidth,trim={3cm 0 2cm 0},clip]{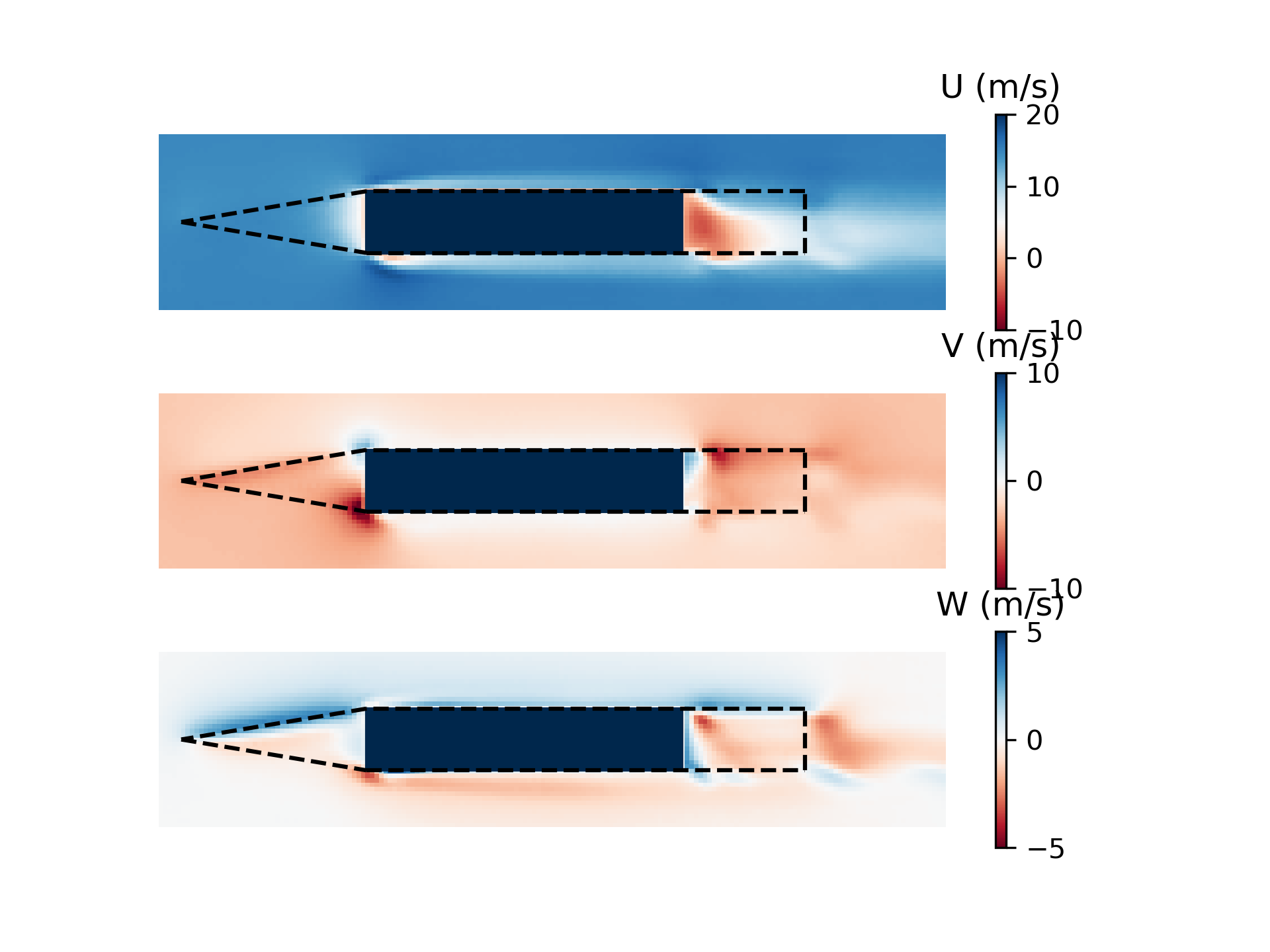}}
	\subfloat[POD prediction]{
		\includegraphics[width=0.33\textwidth,trim={3cm 0 2cm 0},clip]{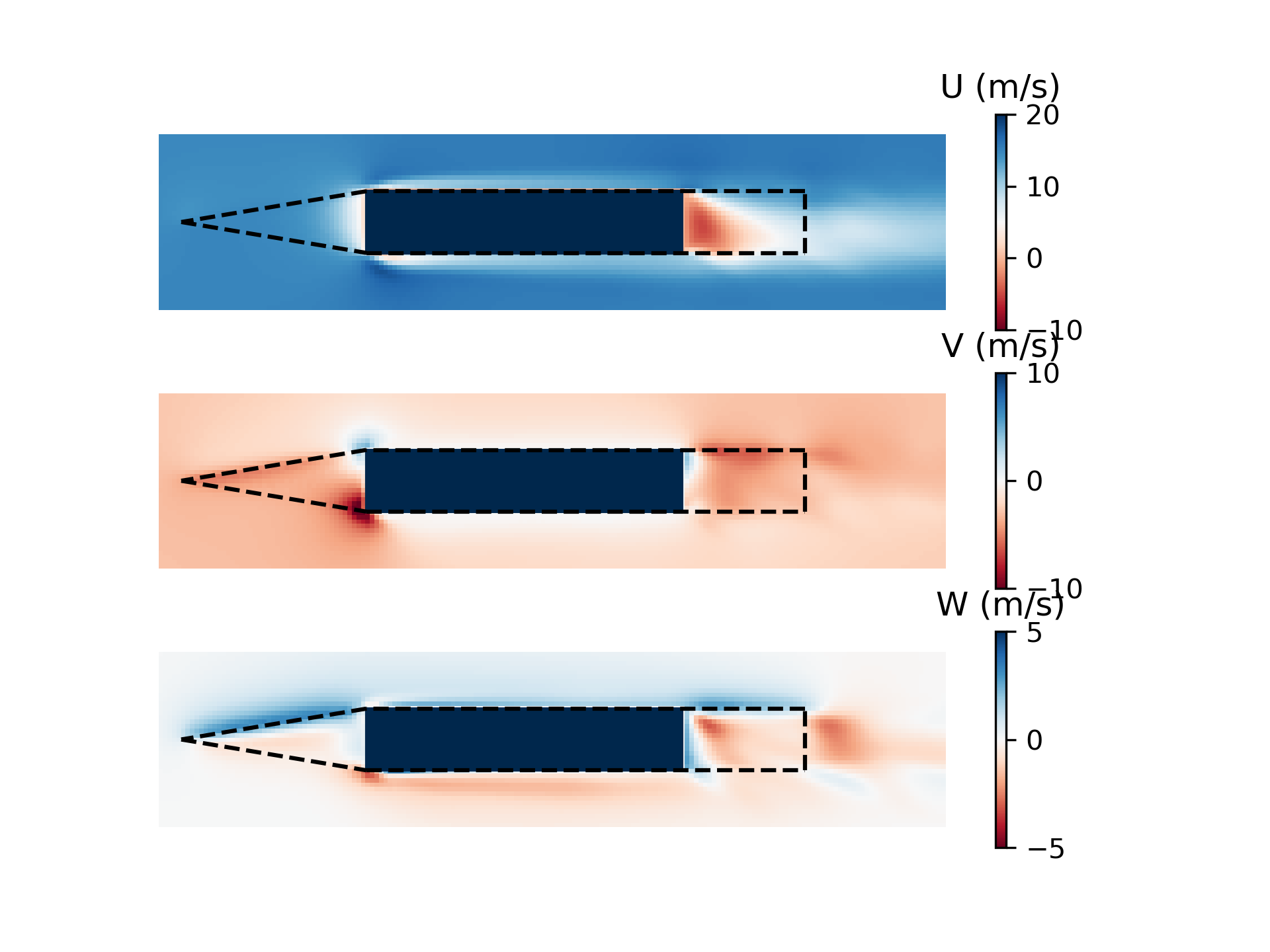}}
	\caption{Transient ship airwake: flow contours from CAE- and POD-based frameworks for $i=10$}\label{fig airwake contour pod} 
\end{figure}

A increase in the reconstruction error is observed especially in the \textcolor{red}{testing stage}. The same point monitor as in Sec.~\ref{sec airwake} is used and the result is shown in Fig.~\ref{fig airwake pod point monitor}. It can be seen that the POD result largely deviates from the truth especially in the first 100 frames where transition from initial towards the new side-slip angle imposes a strong influence on the area over the deck. \textcolor{red}{Error values for this period are listed in addition to those for the full sequence in Table~\ref{table airwake pod}, in order to better illustrate the discrepancy. The gap in performance is clearly visualized in the flow contours for $i=10$ in Fig.~\ref{fig airwake contour pod}, where the POD-based prediction shows a blurred and shifted airwake whereas the CAE-based prediction appears to be almost visually identical to the truth. It should be noted that in Fig.~\ref{fig airwake pod point monitor} final POD-based framework prediction follows the POD reconstructed variables tightly, which illustrates reliable performance of the rest of the framework.}




\section{Latent dimension sensitivity study}\label{sec sensitivity}
Besides the nonlinearity, an important difference between the autoencoder and POD-based model reduction methods is that the latent dimension needs to be pre-determined in the design of the network structure before the training process. On the other hand, the SVD provides a full set of singular vectors, which can be truncated based on {\em a priori} error estimates. As a consequence, the sensitivity of the network performance w.r.t. the latent dimensions in the autoencoders becomes an important feature, and is assessed for the cylinder and ship airwake cases. The shock tube is not considered as  compression is only performed on the number of variables. 

\subsection{Transient flow over a cylinder}
\begin{figure}
	\centering
    \begin{minipage}{0.47\textwidth}
	\centering
    	\includegraphics[width=1\textwidth]{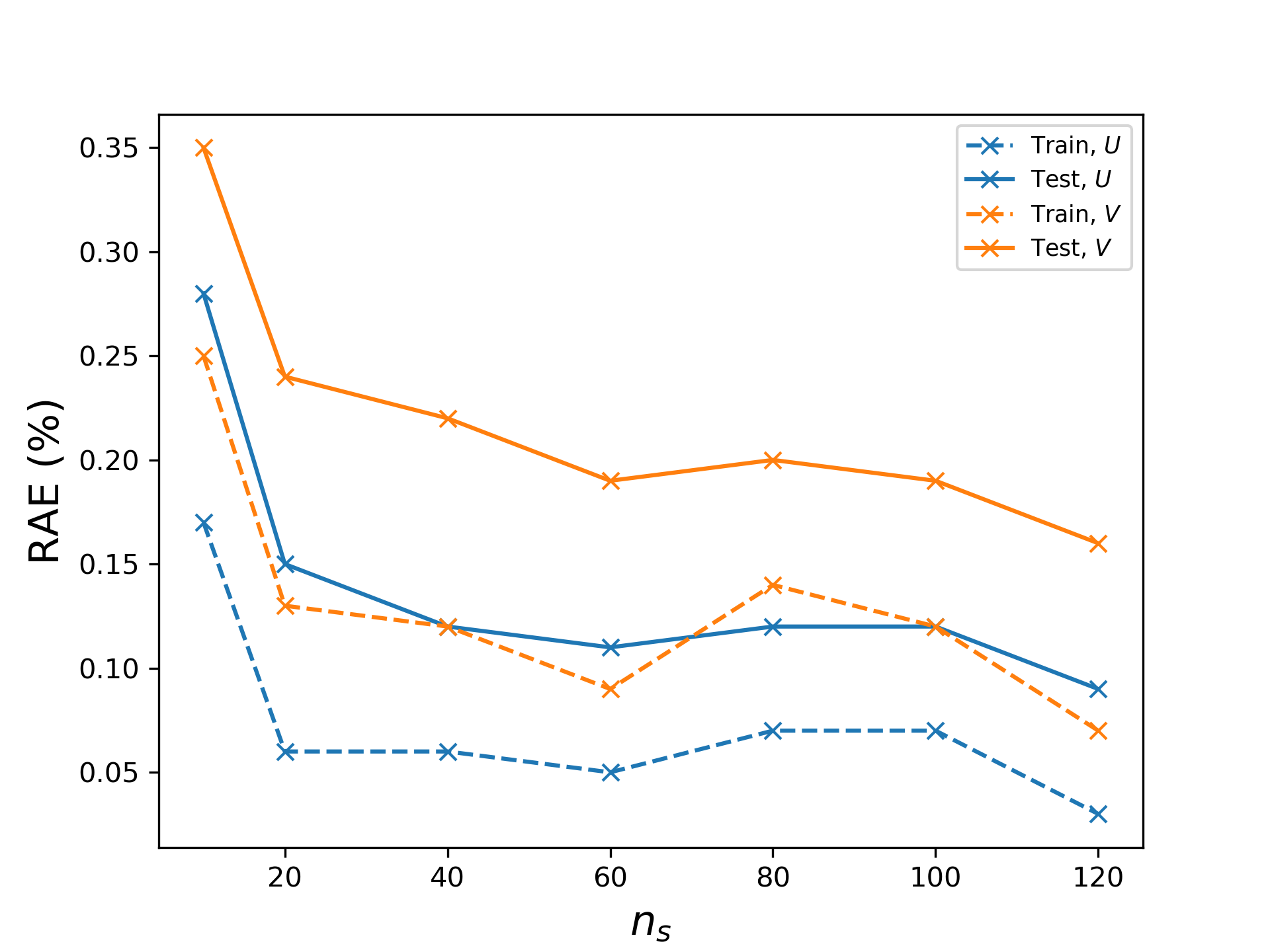}
        \caption{Transient flow over a cylinder: CAE RAE vs $n_s$}\label{fig cylinder CAE error vs ns}
    \end{minipage}\hfill
    \begin{minipage}{0.47\textwidth}
	\centering
    	\includegraphics[width=1\textwidth]{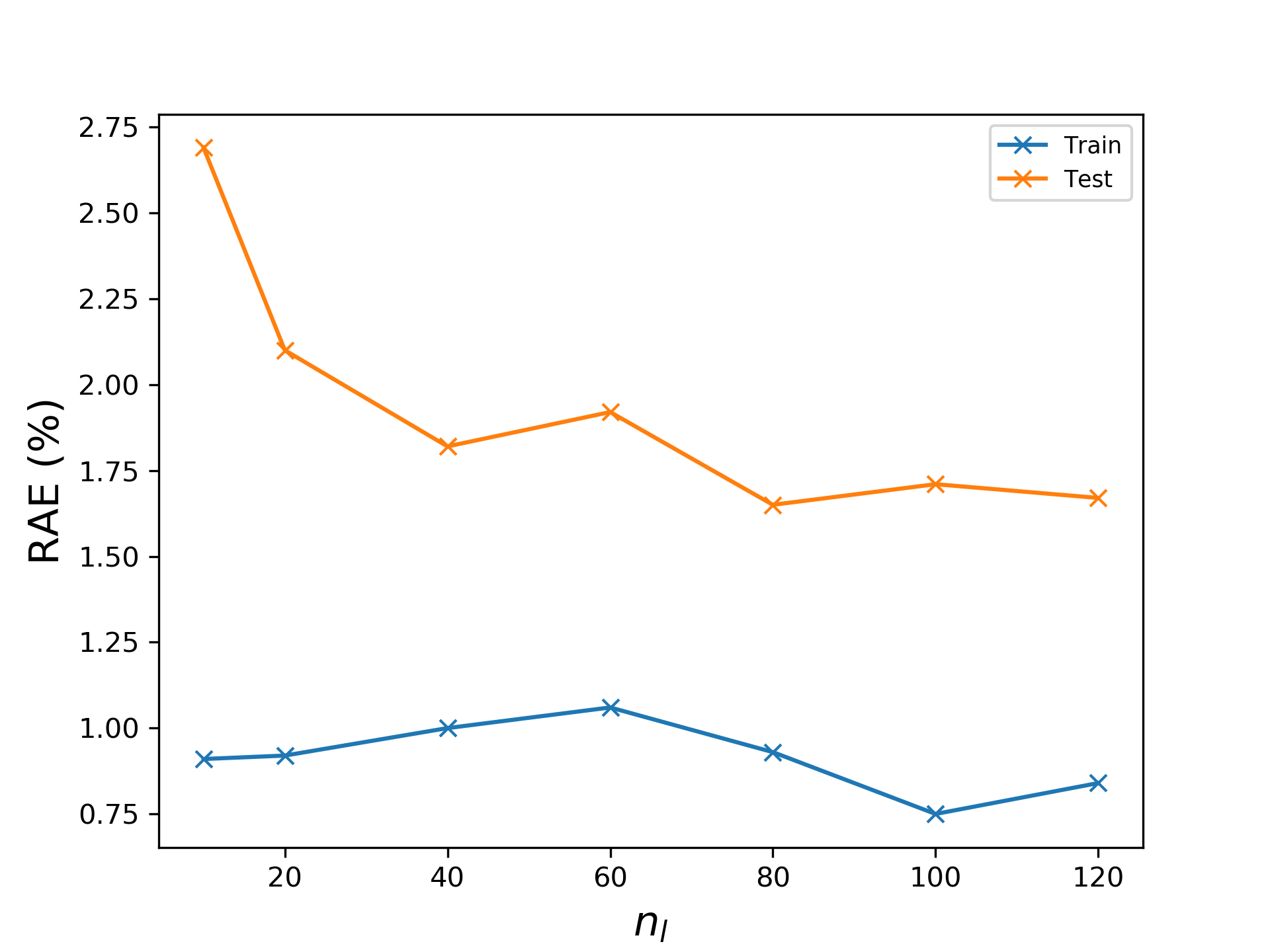}
        \caption{Transient flow over a cylinder: TCAE RAE vs $n_l$}\label{fig cylinder TCAE error vs nl}
    \end{minipage}
\end{figure}

The influence of $n_s$ on the CAE encoding-decoding accuracy is shown in Fig.~\ref{fig cylinder CAE error vs ns}. It can be seen that both training and testing errors experience a sharp drop for $n_s \le 30$, following which the slope starts to flatten and the decay saturates for $n_s \ge 60$. For the TCAE result in Fig.~\ref{fig cylinder TCAE error vs nl}, the decay in training error is almost negligible, yet the testing error drops rapidly for  $n_l \le 80$. 

Due to the rapid decay in leading modes in both levels of autoencoders, the spatio-temporal sequence can be represented using a significantly reduced set of latent variables.

\subsection{Transient ship airwake}
Compared to the previous case, the CAE testing error saturates at a few modes. This is a consequence of the fact that when the inflow side-slip angle changes, the similarity between the training and testing data is very limited in both large and small scales physics. Thus, including extensive latent dimensions does not reduce the testing error efficiently. Due to the same reason, the decay in the TCAE error is also slow. Nevertheless, the testing RAE is  below 0.5\% in the CAE and 5\% in the TCAE, providing a highly efficient model reduction as shown in Sec.~\ref{sec airwake}.

\begin{figure}
	\centering
    \begin{minipage}{0.47\textwidth}
	\centering
    	\includegraphics[width=1\textwidth]{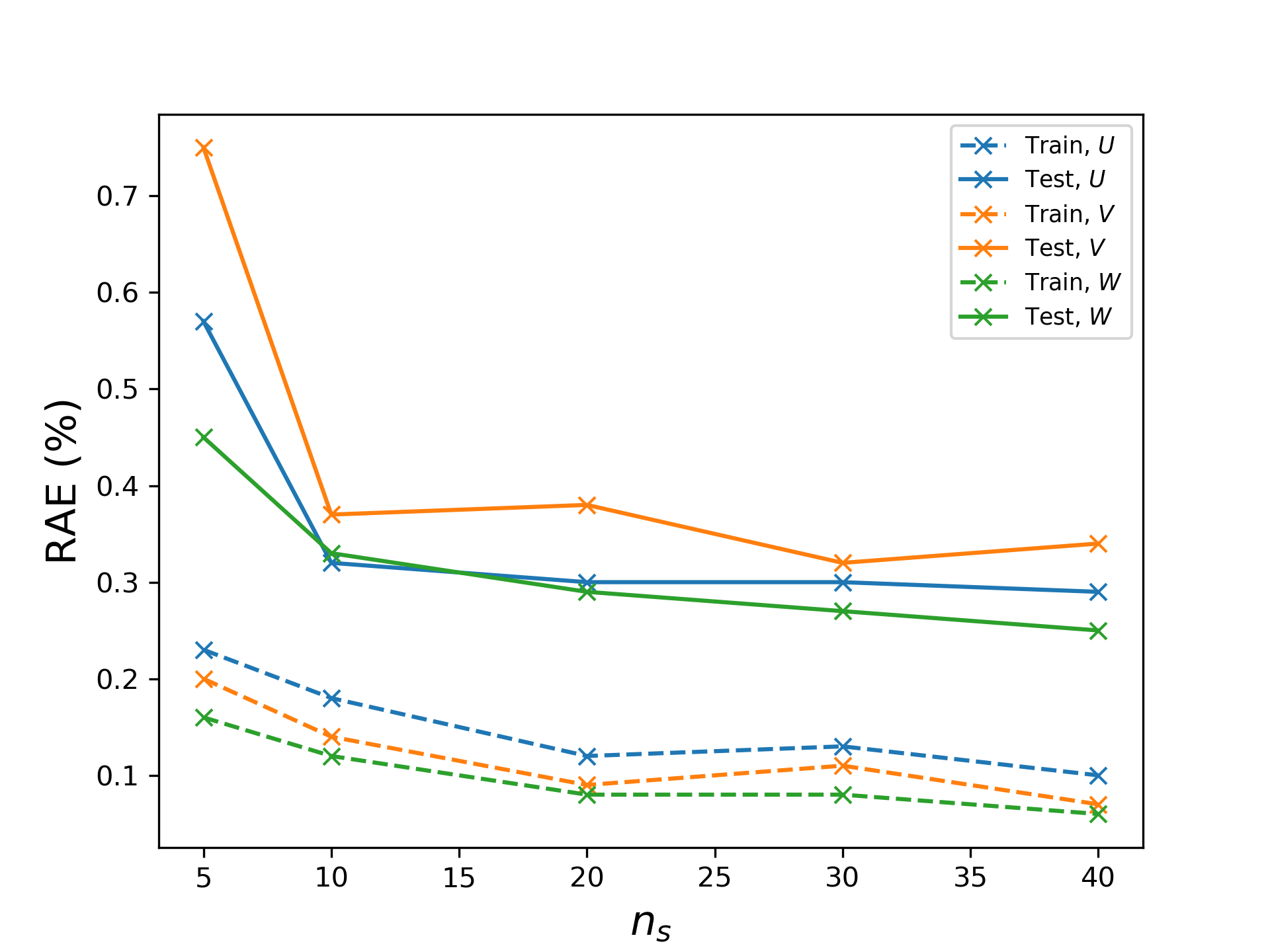}
        \caption{Transient ship airwake: CAE RAE vs $n_s$}\label{fig airwake CAE error vs ns}
    \end{minipage}\hfill
    \begin{minipage}{0.47\textwidth}
	\centering
    	\includegraphics[width=1\textwidth]{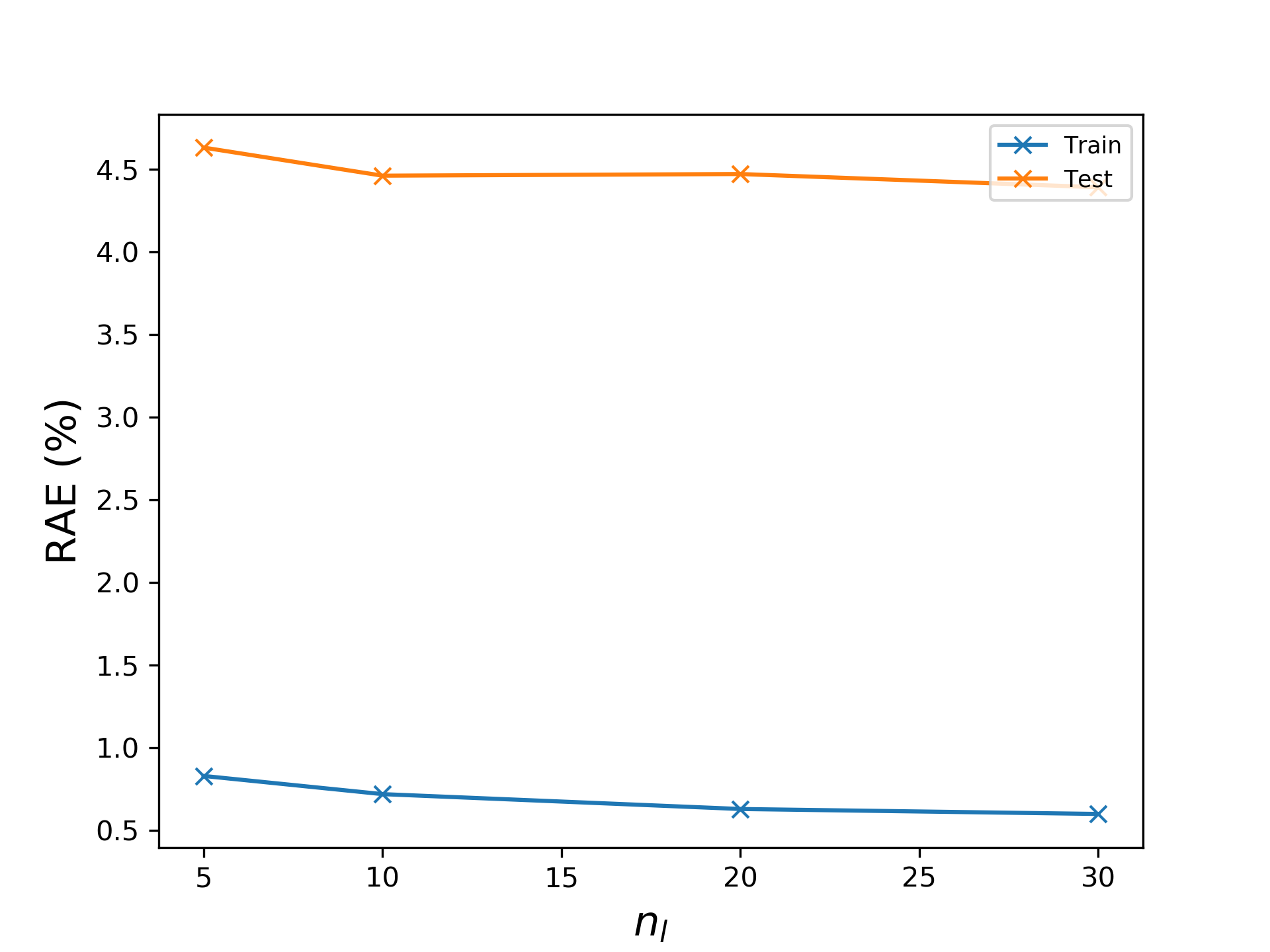}
        \caption{Transient ship airwake: TCAE RAE vs $n_l$}\label{fig airwake TCAE error vs nl}
    \end{minipage}
\end{figure} 
\section{Training convergence history}\label{sec convergence}
\begin{figure}
	\centering
    \begin{minipage}{0.47\textwidth}
	\centering
    	\includegraphics[width=1\textwidth]{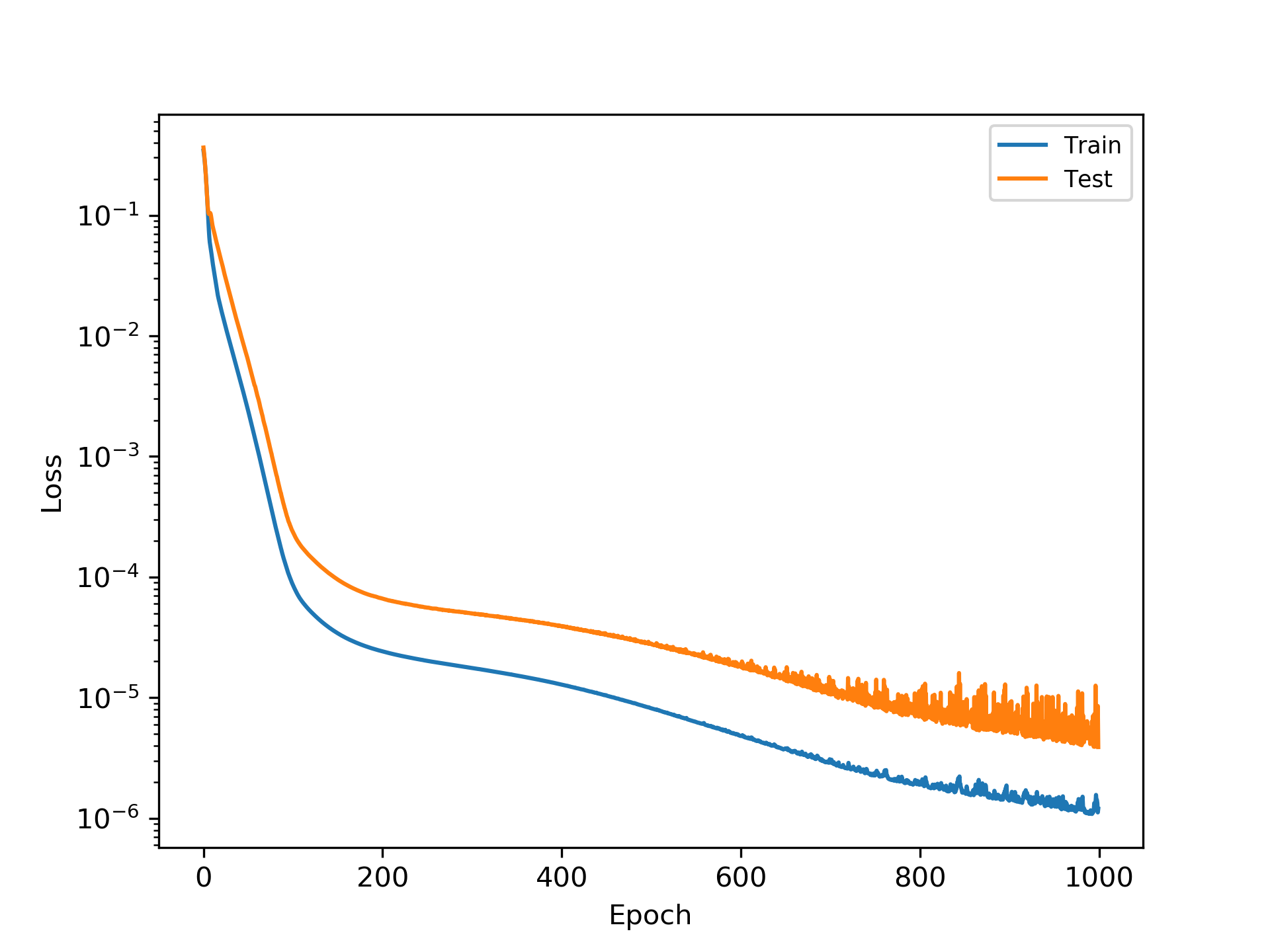}
        \caption{Discontinuous compressible flow: convergence history of CAE}\label{fig sod cae hist}
    \end{minipage}\hfill
    \begin{minipage}{0.47\textwidth}
	\centering
    	\includegraphics[width=1\textwidth]{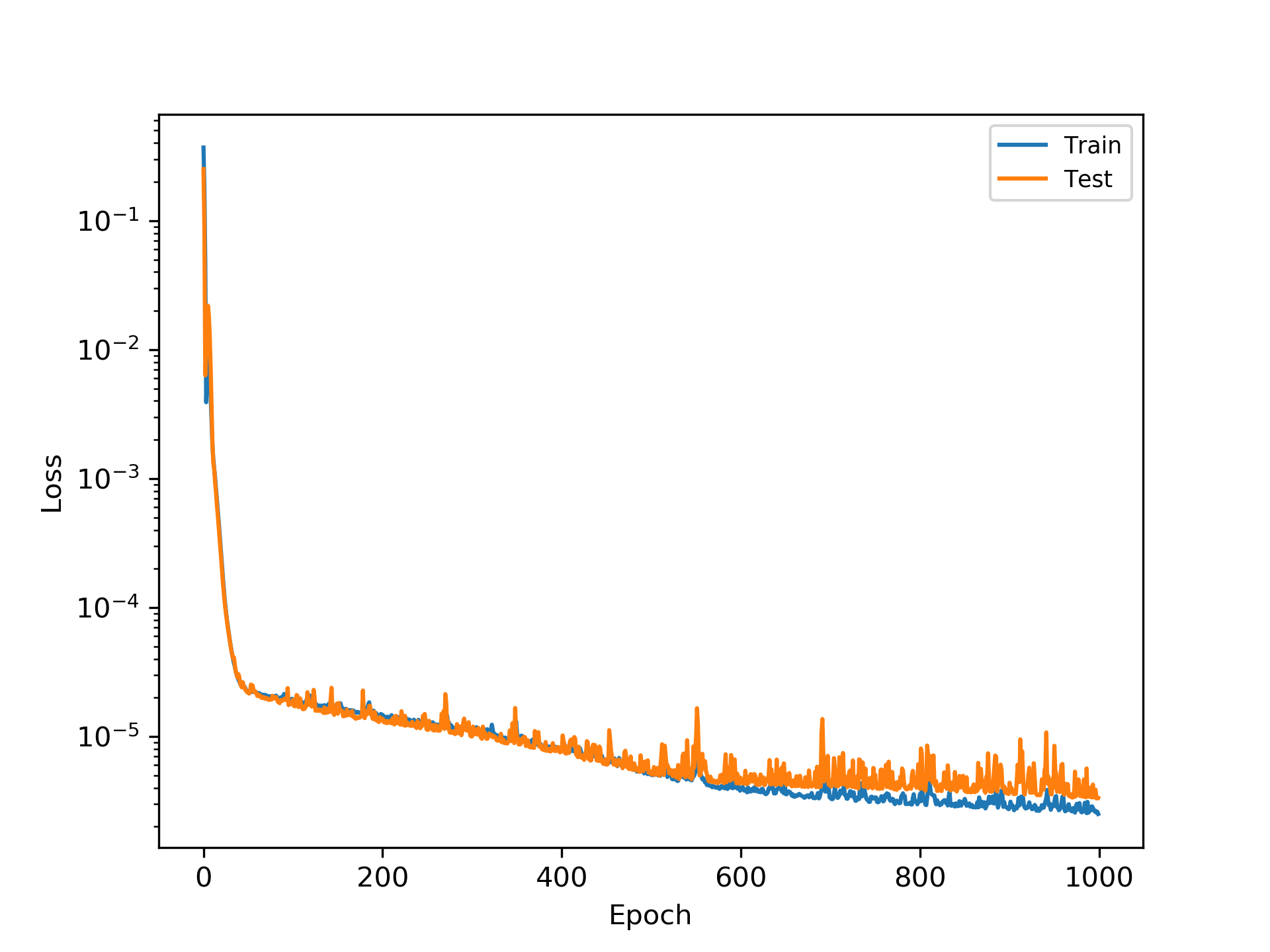}
        \caption{Discontinuous compressible flow: convergence history of TCN}\label{fig sod tcn hist}
    \end{minipage}
\end{figure}

\begin{figure}
	\centering
    \begin{minipage}{0.47\textwidth}
	\centering
    	\includegraphics[width=1\textwidth]{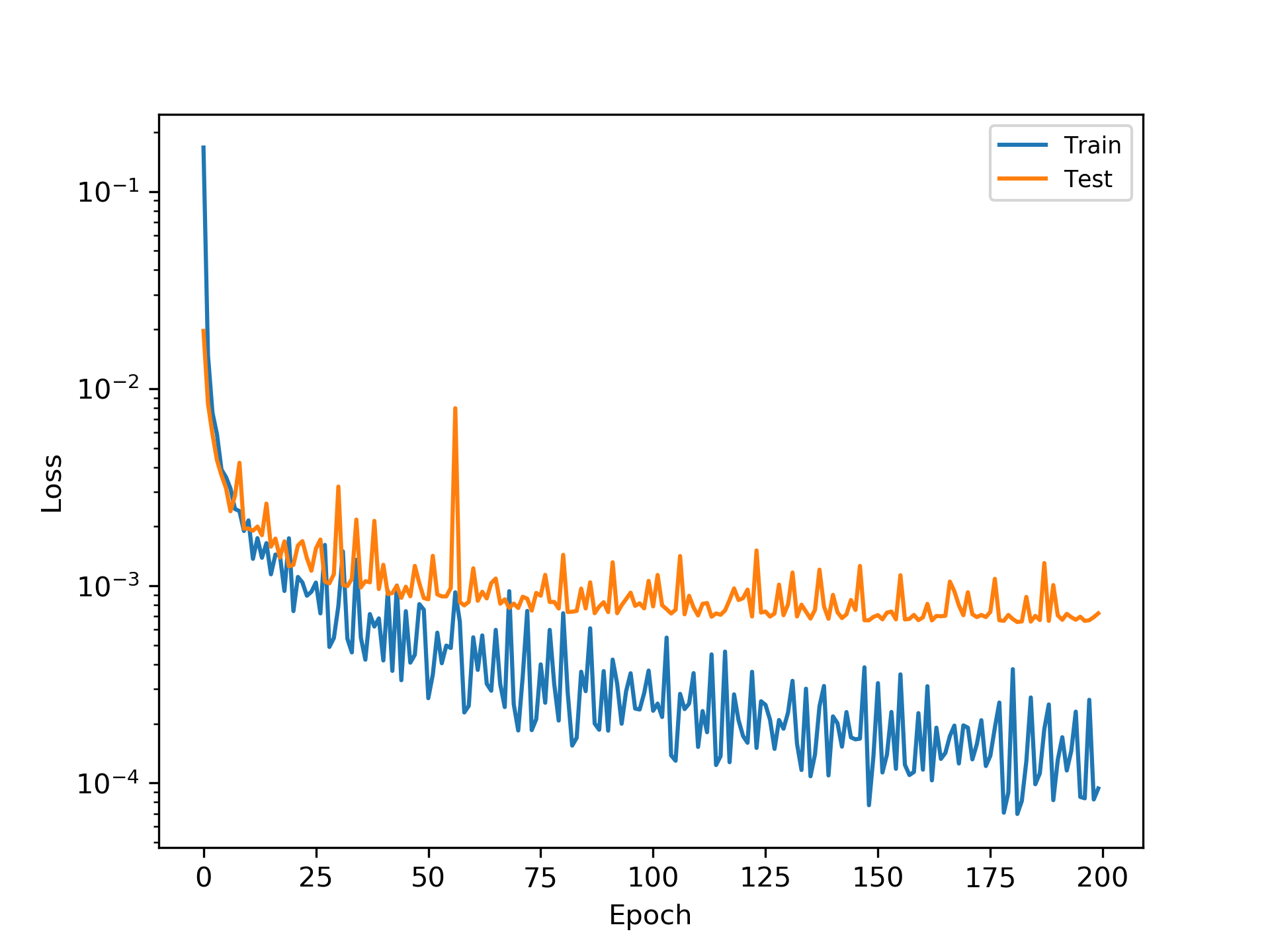}
        \caption{Transient flow over a cylinder: convergence history of CAE}\label{fig cylinder cae hist}
    \end{minipage}\hfill
    \begin{minipage}{0.47\textwidth}
	\centering
    	\includegraphics[width=1\textwidth]{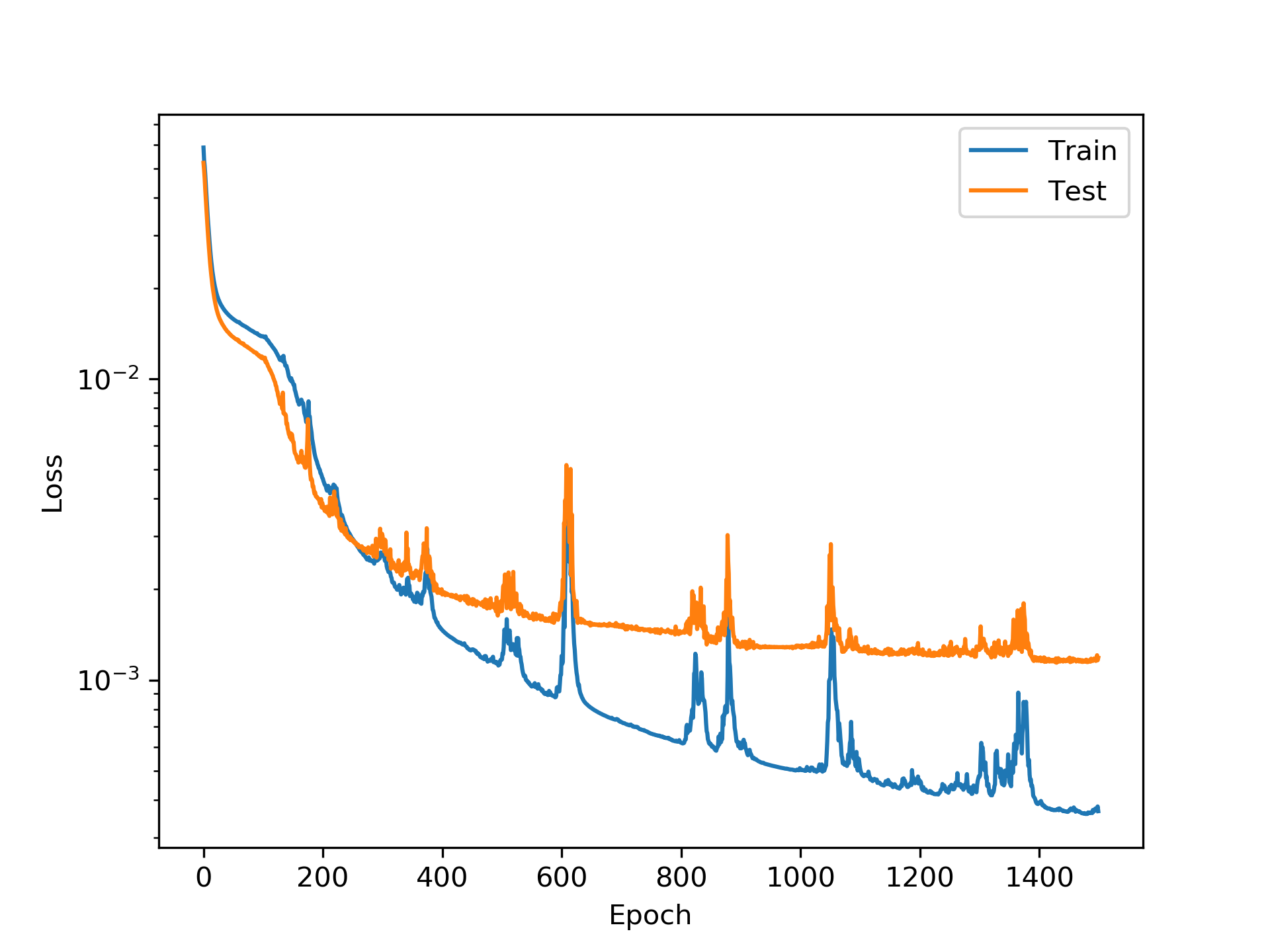}
        \caption{Transient flow over a cylinder: convergence history of TCAE}\label{fig cylinder tcae hist}
    \end{minipage}
\end{figure}

\begin{figure}
	\centering
    \begin{minipage}{0.47\textwidth}
	\centering
    	\includegraphics[width=1\textwidth]{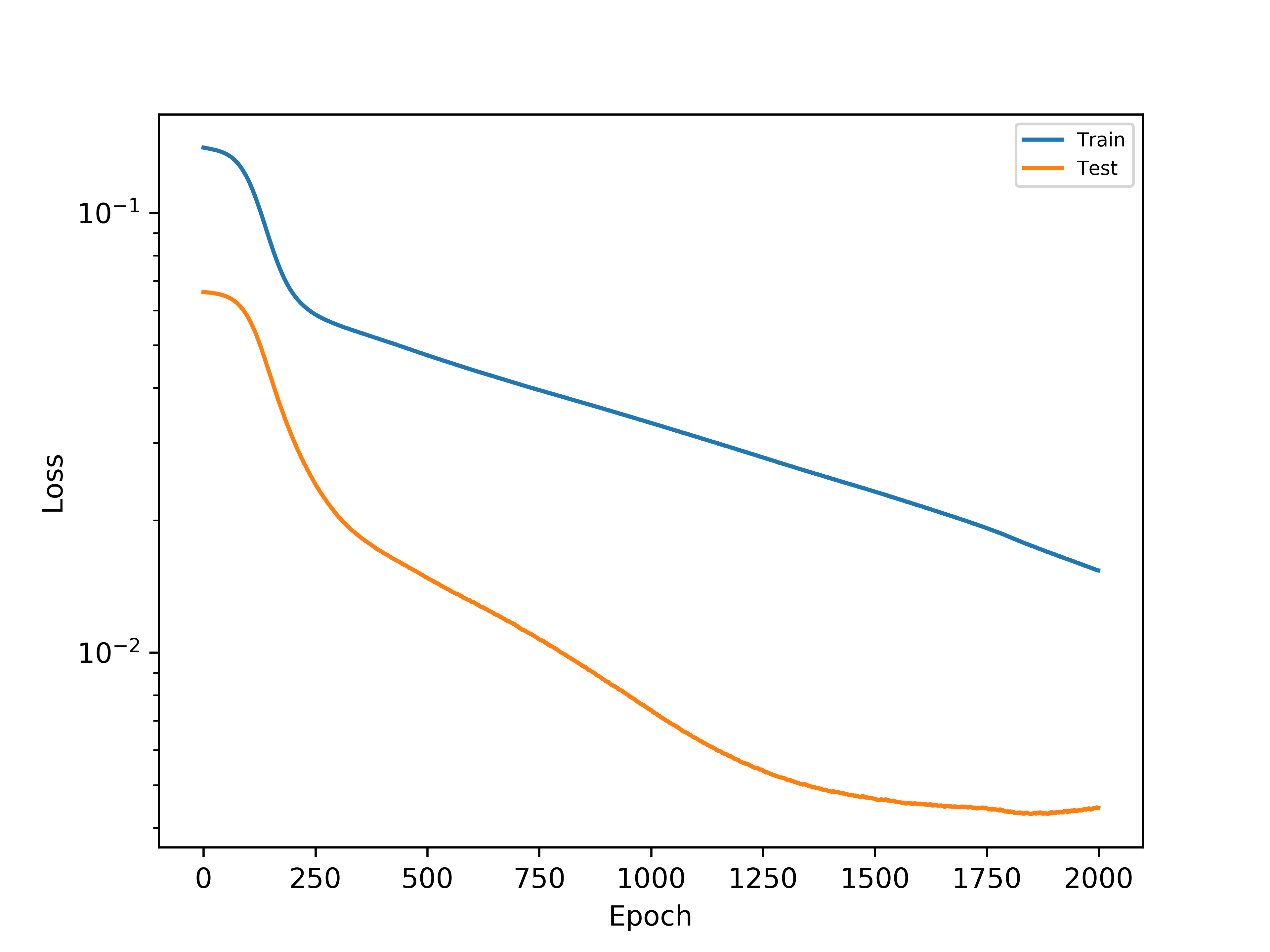}
        \caption{Transient flow over a cylinder: convergence history of MLP}\label{fig cylinder mlp hist}
    \end{minipage}\hfill
    \begin{minipage}{0.47\textwidth}
	\centering
    	\includegraphics[width=1\textwidth]{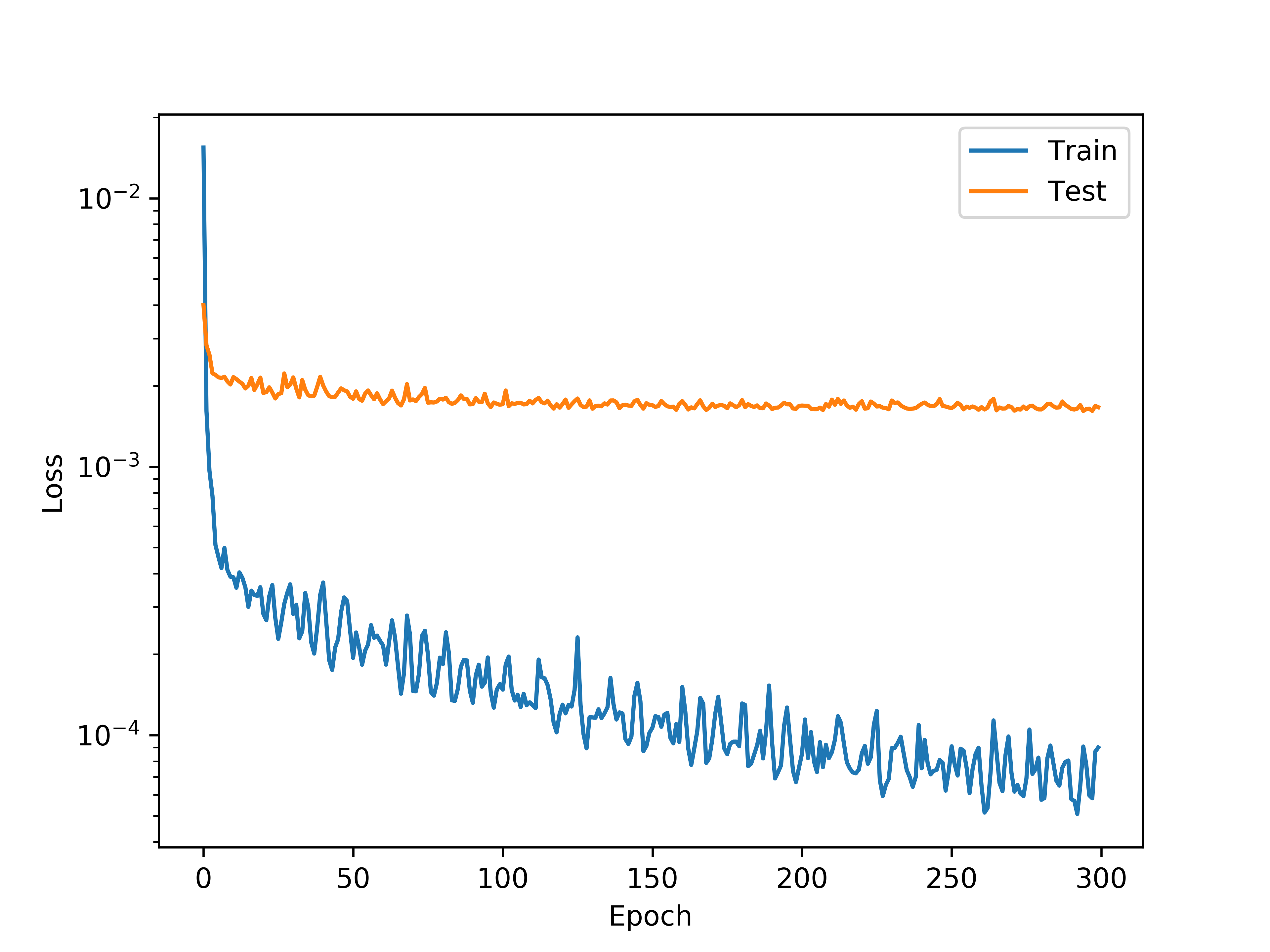}
        \caption{Transient flow over a cylinder: convergence history of TCN}\label{fig cylinder tcn hist}
    \end{minipage}
\end{figure}

\begin{figure}
	\centering
    \begin{minipage}{0.47\textwidth}
	\centering
    	\includegraphics[width=1\textwidth]{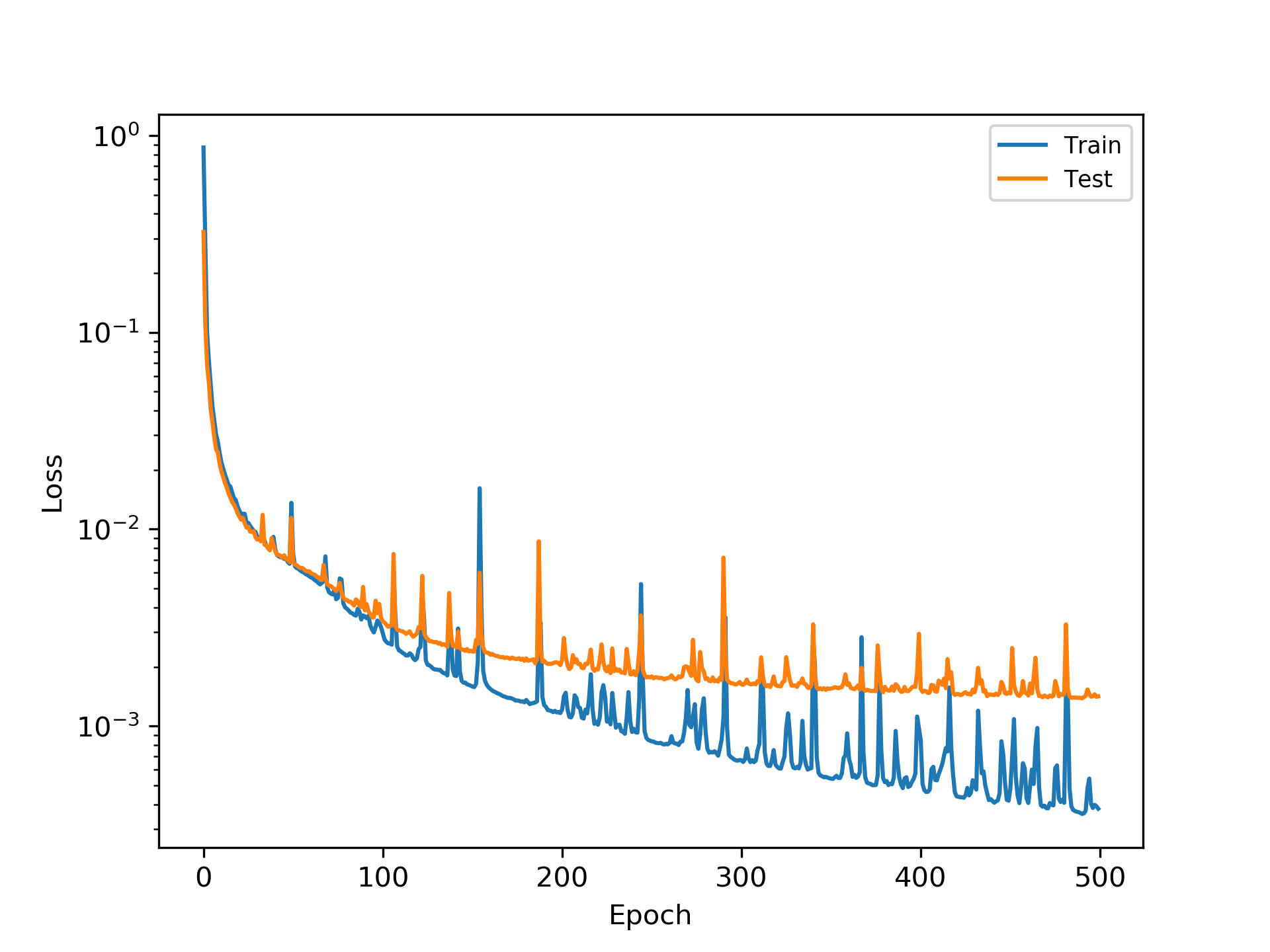}
        \caption{Transient ship airwake: convergence history of CAE}\label{fig airwake cae hist}
    \end{minipage}\hfill
    \begin{minipage}{0.47\textwidth}
	\centering
    	\includegraphics[width=1\textwidth]{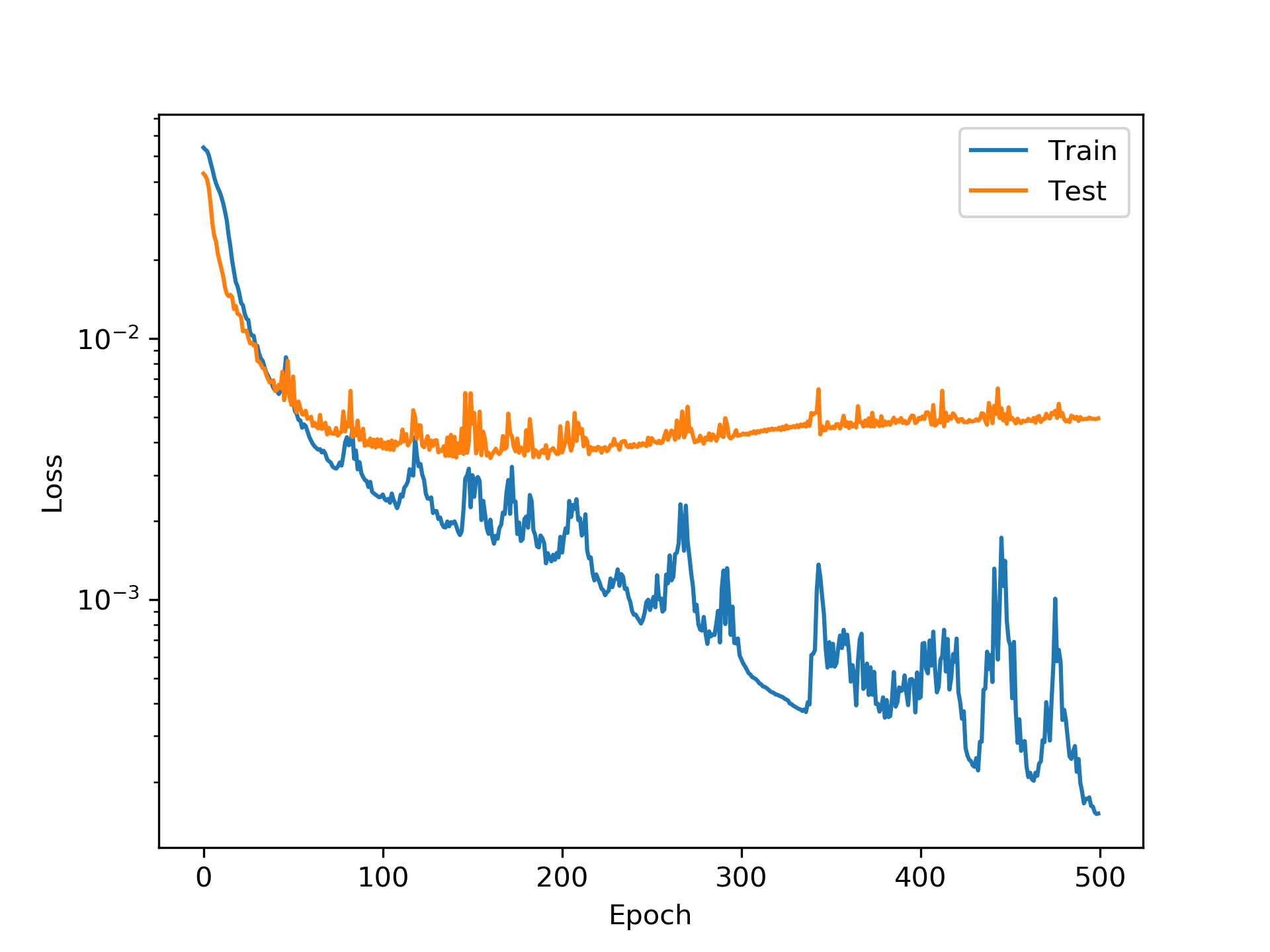}
        \caption{Transient ship airwake: convergence history of TCAE}\label{fig airwake tcae hist}
    \end{minipage}
\end{figure}

\begin{figure}
	\centering
	\subfloat{
	\includegraphics[width=0.47\textwidth]{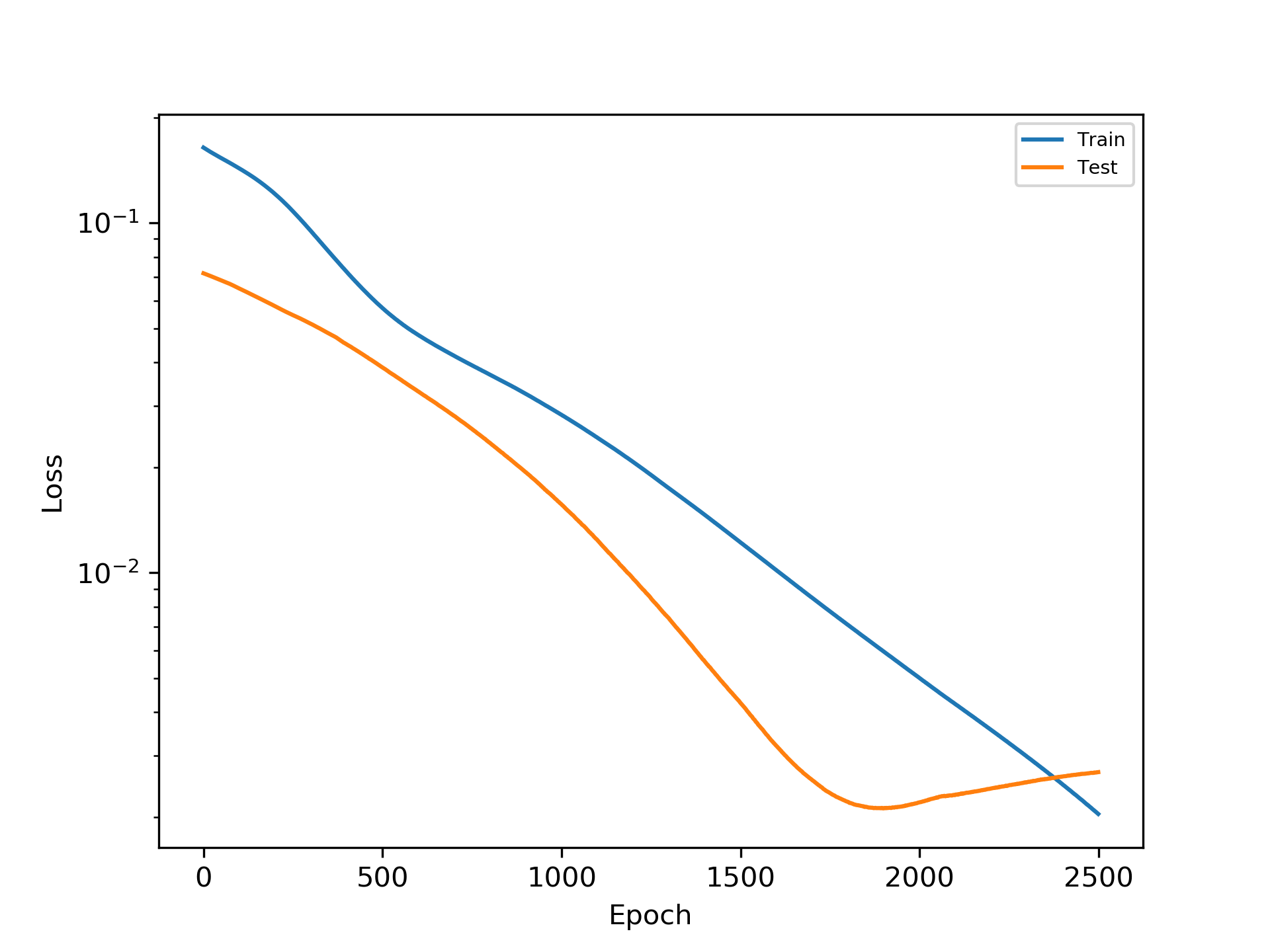}}
    \caption{Transient ship airwake: convergence history of MLP}\label{fig airwake MLP hist}
\end{figure}

Sample convergence histories in the training of different levels of the framework are presented in Fig.~\ref{fig sod cae hist} to \ref{fig airwake MLP hist}. 
It should be noted that the loss functions are evaluated based on scaled inputs to the networks, thus their values are not to be compared with the MSE given in Sec.~\ref{sec tests}. 

It is observed that the gap between the final training and testing loss increases with the departure of  the dynamics  between different parameters or time periods. In the discontinuous compressible flow case, the testing loss follows the training loss closely due to the similarity in local dynamics. In contrast, the testing loss saturates noticeably earlier in the other two cases. This is especially clear in the ship airwake TCAE (Fig.~\ref{fig airwake tcae hist}) due to the significant differences in the patterns of dynamics. 

Despite the slower decay in testing error, all errors are below \num{1e-2} in our tests. In most cases, more than 4 orders of convergence in the training loss is achieved.


\end{document}